\documentclass[11pt]{article}
\newcommand{\anon}{1}
\usepackage[utf8]{inputenc}
\usepackage[T1]{fontenc}
\usepackage{fullpage,amsmath,amssymb,bm,url,mathrsfs}
\usepackage{algorithmic}
\usepackage[ruled,linesnumbered]{algorithm2e}
\usepackage{amsthm}
\usepackage{hyperref}
\usepackage{makecell}
\usepackage{tablefootnote}
\usepackage{comment}
\usepackage{soul}
\usepackage{graphics,graphicx}
\usepackage{subcaption,caption,cleveref}
\hypersetup{
    colorlinks=true,
    linkcolor=blue,
    citecolor=blue,
    filecolor=magenta,      
    urlcolor=cyan,
}

\usepackage[square,sort,comma,numbers]{natbib}
\usepackage{xcolor}
\usepackage{dsfont}
\usepackage{bbm}
\usepackage{enumitem}
\usepackage{blkarray}
\usepackage{amsfonts}
\usepackage[normalem]{ulem}
\usepackage{amssymb}

\makeatletter
\def\blfootnote{\gdef\@thefnmark{}\@footnotetext}

\let\emptyset\varnothing

\def\frakD{{\mathfrak D}}

\def\calE{{\mathcal E}}

\def\calG{{\mathcal G}}

\def\calM{{\mathcal M}}
\def\calN{{\mathcal N}}

\def\calR{{\mathcal R}}
\def\calS{{\mathcal S}}
\def\calT{{\mathcal T}}

\def\calZ{{\mathcal Z}}

\def\EE{{\mathbb E}}

\def\II{{\mathbb I}}

\def\PP{{\mathbb P}}

\def\RR{{\mathbb R}}

\def\a{{\boldsymbol a}}

\def\\alpha{{\boldsymbol \alpha}}

\def\x{{\boldsymbol x}}

\def\z{{\boldsymbol z}}
\def\bmu{{\boldsymbol \mu}}

\DeclareMathOperator*{\argmin}{arg\,min}

\def\calE{{\cal  E}} 
 
\def\calG{{\cal  G}}

\def\calM{{\cal  M}} 
\def\calN{{\cal  N}}

\def\calR{{\cal  R}} 
\def\calS{{\cal  S}} 
\def\calT{{\cal  T}}

\def\calZ{{\cal  Z}}

\newcommand{\bfm}[1]{\ensuremath{\mathbf{#1}}}

   \def\bB{\bfm B}

   \def\bE{\bfm E}  \def\EE{\mathbb{E}}

   \def\bI{\bfm I}  \def\II{\mathbb{I}}

   \def\bP{\bfm P}  \def\PP{\mathbb{P}}
     
     \def\RR{\mathbb{R}}

   \def\bX{\bfm X}  
     
   \def\bZ{\bfm Z}  
                   
  \def\bTheta{\bfm \Theta}

\def\bSigma{\bfm \Sigma}

\def\SNR{\textsf{SNR}}

\def\hat{\widehat}
\def\wt{\widetilde}

\newcommand\inp[2]{\left\langle #1, #2 \right\rangle}

\newcommand\op[1]{\left\|#1\right\|}

\newcommand\brac[1]{\left(#1\right)}
\newcommand\bbrac[1]{\big(#1\big)}
\newcommand\sqbrac[1]{\left[#1\right]}
\newcommand\ebrac[1]{\left\{#1\right\}}
\newcommand\ab[1]{\left|#1\right|}

\SetKwInput{KwData}{Input}
\SetKwInput{KwResult}{Output}

\def\TC{\textsf{TC}}
\def\DP{\textsf{DP}}
\def\ITL{\textsf{ITL}}
\def\ATC{\textsf{ATC}}
\newtheorem{theorem}{Theorem}

\newtheorem{proposition}{Proposition}
\newtheorem{remark}{Remark}

\newtheorem{assumption}{Assumption}
\newtheorem{lemma}{Lemma}
\newtheorem{corollary}{Corollary}

\theoremstyle{remark}
\newtheorem{exmp}{Example}[section]

\newcommand{\Prob}{\mathbb{P}}

\newcommand{\cM}{\mathcal{M}}

\newcommand{\rd}{\mathrm{d}}
\newcommand{\sgn}{\mathrm{sgn}}

\begin{document}

\def\spacingset#1{\renewcommand{\baselinestretch}%
{#1}\small\normalsize} \spacingset{1}


\if1\anon
{
  \title{\bf Adaptive Transfer Clustering: A Unified Framework}\blfootnote{Author names are sorted alphabetically.}
\author{Yuqi Gu\thanks{Department of Statistics and Data Science Institute, Columbia University. Email: \texttt{yuqi.gu@columbia.edu}.}
\and Zhongyuan Lyu\thanks{Discipline of Business Analytics, University of Sydney Business School. Email: \texttt{zhongyuan.lyu@sydney.edu.au}.}
\and Kaizheng Wang\thanks{Department of IEOR and Data Science Institute, Columbia University. Email: \texttt{kaizheng.wang@columbia.edu}.}
}
  \maketitle
} \fi

\if0\anon
{
  \bigskip
  \bigskip
  \bigskip
  \begin{center}
    {\LARGE\bf Adaptive Transfer Clustering: A Unified Framework}
\end{center}
  \medskip
} \fi

\begin{abstract}

We propose a general transfer learning framework for clustering given a main dataset and an auxiliary one about the same subjects. The two datasets may reflect similar but different latent grouping structures of the subjects. We propose an adaptive transfer clustering (\ATC) algorithm that automatically leverages the commonality in the presence of unknown discrepancy, by optimizing an estimated bias-variance decomposition. It applies to a broad class of statistical models including Gaussian mixture models, stochastic block models, and latent class models. A theoretical analysis proves the optimality of \ATC~under the Gaussian mixture model and explicitly quantifies the benefit of transfer. Extensive simulations and real data experiments confirm our method's effectiveness in various scenarios.
\end{abstract}


\section{Introduction}
In recent years, data collection from multiple sources or views, each offering unique insights into the underlying structure, has become increasingly common. As a result,  transfer learning has gained prominence as a powerful framework in machine learning \citep{pan2009survey}.  While transfer learning has been widely applied to supervised settings \citep{reeve2021adaptive} and semi-supervised settings \citep{chattopadhyay2012multisource, li2013learning, zuo2018fuzzy}, its application to unsupervised settings still remains at the infant stage. In this background, \textit{clustering} has emerged as a central unsupervised task with applications in 
 medical imaging \citep{paul2020random,jiang2020novel, xu2024probabilistic}, genetics \citep{gabasova2017clusternomics, MSD24, wang2024transfer, shen2009integrative}, sentiment analysis \citep{zhang2016transfer, dai2021unsupervised}, to name just a few.

Existing transfer learning approaches for clustering primarily focus on the feature space, where the target domain  and source domain are different but related through shared or similar parameters. Gaussian Mixture Model (GMM) \citep{pearson1894contributions} is one of the most canonical models for transfer clustering, which itself plays a fundamental role in statistics.  A line of works \citep{wang2019transfer, TWF22}  on GMMs  apply EM-type algorithms that leverage similar mean or covariance structures in source domain to improve the clustering performance of target data. Specifically, \cite{TWF22} provides  theoretical guarantees by assuming closeness of discriminant coefficients in Euclidean distance.

Nonetheless, these works typically assume  similar parameter structures in the source and the target domains, which contain different subjects.  Little attention has been given to the cases where the target and the source data consist of different aspects of the same subjects, which have similar but different labels and might even follow different models.
Such problems naturally arise in many practical settings. For instance, in social science we may be interested in clustering a group of people based on their friendship network and attributes (e.g., age, education, occupation). However, the network and attributes often exhibit different signal strengths and may not reflect the same community structure.  Another example  in neuroimaging, as noted by \cite{paul2020random}, is that  the community structures within brain networks may differ between subjects due to their physiological differences, responses to changing environmental conditions, and variations in imaging instruments. A related work \citep{chen2022global}  investigates   the fundamental limits of  community detection in multi-layer networks under label flipping for individual networks. However, their method is not adaptive to an unknown flipping probability. 
There is a crucial need for procedures that can adaptively leverage useful information from the source domain to assist clustering in the target domain.

\paragraph*{Main contributions}  Motivated by the above need,  we propose a general framework for adaptive transfer clustering. Our target data $\bX_0$ and source data $\bX_1$ reflect different features of the same set of $n$ subjects, where each dataset has $K$ latent groups. In this framework, $\bX_0$ and $\bX_1$ may come from different mixture models, where a discrepancy parameter $\varepsilon$ indicates the proportion of mismatched latent labels $\bZ^*_0,\bZ^*_1\in[K]^n$. When $\varepsilon$ is small, the two grouping structures are similar but different. Our goal is to estimate $\bZ_0^*$ from $(\bX_0,\bX_1)$. The key challenge is how to adaptively leverage the information in $\bX_1$ to cluster $\bX_0$, without knowing the discrepancy $\varepsilon$. Intuitively, when $\varepsilon=0$, i.e., the labels are perfectly matched, we should ``pool'' the data together; when $\varepsilon$ is large enough, i.e., little label information about $\bX_0$ can be inferred from $\bX_1$, we may discard $\bX_1$ and perform clustering on the target data $\bX_0$ only. Inspired by this, we develop a model-based method with a penalty  term encouraging the similarity of  $\bZ^*_0$ and $\bZ^*_1$.  Informally, we try to minimize the following objective with $\lambda>0$:
\begin{align}\label{eq:model-intro}
-\text{log posterier of }(\bZ_0\mid \bX_0)-\text{log posterier of }(\bZ_1\mid \bX_1)+\lambda\cdot \text{penalty}\brac{\bZ_{0},\bZ_1}.
\end{align}
To summarize, our contributions are two-fold: 
\begin{itemize}[leftmargin=*]
    \item Methodology: Our framework is general to incorporate arbitrary mixture distributions of target and source data, making it applicable to a broad class of  models including Gaussian mixture models, latent class models, and  contextual stochastic block models. We design an  adaptive procedure, named \textit{Adaptive Transfer Clustering} or $\ATC$ for short, that can select the crucial parameter $\lambda$ in \eqref{eq:model-intro} agonistic to the level of discrepancy $\varepsilon$. The key ingredient of achieving  adaptivity is a combination of Goldenshluger-Lepski method \citep{goldenshluger2008universal}  and parametric bootstrap. 
    \item Theory: We establish a sharp clustering error rate for \ATC~in cases where both the target and source data are generated from two-component $d$-dimensional GMMs. Denote by $\textsf{SNR}$  the signal-to-noise ratio  such that the best estimator using  only the target data  achieves a rate of  $\exp\brac{-\textsf{SNR}\brac{1+o(1)}}$ as $\textsf{SNR}\rightarrow\infty$ \citep{loffler2019optimality, lu2016statistical,gao2022iterative}. We show that the optimal rate  in the transfer learning setting is
\begin{align*}
    \exp\brac{-\textsf{SNR}\min\ebrac{1+\frac{\log\brac{1/\varepsilon}}{4\textsf{SNR}},2}\brac{1+o(1)}},
\end{align*}
and it can be achieved by our {\ATC}  procedure without knowing $\varepsilon$. This rate is always better than the aforementioned target-only rate.
\end{itemize}

\paragraph*{Related work} 
Our work is broadly related to the emerging area of unsupervised learning in the transfer learning setting \citep{peng2021integration, MSD24,DMa24, jalan2024transfer}. Below, we outline a few related topics with a non-exhaustive review of relevant literature.
\begin{itemize}[leftmargin=*]
    \item \textbf{Multi-view clustering:}  Our work connects to the paradigm of multi-view clustering, which aims to cluster multiple datasets  representing different aspects (views) of the same subjects \citep{bickel2004multi}. Existing works provide 
    discriminative approaches that optimize carefully designed objective functions to maximize intra-cluster similarity and minimize inter-cluster similarity \citep{kumar2011co, wang2021integrative}, or generative approaches based on EM-type algorithms or Bayesian framework \citep{lock2013bayesian, kirk2012bayesian, gao2022testing2, gao2020clusterings}. See \cite{chao2021survey} for a survey.  We will discuss  comparisons   in detail in \Cref{sec:method}.
    \item \textbf{Network community detection with side information:} Our work is  related to network community detection literature involving multi-layer networks \citep{ma2023community, chen2022global, huang2023spectral, cohen2024multi, jing2021community, gollini2016joint} and  contextual networks \citep{yan2021covariate, binkiewicz2017covariate, mele2019spectral, zhang2016community, weng2022community, ma2017exploration, abbe2022, xu2023covariate, james2024learning, braun2022iterative, lu2023contextual, deshpande2018contextual, hu2024network}.  In the multi-layer network setting, recent works such as \cite{de2024mixture, cohen2024multi} propose multi-view stochastic block models that assume homogeneous label structures across views.   \cite{de2024mixture} uses a variational Bayesian EM algorithm for parameter estimation and clustering. \cite{cohen2024multi} provides  algorithmic guarantees for weak recovery and exact recovery for a special case when each network only consists of two clusters. In the contextual network setting, \cite{braun2022iterative} derives an exponential error for Lloyd-type algorithm for contextual stochastic block model (SBM) by assuming same clustering structures. \cite{james2024learning} incorporates joint and individual structures in both the network and the covariates, and proposes to use a spectral method followed by a refinement step. 
 
    \item \textbf{Testing for clustering structures:} There is also a growing body of theoretical and applied work investigating the testing problems for the common structures across different sources or views \citep{gao2022testing2, gao2020clusterings, gao2022testing, yuan2024testing, MSD24, simpson2013permutation, fujita2014non}. A closely related  work is  \cite{gao2022testing}, which proposes to test whether two datasets from GMMs share a common clustering structure and derives a sharp detection boundary. However, for the task of clustering, it remains unknown whether a naÃƒÂ¯ve dichotomous strategy, based on their testing procedure, is optimal. See \Cref{sec:warm-up} for discussion on difference between clustering (label estimation) and testing.
\end{itemize}

\paragraph*{Organization} \Cref{sec:warm-up} introduces the two-component symmetric univariate Gaussian mixture model as a warm-up example and presents the adaptive transfer clustering algorithm together with theoretical guarantees. \Cref{sec:method} extends the methodology to a general framework applicable to other models. \Cref{sec:theory} provides theoretical analysis within the general framework, with an application  to the two-component symmetric $d$-dimensional Gaussian mixture model fully adaptive to unknown parameters.  \Cref{sec:realdata} demonstrates the application of the our method to three real-world datasets. \Cref{sec:disc} concludes the paper and discusses future directions. Extensive simulations validates the effectiveness of the proposed algorithm and all proofs of theoretical results are included in the Appendix.

\paragraph*{Notation} Throughout the paper, the constants $c_0,C_0,c_1,C_1,\cdots$ may vary from line to line.  We  use plain letters $x,z,X,Z,\cdots$ to denote scalers and  boldface $\x,\z, \bX,\bZ,\cdots $ to denote either vectors or matrices. We write $[n]:=\ebrac{1,2,\cdots,n}$ and $\text{sgn}\brac{x}:=x/\ab{x}$ for $x\in\RR$, or $\text{sgn}\brac{\x}:=\brac{\text{sgn}\brac{x_1},\cdots,\text{sgn}\brac{x_n}}^\top $ for $\x\in\RR^n$. We denote the $\ell_2$ vector norm as $\op{\cdot}:=\op{\cdot }_{2}$. For nonnegative sequences $a_n$ and $b_n$, we write $a_n\lesssim b_n$ or $b_n\gtrsim a_n$ or $a_n=O(b_n)$ if there exists a universal constant $C>0$ such that $a_n\le Cb_n$. We write $a_n\asymp b_n$ if both $a_n\lesssim b_n$ and $b_n\lesssim a_n$, and  $a_n=o\brac{b_n}$ or $b_n=\omega\brac{a_n}$ if $a_n=O\brac{c_nb_n}$ for some $c_n\rightarrow0$. In most cases, we omit the subscript of $n$ when it is clear from context.

\section{Warm-up: Transfer Learning for GMM}\label{sec:warm-up}

In this section, we consider the transfer learning problem for one-dimensional, two-component symmetric Gaussian mixture model as a warm-up example.

\subsection{Problem Setup}

Let $\{X_{0,i}\}_{i=1}^n \subseteq \RR$ be i.i.d.~samples from a one-dimesional two-component symmetric Gaussian mixture model
\begin{align*}
X_{0, i} \sim \frac{1}{2} N( \mu , \sigma^2 ) + \frac{1}{2} N(-  \mu , \sigma^2 )
\end{align*}
with parameters $\mu \in \RR$ and $\sigma > 0$. Assume $\mu \geq 0$ for the sake of identifiability. There exist i.i.d.~Rademacher latent variables $\{ Z_{0, i}^* \}_{i=1}^n \subseteq \{ \pm 1 \}$ such that $X_{0, i} | Z_{0, i}^* \sim N( Z_{0, i}^* \mu , \sigma^2 )$. The goal of \emph{clustering} is to recover $\bZ_0^* = ( Z_{0, 1}^* , \cdots , Z_{0, n}^*  )$ from $\bX_0 = ( X_{0, 1} , \cdots , X_{0, n}  )$. 

Define the normalized Hamming distance between any pair of label vectors:
\begin{align*}
\ell ( \bZ,  \bZ^\prime ) = \frac{1}{n}\sum_{i=1}^n\II \brac{Z_i\ne  Z^\prime_i} , \qquad 
\forall \bZ,  \bZ^\prime \in \{ \pm 1 \}^n.
\end{align*}
For any estimate $\hat\bZ$ of $\bZ_0^*$, the quantity $\ell ( \hat\bZ, \bZ_0^* ) $ is often referred to as the  misclassification 
rate. It is easily seen that $\sgn ( \bX_{0} )$ is a maximum a posteriori (MAP) estimate, which achieves the minimum expected misclassification rate $\Phi\brac{-\mu/\sigma}$, where $\Phi(\cdot)$ is the c.d.f. of standard normal distribution. Define $\textsf{SNR}:=\mu^2/\brac{2\sigma^2}$. For our interest, we consider the diverging signal-to-noise  ratio regime where  $\textsf{SNR}\rightarrow\infty$. On the other hand, we consider $\textsf{SNR}<\log n$ as otherwise $\sgn ( \bX_{0} )$ would lead to exact clustering, a less interesting case. This can be seen from the following relations  which holds for $\textsf{SNR}>\log n$ and large $n$:
$$\EE\ell\brac{\sgn ( \bX_{0} ),\bZ^*}= \Phi\brac{-\sqrt{2\textsf{SNR}}}\le \exp\brac{-\textsf{SNR}}<\frac{1}{n}.$$
The expected number of misclassified samples is less than $1$, indicating exact clustering with high probability. 

Suppose we also collect another batch of i.i.d.~samples $\ebrac{X_{1,i}}_{i=1}^n$ from the mixture distribution $\frac{1}{2} N( \mu , \sigma^2 ) + \frac{1}{2} N(-  \mu , \sigma^2 )$, whose latent variables $\{ Z_{1, i}^* \}_{i=1}^n $ are possibly similar to $\{ Z_{0, i}^* \}_{i=1}^n$ in the following sense:
\begin{align}
\PP ( Z_{1, i}^* \neq Z_{0, i}^* ) \leq \varepsilon .
\end{align}
Here $\varepsilon\in[0,1/2]$ is an unknown small number that controls the discrepancy\footnote{It worth noting that when $\varepsilon> 1/2$, this problem can be reformulated as considering $\ebrac{X_{0,i}}_{i=1}^n$ and $\ebrac{-X_{1,i}}_{i=1}^n$ with discrepancy parameter $\varepsilon^\prime=1-\varepsilon$, hence it suffices to restrict ourselves to $\varepsilon\in[0,1/2]$.}. We refer to $\bX_0$ and $\bX_1 = ( X_{1, 1} , \cdots , X_{1, n}  )$ as the \emph{target} and the \emph{source} data, respectively. With the aid of source data, the goal of \emph{transfer clustering} is to recover $\bZ_0^*$ from $(\bX_0,\bX_1)$. When $\varepsilon=0$, the target data and source data share the same labels, whereas when $\varepsilon=1/2$, the clustering structures between the target data and source data, in the worst scenario, are uninformative. For ease of representation, we assume $(\mu,\sigma)$ are known throughout this warm-up section, and our general methodology (Section~\ref{sec:method}) and theoretical results (Section~\ref{sec:theory}) hold without this assumption. The following are two common estimation strategies  which we call \emph{independent task learning} ({\ITL}) and  \emph{data  pooling} ({\DP}), respectively.

\begin{exmp}[Independent task learning]
    If we only have access to target data $\bX_0$, the optimal procedure for estimating $\bZ^*_{0}$ is given by $\check \bZ^{\ITL}_{0}$, where 
\begin{align*}
    \check Z^{\ITL}_{0,i}&\in  \argmin_{u\in\{\pm1\}}\left \{\frac{(X_{0,i}-u\mu)^2}{\sigma^2}\right \}=\text{sgn}\brac{X_{0,i}},\quad  \forall i\in[n].
\end{align*}
\end{exmp}
\begin{exmp}[Data pooling]
     If we ignore the potential label discrepancy between target data  $\bX_0$ and source data  $\bX_1$, we can estimate $\bZ^*_0$ as if $(\bX_0,\bX_1)$ is from a $2$-dimensional symmetric Gaussian mixture via $\check\bZ_0^{\DP}$, where 
\begin{align*}
\check Z^{\DP}_{0,i}&\in  \argmin_{u\in\{\pm1\}}\left \{\frac{(X_{0,i}-u\mu)^2}{\sigma^2}+\frac{(X_{1,i}-u\mu)^2}{\sigma^2}\right \}=\text{sgn}\brac{X_{0,i}+X_{1,i}},\quad  \forall i\in[n].
\end{align*}
\end{exmp}
We could obtain the following expected misclassification rate for $\check\bZ_0^{\ITL}$ and $\check\bZ_0^{\DP}$, whose proof is deferred to Appendix \ref{sec-prop:ind-dp-gmm-proof}.

\begin{proposition}\label{prop:ind-dp-gmm} 
For each $i\in[n]$, we have
\[
\PP \bbrac{\check Z^{\ITL}_{0,i}\ne  Z^*_{0,i}} = \Phi \brac{-\frac{\mu}{\sigma}}
\qquad\text{and}\qquad
0 \leq \PP\bbrac{\check Z^{\DP}_{0,i}\ne  Z^*_{0,i}} - \Phi\brac{-\frac{\sqrt{2}\mu}{\sigma}} \leq \frac{\varepsilon}{2}.
\]
Consequently, we have
\begin{align*}
    \PP \bbrac{\check Z^{\ITL}_{0,i}\ne  Z^*_{0,i}}\lesssim\exp\brac{-\textsf{SNR}}\qquad\text{and}\qquad  \PP\bbrac{\check Z^{\DP}_{0,i}\ne  Z^*_{0,i}} \lesssim \exp\brac{-2\cdot\textsf{SNR}}+\varepsilon.
\end{align*}
\end{proposition}

A natural question is whether we can do better than $\check\bZ_0^{\ITL}$ and $\check\bZ_0^{\DP}$ in the general case. Next, we propose a transfer clustering procedure to answer this question affirmatively.

\subsection{Transfer Learning Procedure}\label{subsec:tl-procedure}

Consider a special case where $\PP ( Z_{1,i}^* \neq Z_{0,i}^* | Z_{0,i}^* ) = \varepsilon$. If $\varepsilon$ is known, then the joint MAP of $( Z_{0,i}^* , Z_{1,i}^*)$ is given by 
\begin{align}\label{eq:MAP}
 \argmin_{ ( u,v ) \in\{\pm1\}^2 }\left \{\frac{(X_{0,i}-u\mu)^2}{2\sigma^2}+\frac{(X_{1,i}-v\mu)^2}{2\sigma^2}
+ \log \bigg(
\frac{1 - \varepsilon}{\varepsilon}
\bigg) 
 \II\brac{u\ne v}\right \}.
\end{align}
The term $\log\brac{\varepsilon^{-1}\brac{1-\varepsilon}}\II\brac{u\ne v}$, which involves unknown parameter $\varepsilon$, serves as a penalty term. Motivated by \eqref{eq:MAP}, we  consider the following procedure to leverage the information from source data. For any $\lambda>0$, define 
\begin{align}\label{eq:optimize-prob}
    \brac{\check Z^\lambda_{0,i},\check Z^\lambda_{1,i}}&\in \argmin_{u,v\in\{\pm1\}}\left \{\frac{(X_{0,i}-u\mu)^2}{2\sigma^2}+\frac{(X_{1,i}-v\mu)^2}{2\sigma^2}+\lambda\II\brac{u\ne v}\right \}\notag\\
    &=\argmin_{u,v\in\{\pm1\}}\left \{-\mu \brac{u X_{0,i}+vX_{1,i}}+\lambda\sigma^2\II\brac{u\ne v}\right \},\quad  \forall i\in[n]
\end{align}
We denote $(\check\bZ^\lambda_0,\check\bZ^\lambda_1)=\TC\brac{\bX_0,\bX_1;\lambda,\mu,\sigma}$ for the procedure defined by \eqref{eq:optimize-prob}. Notice that $\check \bZ^\lambda_{0}$ reduces to the independent task learning estimator when $\lambda=0$ and the data pooling estimator when $\lambda=\infty$, i.e., $\check \bZ^0_{0}=\check \bZ^{\ITL}_{0}$ and $\check \bZ^\infty_{0}=\check \bZ^{\DP}_{0}$. As such, $\check \bZ^\lambda_{0}$ can be viewed as a ``soft'' version of data pooling between $\check \bZ^{\ITL}_{0}$ and $\check \bZ^{\DP}_{0}$. The following theorem characterizes the performance of $\check \bZ^\lambda_{0}$, whose proof is deferred to \Cref{sec-prop:mis-error-lam-proof}.

\begin{theorem}\label{thm:mis-error-lam}
For $s > 0$, $\varepsilon \in [0, 1/2]$ and $\lambda \geq 0$, define
\begin{align*}
    \cM ( s , \varepsilon, \lambda ):= \Phi ( - \sqrt{2} s ) +
\Phi \bigg(
-s-\frac{\lambda}{2s}
\bigg)
\Phi
\bigg(
s - \frac{\lambda}{2 s }
\bigg)
+ 2 \varepsilon \Phi \bigg(
- s +\frac{\lambda}{2 s }
\bigg).
\end{align*}
The estimate $\check Z^\lambda_{0,i}$ in \eqref{eq:optimize-prob} satisfies
\[
\PP ( \check Z^{\lambda}_{0,i}\ne  Z^*_{0,i} )
\leq \cM \brac{{\mu}/{\sigma} , \varepsilon, \lambda } , \qquad \forall i \in [n].
\]
\end{theorem}

When $\varepsilon=0$, labels are perfectly matched and the error is purely caused by noise. In this case, the error bound given by \Cref{thm:mis-error-lam} is $\psi\brac{\lambda}:=\cM \brac{{\mu}/{\sigma} , 0, \lambda }$. When $\varepsilon>0$, the label discrepancy leads to an additional term $\phi\brac{\lambda}:=2\varepsilon\Phi\brac{-\mu/\sigma+{\lambda}\brac{2\mu/\sigma}^{-1}}$. Therefore, $\psi\brac{\lambda}$ and $\phi\brac{\lambda}$ can be interpreted as stochastic error and systematic error, respectively.  Note that  $\psi\brac{\lambda}$ is non-increasing with $\lambda$ and $\phi\brac{\lambda}$ is non-decreasing with $\lambda$, and the relation $\cM \brac{{\mu}/{\sigma} , \varepsilon, \lambda }=\psi\brac{\lambda}+\phi\brac{\lambda}$ can be viewed as the bias-variance decomposition. In \Cref{thm:gmm-lb} below, we will show that  choosing $\lambda=\log\brac{\varepsilon^{-1}\brac{1-\varepsilon}}$ as in \eqref{eq:MAP} optimizes $\calM\brac{\mu/\sigma,\varepsilon,\lambda}$ up to a constant factor, and achieve the bias-variance trade-off. Formally, we have the following theorem regarding $\check\bZ_0^\lambda$, whose proof can be found in Appendix \ref{pf-thm:oracle-gmm}.

\begin{theorem}\label{thm:oracle-gmm}
    Define $\alpha:={\log\brac{1/\varepsilon}}/\brac{4\textsf{SNR}}$,
then there exists some universal constant $C_0>0$ such that 
     \begin{align*}
        \inf_{\lambda\ge 0}\EE\ell\brac{\check \bZ^{\lambda}_0,\bZ^*_0}\le C_0\exp\brac{-\textsf{SNR}\cdot \min\ebrac{\brac{1+\alpha}^2, 2}}.
     \end{align*}

\end{theorem}
By \Cref{thm:oracle-gmm}, we can see that $\alpha$ characterizes the  informativeness  of  source data. Denote by $\lambda^*$ the quantity such that $\EE\ell\brac{\check \bZ^{\lambda^*}_0,\bZ^*_0}\asymp\inf_{\lambda\ge 0}\EE\ell\brac{\check \bZ^{\lambda}_0,\bZ^*_0}$.  In view of \Cref{prop:ind-dp-gmm}, we have 
\begin{align*}
    \EE\ell\brac{\check \bZ^{\ITL}_0,\bZ^*_0}\lesssim \exp\brac{-\textsf{SNR}}\qquad\text{and}\qquad\EE\ell\brac{\check \bZ^{\DP}_0,\bZ^*_0}\lesssim \exp\brac{-\textsf{SNR}\cdot \min\ebrac{4\alpha, 2}}.
\end{align*}
We can thereby compare the performance of $\check\bZ^{\lambda^*}$ with the $\ITL$ estimator $\check\bZ^{\ITL}$ and the DP estimator $\check\bZ^{\DP}$, summarized in \Cref{tab:Comparison}. For simplicity, we consider  $\textsf{SNR}=o\brac{\log n}$.
\begin{itemize}[leftmargin=*]
    \item $\check \bZ^{\lambda^*}_{0}$  beats $\check \bZ^{\ITL}_{0}$ for $\alpha\in(0,\infty]$; $\check \bZ^{\lambda^*}_{0}$  beats $\check \bZ^{\DP}_{0}$ for $\alpha\in(0,1/2)$, and  performs as well as $\check \bZ^{\DP}_{0}$ for  $\alpha\in[1/2,\infty)$.
    \item When  $\alpha={\sqrt{2}-1}$, $\check \bZ^{\lambda^*}_{0}$ can achieve a rate of 
    $\exp\brac{-2\cdot \textsf{SNR}\brac{1+o(1)}}$, which is an ideal rate  when there is no mismatch ($\varepsilon=0$), while $\check \bZ^{\DP}_{0}$  can only achieve a rate of  $$\exp\brac{-4\brac{\sqrt{2}-1}\cdot \textsf{SNR}\brac{1+o(1)}}\approx \exp\brac{-1.66\cdot \textsf{SNR}\brac{1+o(1)}}.$$
\end{itemize}

\begin{table}[!tb]
\def\arraystretch{1.4}%
\centering
\begin{tabular}{c|c|c|c}
\hline
Range of $\varepsilon$   & $<e^{-2\textsf{SNR}}$ & $e^{-2\textsf{SNR}}\sim e^{-4\brac{\sqrt{2}-1}\textsf{SNR}}$ & $>e^{-4\brac{\sqrt{2}-1}\textsf{SNR}}$  \\ \hline
 Range of $\alpha$   & $[1/2,\infty)$ & $[\sqrt{2}-1,1/2)$ & $(0,\sqrt{2}-1)$ \\ \hline
\TC  &  \textbf{optimal as if} $\varepsilon=0$    &  \textbf{optimal as if} $\varepsilon=0$     &  \textbf{optimal}                            \\ \hline
\DP & \textbf{optimal as if} $\varepsilon=0$     &  sub-optimal     &  sub-optimal                            \\ \hline
\ITL &  sub-optimal     & sub-optimal     &   sub-optimal                           \\ \hline
\end{tabular}
 \caption{Comparison of \TC, \DP~and \ITL~under the regime  $\textsf{SNR}=o\brac{\log n}$. } 
    \label{tab:Comparison}
\end{table}


It is also informative to compare our transfer clustering procedure with the hypothesis testing framework studied in \citet{gao2022testing}. See \Cref{sec:testing} for a detailed discussion.

We certify the optimality of transfer learning procedure \eqref{eq:optimize-prob} by establishing a matching lower bound for estimating $\bZ_0^*$, whose proof is deferred to Appendix \ref{sec-thm:gmm-lb-proof}.

\begin{theorem}[Lower bound]\label{thm:gmm-lb}
Assume that $\PP ( Z^*_{0,i} \ne Z^*_{1,i} | Z^*_{0,i} ) = \varepsilon$. Let  $\cM$ be the function defined in \Cref{thm:mis-error-lam}. For any estimate $\hat\bZ_0$ of $\bZ_0^*$, we have
\[
\PP (  \hat Z_{0, i} \neq Z_{0, i}^* ) \geq \frac{1}{4}
\cM \brac{ \frac{\mu }{ \sigma} , \varepsilon, \log \brac{
\frac{1 - \varepsilon}{\varepsilon}
}}
 , \qquad \forall i \in [n].
\]
The right-hand side also lower bounds $\EE \ell (  \hat \bZ_{0} , \bZ_{0}^* )$.
\end{theorem}

\subsection{Adaptivity: Goldenshluger-Lepski Method}

In view of  \Cref{thm:mis-error-lam} and \Cref{thm:gmm-lb}, the optimal choice of $\lambda$ for our \TC~estimator should be $\log\brac{\varepsilon^{-1}\brac{1-\varepsilon}}$. The  key challenge we are going to tackle is the unknown $\varepsilon$.   
For clarity of presentation, we assume $\brac{\mu,\sigma}$ is given by an oracle. In principle, estimating $\brac{\mu, \sigma}$ is much easier than estimating $\varepsilon$.

A direct method is to first estimate $\varepsilon$ by $\hat \varepsilon:=\ell\brac{\check \bZ^{\ITL}_0,\check \bZ^{\ITL}_1}$ and then use  $\log\brac{\hat\varepsilon^{-1}\brac{1-\hat\varepsilon}}$ as a  plug-in estimator of $\log\brac{\hat\varepsilon^{-1}\brac{1-\hat\varepsilon}}$. However, the following lemma shows that the direct plug-in method may not work for $\varepsilon$ when $\varepsilon$ is very small\footnote{A possible approach for estimating small $\varepsilon$ for GMM is given by \cite{cai2007estimation}, which, however, heavily relies on the Gaussianity assumption in the model. Our method can be easily adapted to other distributions, see \Cref{alg:GL} and \Cref{alg:bootstrap-quantile}. }:
\begin{lemma}\label{lem:naive-eps} 
Assume that $\PP ( Z^*_{0,i} \ne Z^*_{1,i} | Z^*_{0,i} ) = \varepsilon=n^{-\beta}$ for $\beta\in(0,1)$. If 
\begin{align*}
    \beta>\min\ebrac{\frac{\textsf{SNR}}{\log n},~\frac{1}{2}\brac{\frac{\textsf{SNR}}{\log n}+1}},
\end{align*}
then we  have $\EE\sqbrac{\brac{\hat\varepsilon/\varepsilon}-1}^2\ge c_0$
for some universal constant $c_0>0$.
\end{lemma}
Consider the case when $\textsf{SNR}=o\brac{\log n}$, then \Cref{lem:naive-eps} tells us $\hat \varepsilon$ fails to consistently estimate $\varepsilon$ if $\varepsilon=o\brac{n^{-1/2}}$. Consequently, $\log\brac{\hat\varepsilon^{-1}\brac{1-\hat\varepsilon}}$ is a not consistent estimator of $\log\brac{\varepsilon\brac{1-\varepsilon}}$.
To bridge this gap, we develop a novel adaptive method that bypasses the estimation of $\varepsilon$ to choose $\lambda$, inspired by the Goldenshluger-Lepski method \citep{goldenshluger2008universal}. The high-level idea is outlined below.

 Consider a grid set $\Lambda=\ebrac{0=\lambda_1<\lambda_2<\cdots<\lambda_M=\log n}$. Recall  the bias-variance decomposition $\cM \brac{{\mu}/{\sigma} , \varepsilon, \lambda }=\psi\brac{\lambda}+\phi\brac{\lambda}$, the key to our adaptive procedure is to construct estimates of $\psi$ and $\phi$ on $\Lambda$. To estimate $\psi$, it boils down to characterizing the stochastic error purely incurred by the noise, i.e., $\calM\brac{\mu/\sigma,0,\lambda}$. Therefore, we need to construct samples from the distribution where $\varepsilon=0$. Note that we can write $\bX_0=\mu\bZ^*_0+\bE_0$ and $\bX_1=\mu\bZ^*_1+\bE_1$, where $\bE_0,\bE_1$ have i.i.d. $N(0,\sigma^2)$ entries. Suppose  we have access to  an imaginary sample  $\bar\bX_0=\mu \bZ^*_0+\bE_1$, and let $(\bar\bZ^\lambda_0,\bar\bZ^\lambda_1)=\TC\brac{\bX_0,\bar\bX_0;\lambda,\mu,\sigma}$ for $\lambda\in\Lambda\cup\ebrac{\infty}$. $\bar\bX_0$ is a ``hybrid'' sample of target and source, in the sense that it has the same label as $\bX_0$, while the same noise as $\bX_1$. A benchmark estimator should be $\bar\bZ^\infty_0$, as $\bX_0$ and $\bar\bX_0$ share the same labels and $\ell\brac{\bar\bZ^\infty_0,\bZ_0^*}$ is the best error rate we can achieve when $\varepsilon=0$. It suffices to characterize $\ell\brac{\check \bZ_0^\lambda,\bar\bZ^\infty_0}$ due to the triangle inequality. Moreover, we have
 \begin{align*}
    \ell\brac{\check \bZ_0^\lambda,\bar\bZ^\infty_0}\le \ell\brac{\bar \bZ_0^\lambda,\bar\bZ^\infty_0}+\ell\brac{\check \bZ_0^\lambda,\bar\bZ^\lambda_0}.
 \end{align*}
It can be shown that with high probability,  $\ell\brac{\bar \bZ_0^\lambda,\bar\bZ^\infty_0}\lesssim \psi\brac{\lambda}$ 
and  $\ell\brac{\check \bZ_0^\lambda,\bar\bZ^\lambda_0}\lesssim \phi\brac{\lambda}$. 
We can then simulate from the distribution of $\brac{\bX_0,\bar\bX_0}$,  generate bootstrap samples $\ebrac{\ell\brac{\bar \bZ^{\lambda,b}_{0}, \bar \bZ^{\infty,b}_{0}}, b\in[B]}$, and construct an estimator $\hat\psi\brac{\lambda}$ for $\psi\brac{\lambda}$ by its sample quantile. This procedure is detailed in Algorithm \ref{algo:bootstrap-gmm}. 

To estimate $\phi\brac{\lambda}$, a natural idea from  Goldenshluger-Lepski method \citep{goldenshluger2008universal} is to use the  quantity
\begin{align}\label{eq:bias-proxy}
    \max_{\lambda^\prime\in\Lambda,\lambda^\prime<\lambda}\sqbrac{\ell\brac{\check \bZ_0^{\lambda},\check \bZ_0^{\lambda^\prime}}-\hat\psi\brac{\lambda}-\hat\psi\brac{\lambda^\prime}}_+,
\end{align}
which serves as an estimate of $\phi\brac{\lambda}$ by eliminating two estimates of stochastic error. The aforementioned G-L method is a powerful tool for adaptive nonparametric estimation.  For our purposes, we can further simplify \eqref{eq:bias-proxy}  and  use the following estimator for $\phi\brac{\lambda}$:
\begin{align*}
\hat\phi\brac{\lambda}:=\max_{\lambda^\prime\in\Lambda,\lambda^\prime<\lambda}\sqbrac{\ell\brac{\check \bZ_0^{\lambda},\check \bZ_0^{\lambda^\prime}}-\hat\psi\brac{\lambda^\prime}}_+.
\end{align*} 
{\color{black}The maximum of pairwise comparisons, after subtracting a bootstrap majorant $\hat\psi(\cdot)$, serves as a data-driven proxy for the unknown bias $\phi(\cdot)$, following the adaptation of GoldenshlugerÃ¢â‚¬â€œLepski in \cite{page2021goldenshluger}.}
The final choice of $\lambda$ is given by  $\hat\lambda\in \argmin_{\lambda\in\Lambda}\ebrac{\hat\phi\brac{\lambda}+\hat\psi\brac{\lambda}}.$
This leads to our adaptive estimator
$\ATC\brac{\bX_0,\bX_1;\Lambda,\mu,\sigma}
=
\hat\bZ_0^{\hat\lambda}$,
where $\ATC$ is a shorthand for \underline{A}daptive \underline{T}ransfer \underline{C}lustering.
\begin{algorithm}[!tbp]
\footnotesize
\caption{Bootstrap-Quantile for $\psi$ in two-component GMM}\label{algo:bootstrap-gmm}
    \KwData{ Parameters $\mu,\sigma>0$,  $\zeta\in(0,1)$. }
    \For{$b=1,\cdots,B$}{
        Generate $Z^{*,b}_{0,i}\overset{i.i.d.}{\sim} \text{Unif}\ebrac{\pm 1}$ for $i\in[n]$.

Generate $X_{0,i}^{b}\overset{i.i.d.}{\sim}N\brac{Z^{*,b}_{0,i}\mu,\sigma^2}$ and $X_{1,i}^{b}\overset{i.i.d.}{\sim}N\brac{Z^{*,b}_{0,i}\mu,\sigma^2}$ for $i\in[n]$ independently.

For each $\lambda\in\Lambda\cup\ebrac{\infty}$, compute  $\big(\bar\bZ_0^{\lambda,b},\bar\bZ_1^{\lambda,b}\big):=\TC\brac{\bX^b_0,\bX^b_1;\lambda,\mu,\sigma}$.
    }
For each $\lambda\in\Lambda$, compute $\hat\psi\brac{\lambda}:=1.01\hat Q_{\zeta/2}^{\lambda}$, where  $\hat Q_{\zeta/2}^{\lambda}$ is $\brac{1-\zeta/2}$  {quantile} of $\ebrac{\ell\brac{\bar \bZ^{\lambda,b}_{0}, \bar \bZ^{\infty,b}_{0}}, b\in[B]}$.
\KwResult{$\ebrac{\hat\psi\brac{\lambda},\lambda\in\Lambda}$}
\end{algorithm}
The following theorem provides an error bound for $\ATC$, whose proof can be found in Appendix \ref{sec-thm:gl-gmm-proof}.

\begin{theorem}\label{thm:gl-gmm}
    Let $\zeta\in(0,1)$, and assume that $B\ge C_0\zeta^{-2}\log\brac{M\zeta^{-1}}$ for some universal constant $C_{0}>0$, then with probability at least $1-\zeta$, we have
    \begin{align*}
    \ell\brac{\check  \bZ_0^{\hat \lambda}, \bZ_0^*}\le C_{1}\sqbrac{\min_{\lambda\in\Lambda}\cM \brac{\frac{\mu}{\sigma} , \varepsilon, \lambda } +\frac{\log\brac{M\zeta^{-1}}}{n}}.
\end{align*}
for some universal constant $C_{1}>0$ determined  by  $C_0$.
\end{theorem}

Due to the smoothness of $\Phi$, it follows that  that $\ATC$ can achieve the optimal clustering error rate $\cM \left( {\mu}/{\sigma}, \varepsilon, \log \left({\varepsilon}^{-1}\brac{1 - \varepsilon} \right) \right)$, up to a negligible additive term ${\log(M\zeta^{-1})}/n$, provided that the grid in $\Lambda$ is sufficiently dense. Formally, we have the following result, whose proof can be found in Appendix \ref{sec-col:optimality-gmm-proof}.


\begin{corollary}\label{col:optimality-gmm}
    Suppose the conditions of \Cref{thm:gl-gmm} hold.  Let $M=\lceil \log^2 n \rceil$ and $\lambda_{j}=jM^{-1} \log n$ for $j=0,\cdots,M$, then there exists some universal constant $C_2>0$ determined  by $C_1$ such that  with probability at least $1-\zeta$,
        \begin{align*}
    \ell\brac{\check  \bZ_0^{\hat \lambda}, \bZ_0^*}\le C_{2}\sqbrac{\cM \brac{\frac{\mu}{\sigma} , \varepsilon, \log \brac{\frac{1 - \varepsilon}{\varepsilon}}} +\frac{\log\brac{M\zeta^{-1}}}{n}}.
\end{align*}
Moreover, if  $\textsf{SNR}<\brac{\frac{1}{2}-c}\log n$ for some constant $c\in(0,1)$, then there exists some universal constant $C_3>0$ determined  by  $C_2$ such that with probability at least $1-n^{-10}$,
    \begin{align}\label{eq:adap-upper-bound}
    \ell\brac{\check  \bZ_0^{\hat \lambda}, \bZ_0^*}\le C_{3}\cM \brac{\frac{\mu}{\sigma} , \varepsilon, \log \brac{\frac{1 - \varepsilon}{\varepsilon}}}.
\end{align}
\end{corollary}
\Cref{col:optimality-gmm} delivers that $\ATC$ estimator achieves the error rate that matches the lower bound in \Cref{thm:gmm-lb}.

\section{Methodology for General Mixture Models}\label{sec:method}
In this section, we present our general framework for transfer learning in mixture models. 
\subsection{Problem Setup}
In the two-component symmetric GMM, our data can be written as $\bX_0=\mu \bZ_0^*+\sigma\bE_0$ and $\bX_1=\mu \bZ_1^*+\sigma\bE_1$, where $\bE_0$ and $\bE_1$ are independent $N\brac{0,\bI_n}$. We now extend it  to a more general data generating mechanism. 
Consider two generative models $\calG_0$ and $\calG_1$ that generate the target data $X$ and source data $Y$ respectively:
\begin{align*}
    \bX_0 = \calG_0\brac{\bZ_0^*,\bTheta_0^*,\bE_0},\quad \bX_1 = \calG_1\brac{\bZ_1^*,\bTheta_1^*,\bE_1},
\end{align*}
where $\bZ_0^*,\bZ_1^*\in[K]^n$ are latent labels of our interest, $\bTheta_0^*,\bTheta_1^*$ are nuisance parameters, and $\bE_0,\bE_1$ are random seeds.  Our goal is to estimate $\bZ_0^*$, whose performance is measured by  
\begin{align*}
\frakD ( \bZ,  \bZ^\prime ): = \frac{1}{n}\sum_{i=1}^n\II \brac{Z_i\ne  Z^\prime_i} , \qquad 
\forall \bZ,  \bZ^\prime \in [K]^n.
\end{align*}
Denote by $L_m\brac{\bZ,\bX;\bTheta}$ the posterior density of either target data ($m=0$)  or source data ($m=1$). We list several illustrative examples on the generating mechanism of target and source data.

\medskip
\begin{exmp}[Gaussian mixture model, warm-up case in \Cref{sec:warm-up}]
    A primary example of theoretical interest is the two-component symmetric Gaussian mixture model  discussed in \Cref{sec:warm-up}. For $m\in\ebrac{0,1}$, we have $\bTheta_{m}^*=\brac{\mu,\sigma}$,  and $ \calG_m\brac{\bZ,\brac{\mu,\sigma},\bE}=\mu\bZ+\sigma\bE$.
      We thus have $L_0=L_1=L_{\textsf {GMM}_2}$ with
    \begin{align*}
        \log L_{\textsf {GMM}_2}\brac{\bZ,\bX;\brac{\mu,\sigma}}=-\sum_{i=1}^n\sqbrac{\frac{(X_{0,i}-\mu)^2}{2\sigma^2}+\log\brac{1/2}}.
    \end{align*}
\end{exmp}

\begin{exmp}[Gaussian mixture model, general case]
     \sloppy We can extend the warm-up case to the general scenario as follows.  In this case for $m\in\ebrac{0,1}$,  we have  $\bTheta_m^*=\brac{\ebrac{\pi_{m,k}}_{k=1}^K,\ebrac{\bmu_{m,k}}_{k=1}^K,\ebrac{\bSigma_{m,k}}_{k=1}^K}$,  $\bE_m\in\RR^{n\times d}$ has i.i.d. $N(0,1)$ entries, and $ \calG_m\brac{\bZ,\brac{\ebrac{\pi_{k}}_{k=1}^K,\ebrac{\bmu_{k}}_{k=1}^K,\ebrac{\bSigma_{k}}_{k=1}^K},\bE}\in\RR^{n\times d}$ with its $i$-th row being $\bmu_{Z_{i}}+\bSigma^{1/2}_{Z_{i}}\bE_{i,\cdot}^\top $.
      We thus have $L_0=L_1=L_{\textsf {GMM}_K}$ with
    \begin{align*}
        \log L_{\textsf {GMM}_K}\brac{\bZ,\bX;\brac{\ebrac{\pi_{k}}_{k=1}^K,\ebrac{\bmu_{k}}_{k=1}^K,\ebrac{\bSigma_{k}}_{k=1}^K}}:=\sum_{i=1}^n\sqbrac{\log \calN\brac{\bX_i\mid \bmu_{Z_i},\bSigma_{Z_i}}+\log\pi_{Z_i}},
    \end{align*}
    where $\calN\brac{\cdot\mid\bmu,\bSigma}$ is the multivariate normal density with mean $\bmu$ and covariance $\bSigma$. 
\end{exmp}

\begin{exmp}[Latent class model]
    Another example is the latent class model \cite[LCM;][]{goodman1974exploratory, lazarsfeld1950logical}, which is a popular mixture model for multivariate categorical/binary data, widely used in social sciences  \citep{zeng2023tensor}, genetics and genomics \citep{kiselev2019challenges}, and health sciences \citep{zhang2012latent}. Let  $m\in\ebrac{0,1}$. The data $\bX_m\in\ebrac{0,1}^{n\times d}$ consists of  $d$-dimensional binary response vectors of $n$ subjects. $\ebrac{\bX_{m,i}\in\ebrac{0,1}^d}_{i=1}^n$ are assumed to be  conditionally independent given the latent label vector $\bZ_m^*$. Let $\bP_{m} \in[0,1]^{d\times K}$ be the item parameter matrix,  we have $\PP\brac{X_{m,ij}=1\mid Z^*_{m,i}}= ( \bP_m)_{jZ^*_{m,i}}$ for $i\in[n]$ and $j\in[d]$. In addition, we assume entries of $\bZ_{m}^*$ are i.i.d. categorical random variables with parameters $\ebrac{\pi_{m,k}}_{k=1}^K$ for $m\in\ebrac{0,1}$.  In this case, we have  $\bTheta_m^*=\brac{\ebrac{\pi_{m,k}}_{k=1}^K,\bP_m}$, $\bE_m\in[0,1]^{n\times d}$ has i.i.d. Unif$[0,1]$ entries, and we can write $ \calG_m\brac{\bZ,\brac{\ebrac{\pi_{k}}_{k=1}^K,\bP},\bE}\in\ebrac{0,1}^{n\times d}$  with its $(i,j)$-th entry being  $\II\brac{E_{ij}\le P_{jZ_i}}$. We thus have $L_0=L_1=L_{\textsf {LCM}}$ with
\begin{align*}
        \log L_{\textsf {LCM}}&\brac{\bZ,\bX;\brac{\ebrac{\pi_k}_{k=1}^K,\bP}}\\
        &:=\sum_{i=1}^n\sum_{j=1}^d\sqbrac{X_{ij}\log P_{jZ_i}+\brac{1-X_{ij}}\log \brac{1-P_{jZ_i}}}+\sum_{i=1}^n\log\pi_{Z_i}.
    \end{align*}
\end{exmp}
\begin{exmp}[Contextual Stochastic Block Model] Our general framework also fits the contextual SBM \citep{lu2023contextual, deshpande2018contextual}. We first introduce SBM, which is arguably the most well-studied model in thes network literature \citep{holland1983stochastic, abbe2018community}. Let $\bX_0\in\ebrac{0,1}^{n\times n}$ be the adjacency matrix for a graph of $n$ nodes,  $\bZ_0^*\in[K]^n$ be the community label vector and  $\bB\in[0,1]^{K\times K}$ be the link matrix. For a symmetric SBM, we assume  $\PP\brac{X_{0,ij}=1\mid \bB, \bZ^*_0}=B_{Z^*_iZ^*_j}$ and $X_{ij}=X_{ji}$ and $X_{ii}= 0$ for $i\ne j\in[n]$. Moreover, $\ebrac{\bX_{0,ij}}_{i<j}$ are conditionally independent given $\bZ^*_0$ and $ \bB$, and entries of $\bZ_{0}^*$ are i.i.d. categorical random variables with parameters $\ebrac{\pi_{0,k}}_{k=1}^K$. We have $\bTheta_0^*=\brac{\ebrac{\pi_{0,k}}_{k=1}^K,\bB}$,  and $\calG_0$ can be defined similarly to LCMs. The posterior  $L_{\textsf {SBM}}$ is defined as
\begin{align*}
    \log L_{\textsf{SBM}}&\brac{\bZ,\bX;\brac{\ebrac{\pi_k}_{k=1}^K,\bB}}\\
    &:=\sum_{1\le i<j\le n}\sqbrac{X_{ij}\log B_{Z_iZ_j}+\brac{1-X_{ij}}\log \brac{1-B_{Z_iZ_j}}}+\sum_{i=1}^n\log \pi_{Z_i}.
\end{align*}
 \sloppy In the contextual setting, in addition we have $\ebrac{\bX_{1,i}\in \RR^{d}}_{i=1}^n$  representing the features of  nodes, which follow the Gaussian mixture model with parameter $\bTheta_1^*=\brac{\ebrac{\pi_{1,k}}_{k=1}^K,\ebrac{\bmu_{k}}_{k=1}^K,\ebrac{\bSigma_{k}}_{k=1}^K}$.  We thus have $L_0=L_{\sf SBM}$, $L_1=L_{{\sf GMM}_K}$. It is worth noting that we can exchange the roles of target and source data to have $L_0=L_{{\sf GMM}_K}$ and $L_1=L_{\sf SBM}$, i.e., we treat the feature $\bX_1$ as target and use network $\bX_0$ as source to help us estimate $\bZ_1^*$, which is the latent label vector of $\bX_1$. 

\end{exmp}

\subsection{General Transfer Clustering Methodology}

We will develop an adaptive transfer clustering method for the general framework. For any nuisance parameter $(\hat\bTheta_0,\hat\bTheta_1)$, our transfer learning estimators for the clustering labels are defined as follows:
\begin{align}\label{eq:hatZ0Z1-def}
    \brac{\hat \bZ_0^{\lambda},\hat \bZ_1^{\lambda}}=\argmin_{\bZ_0,\bZ_1\in[K]^n}\ebrac{-\log L_0\brac{\bZ_0,\bX_0;\hat\bTheta_0}-\log L_1\brac{\bZ_1,\bX_1;\hat\bTheta_1}+\lambda n\frakD\brac{\bZ_{0},\bZ_{1}}},
\end{align}
where $\lambda>0$ is a tuning parameter. The procedure \eqref{eq:hatZ0Z1-def}, with slight abuse of notation, is denoted by $\TC\brac{\bX_0,\bX_1;\lambda,\hat\bTheta_0,\hat\bTheta_1}$. Here, instead of assuming known parameters in warm-up example, we need to have an estimator for nuisance parameters  $(\bTheta_0,\bTheta_1)$ (See \Cref{assump:par-est-general}). In LCMs and related latent variable models, this corresponds to the common \textit{item parameter calibration}  setting \citep{kim2006comparative, wainer2000item}, where the parameters have been estimated and calibrated accurately with previous samples. 

Given a grid set $\Lambda\subset[0,\infty)$ and $\hat\psi$ on $\Lambda$, we generalize the  Goldenshluger-Lepski procedure in \Cref{sec:warm-up} to Algorithm \ref{alg:GL}, in order to select the optimal $\lambda$. Here $\hat\psi$ is estimated by using a bootstrap procedure detailed in  Algorithm \ref{alg:bootstrap-quantile}, which can be regarded as a generalization to Algorithm \ref{algo:bootstrap-gmm}. The final estimator  is denoted by
\[
\ATC\brac{\bX_0,\bX_1;\Lambda,\hat\bTheta_0,\hat\bTheta_1}
=
\hat\bZ_0^{\hat\lambda}.
\]
{\color{black}
We provide a flowchart (\Cref{fig:flowchart}) to illustrate the whole procedure of \ATC.
\begin{figure}[!tb]
    \centering
    \includegraphics[width=0.7\linewidth]{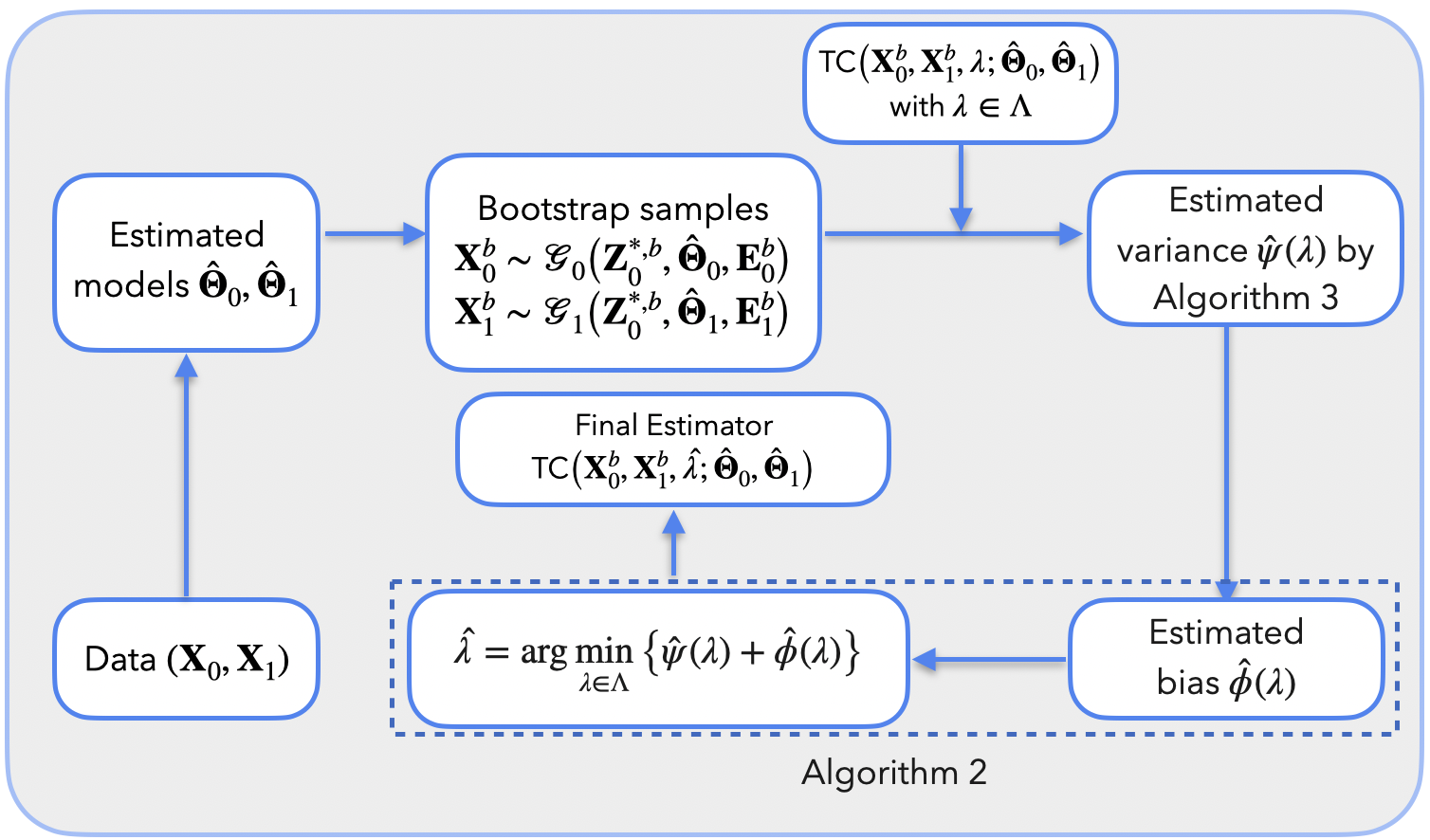}
    \caption{\textcolor{black}{Flowchart of \ATC~ for general distributions.}}
    \label{fig:flowchart}
\end{figure}
}

A natural and commonly asked question is why not directly estimate a joint model with both the source and target data, where $[K] \times [K]$ is the label space for $(Z_{0,i}, Z_{1,i})$. While this might seem like a straightforward approach in the current literature  \citep{gao2020clusterings, gao2022testing2}, there are several reasons why it is less favorable than our proposed method. Firstly, when using a joint model with $[K] \times [K]$ labels, the number of clusters increases significantly to $K^2$, which can be computationally inefficient.  In the interesting regime where $\varepsilon=o(1)$, the sizes of these clusters become extremely unbalanced, making it more difficult to achieve high statistical accuracy. Joint modeling is necessary in \citep{gao2020clusterings, gao2022testing2}, as their focus is on testing the independence of clusters between target and source data.  In contrast, our method aims to improve clustering performance of the target data by estimating only $K$ clusters in the target domain, and using a discrepancy parameter $\varepsilon$ to gauge the discrepancy between the source and target domains. This avoids overfitting to the noise present in smaller clusters. Secondly, a  joint model can lead to difficulties in fitting when combining different types of models for the source and target datasets, e.g., contextual SBM. Our approach circumvents this complexity by separately fitting the models for the source and target data, then aggregating them adaptively. This makes our approach easily adaptable to different data structures and combinable with existing estimators.
\begin{remark}
     Our methodology framework can be readily adapted to other latent variable models beyond clustering (e.g., mixed membership models, factor models). For latent variables $\bZ_0^*,\bZ_1^*\in\calZ^n$ of our interest, where $\calZ$ is some metric space, we only need to introduce an appropriate metric in $\calZ^n$ in place of $\frakD$, and modify the penalty term in \eqref{eq:hatZ0Z1-def}  accordingly. 
\end{remark}
\begin{algorithm}[!tb]
\footnotesize
    \caption{Goldenshluger-Lepski method}\label{alg:GL}
    \KwData{Data $(\bX_0,\bX_1)$, estimated parameter $(\hat\bTheta_0,\hat\bTheta_1)$, grid set $\Lambda$, function $\hat\psi$ on $\Lambda$}
    For each $\lambda\in\Lambda$, compute $\hat \bZ_0^{\lambda}$ according to \eqref{eq:hatZ0Z1-def}.

    For each $\lambda\in\Lambda$, compute $\hat\phi\brac{\lambda}= \max_{\lambda^\prime\in\Lambda,\lambda^\prime<\lambda}\sqbrac{\frakD\brac{\hat \bZ_0^{\lambda},\hat \bZ_0^{\lambda^\prime}}-\hat\psi\brac{\lambda^\prime}}_+$.
    
Let $\hat\lambda\in \argmin_{\lambda\in\Lambda}\ebrac{\hat\phi\brac{\lambda}+\hat\psi\brac{\lambda}}$.

\KwResult{$\hat\lambda$}
\end{algorithm}

\begin{algorithm}[!tb]
\footnotesize
\caption{Bootstrap-Quantile for $\psi$ in general mixture models}\label{alg:bootstrap-quantile}
    \KwData{Estimated parameter $(\hat\bTheta_0,\hat\bTheta_1)$, tuning parameter $\zeta\in(0,1)$ }
    \For{$b=1,\cdots,B$}{
        Generate $Z^{*,b}_{0,i}\overset{i.i.d.}{\sim} \hat \Pi_0$ for $i\in[n]$, where $\hat\Pi_0$ is the estimated parameter for  marginal distribution of $\bZ_0$ included in $\hat\bTheta_0$.

Generate $\bX_0^{b}=\calG_0\brac{\bZ_0^{*,b},\hat\bTheta_0,\bE_0^b}$ and  $\bX_1^{b}=\calG_1\brac{\bZ_0^{*,b},\hat\bTheta_1,\bE_1^b}$.

For each $\lambda\in\Lambda\cup\ebrac{\infty}$, compute  
$\brac{\bar \bZ^{\lambda,b}_{0},\bar \bZ^{\lambda,b}_{1}}=\TC\brac{\bX^b_0,\bX^b_1;\lambda,\hat\bTheta_0,\hat\bTheta_1}$.
    }
For each $\lambda\in\Lambda$, compute $\hat\psi\brac{\lambda}=1.01\hat Q_{\zeta/2}^{\lambda}$, where  $\hat Q_{\zeta/2}^{\lambda}$ is $\brac{1-\zeta/2}$  {quantile} of $\ebrac{\frakD\brac{\bar \bZ^{\lambda,b}_{0}, \bar \bZ^{\infty,b}_{0}}, b=[B]}$. 

\KwResult{$\ebrac{\hat\psi\brac{\lambda},\lambda\in\Lambda}$}
\end{algorithm}

\section{Real Data Applications}\label{sec:realdata}
We evaluate the performance of $\ATC$ on both simulated data and three real-world datasets, including the Lazega Lawyers Network data, TIMSS 2019 educational assessment data, and Business Relation Network data. {\color{black}We emphasize that selecting the best method among a class in unsupervised learning is nontrivial, as true labels are unavailable. It worth noting that \ATC~automatically adapts to the unknown discrepancy without any label information. In this section, we evaluate different methods using external annotations, which are not required for running the methods themselves. In all reported datasets, \ATC~matches or outperforms the strongest baseline (which varies across datasets), with gains that range from modest to substantial depending on the problem.}  
All simulations are deferred to \Cref{sec:numerical} and the analysis of Business Relation Network data is deferred to  \Cref{subsec:BRN} due to space
constraint.  
\if1\anon
{
Our code for reproducing results in this section is available at \url{https://github.com/ZhongyuanLyu/ATC}. 
} \fi
\if0\anon
{
Our code for reproducing results in this section is available on GitHub (link omitted for blind peer review).
} \fi

\subsection{Lazega Lawyers Network}

The Lazega Lawyers Network  originates from a network study of corporate law partnership conducted in a US corporate law firm in 1988-1991 \citep{lazega2001collegial}. {\color{black}The dataset includes  a strong-coworker network (source data $\bX_1$) among the $n=66$ attorneys of this firm.  For each attorney, we selected \textit{years with the firm} as the covariate (target data $\bX_0$),  \textit{status} (either partner or associate) as $\bZ_0^*$, and \textit{office} (either Hartford or Boston) as $\bZ_1^*$.} The misclassification error is $0.151$ when using only the covariate and $0.394$ when using only the network. This suggests that the covariate largely reflects the clustering structure, while the network provides limited information regarding the clustering of attorney status. We treat the covariate as the target data and the network as the source data and apply \ATC~with $1-\zeta\in \ebrac{0.8,0.9,0.95}$. 

\begin{figure}[!tb]
    \centering
    \includegraphics[width=0.45\linewidth]{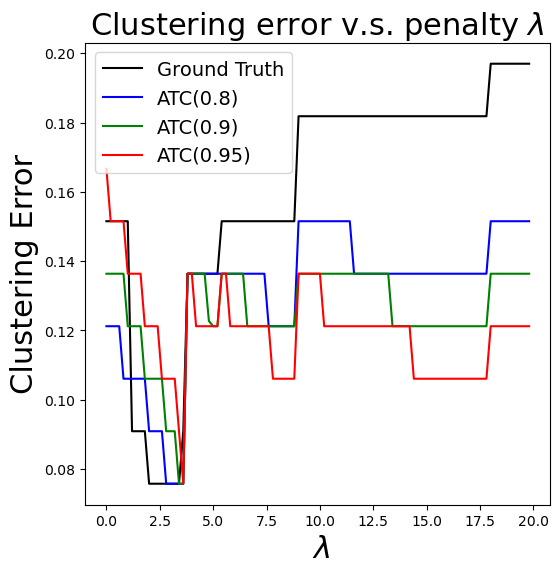}
    \caption{True clustering error (Ground Truth) and estimated excess error ({\sf ATC}) v.s. penalty $\lambda$ for Lawyers Network.  Target: years with the firm (GMM); Source: strong-coworker network (SBM).}
    \label{fig:Layers_lambda}
\end{figure}

In Figure \ref{fig:Layers_lambda}, the ground truth line is the misclassification error of $\TC$ varying with $\lambda$. Other lines are the estimated excess error, i.e. $\hat\phi(\lambda)+\hat\psi(\lambda)$ based on $\ATC$ without using the true labels. Note that these lines represent discrepancy between the transfer learning estimator $\hat\bZ_0^\lambda$ and an oracle estimator $\bar\bZ_0^\infty$, and hence there is an additive constant difference compared to the misclassification error. As shown in Figure \ref{fig:Layers_lambda}, \ATC~can adaptively choose the smallest $\lambda$ that minimizes the true misclassification error across all choices of $\zeta$, which is $0.076$. This indicates that our method effectively integrates information from both the covariate and the network, resulting in improved clustering performance.

For  comparison, we  apply three other methods to this dataset: CASC (covariate-assisted spectral clustering) \citep{binkiewicz2017covariate}, the SDP approach in \citep{yan2021covariate}, and NAC (network-adjusted covariates) \citep{hu2024network}.  As shown in \Cref{tab:lawyers_error}, $\sf ATC$ outperforms other methods in terms of misclassification error and ARI (adjusted random index).
\begin{table}[!htbp]
    \centering
    \resizebox{0.8\linewidth}{!}{  
    \begin{tabular}{ccccccccc}
        \textbf{Type}& \textbf{CASC} & \textbf{SDP} & 
        \textbf{NAC} & \textbf{ATC(0.8)} & \textbf{ATC(0.9)} & \textbf{ATC(0.95)}  \\
        \hline
        Misclassification & 0.348 & 0.106 & 
        0.5 & {\bf 0.076} & {\bf 0.076} & {\bf 0.076}\\
        ARI & 0.078  & 0.615 & -0.002 & {\bf 0.716} & {\bf 0.716} & 0.664
    \end{tabular}
    }
    \caption{Error rates for lawyers network }
    \label{tab:lawyers_error}
\end{table}

\subsection{Trends in International Mathematics and Science Study Data}
We investigate the performance of our method on TIMSS 2019 dataset (Trends in International Mathematics and Science Study)\footnote{The data is publicly available at \url{https://timssandpirls.bc.edu/timss2019}}. We used the student achievement data and  context data at eighth grades in Singapore, which has highest average score across all participating countries. We preprocess the data by selecting $n=259$ students that have access to a common set of items related to science and mathematics. {\color{black} We collect the  feature that represents to what extent the student would agree to the statement ``My teacher tells me I am good at science'', categorized by $1$: Agree a lot; $2$: Agree a little; $3$: Disagree a little; $4$: Disagree a lot. We further label $\ebrac{1}$ as $1$ and $\ebrac{2,3,4}$ as 0, indicating whether the student is good at science or not as our ground truth label $\bZ_0^*$ for target data. Similarly, we obtain the ground truth label $\bZ_1^*$ from ``My teacher tells me I am good at mathematics'' for source data. The target data $\bX_0$ contain students' responses to $d=17$ science-related items, while the source data $\bX_1$ include their responses to $d=17$ mathematics-related items.}
We use the latent class model with two classes for both the source data and the target data.

\begin{figure}[!tb]
    \centering
    \includegraphics[width=0.45\linewidth]{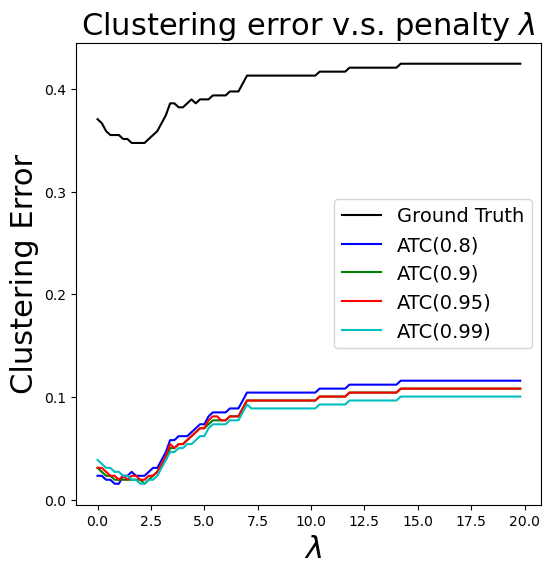}
     \includegraphics[width=0.46\linewidth]{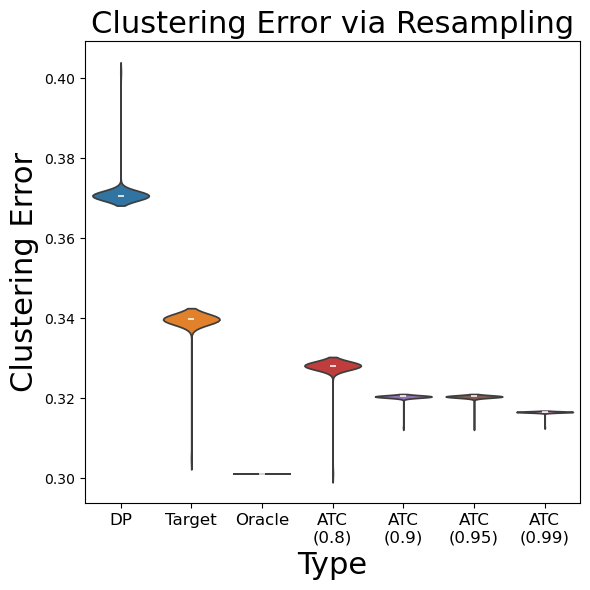}
    \caption{(Left) True clustering error (Ground Truth) and estimated excess error ({\sf ATC}) v.s. penalty $\lambda$ for TIMSS data. Target: responses to science-related items (LCM); Source: responses to math-related items (LCM). (Right) Clustering error via resampling for TIMSS data. Violin plot is based on $N_b=100$ bootstrap (sample with replacement) of original $n=259$  samples.}
    \label{fig:TIMSS}
\end{figure}

Left panel in Figure \ref{fig:TIMSS} presents the performance of \ATC~for    $1-\zeta \in \ebrac{0.8, 0.9, 0.95, 0.99}$. The misclassification error was $0.371$ using  the target data.
In contrast, \ATC~achieves a lower misclassification error of $0.347$ across all choices of $\zeta$.

To further quantify the uncertainty of \ATC, we present a violin plot (right panel in Figure \ref{fig:TIMSS}) showing the misclassification errors based on $N_b=100$ bootstrap samples (with replacement) from the original $n=259$ samples. Here, the  ``oracle" method refers to minimal possible error achieved by {\sf TC} across all $\lambda$'s. The plot clearly demonstrates that our method exhibits greater robustness compared to \DP~and \ITL.

\begin{figure}[!tb]
    \centering
    \includegraphics[width=1\linewidth]{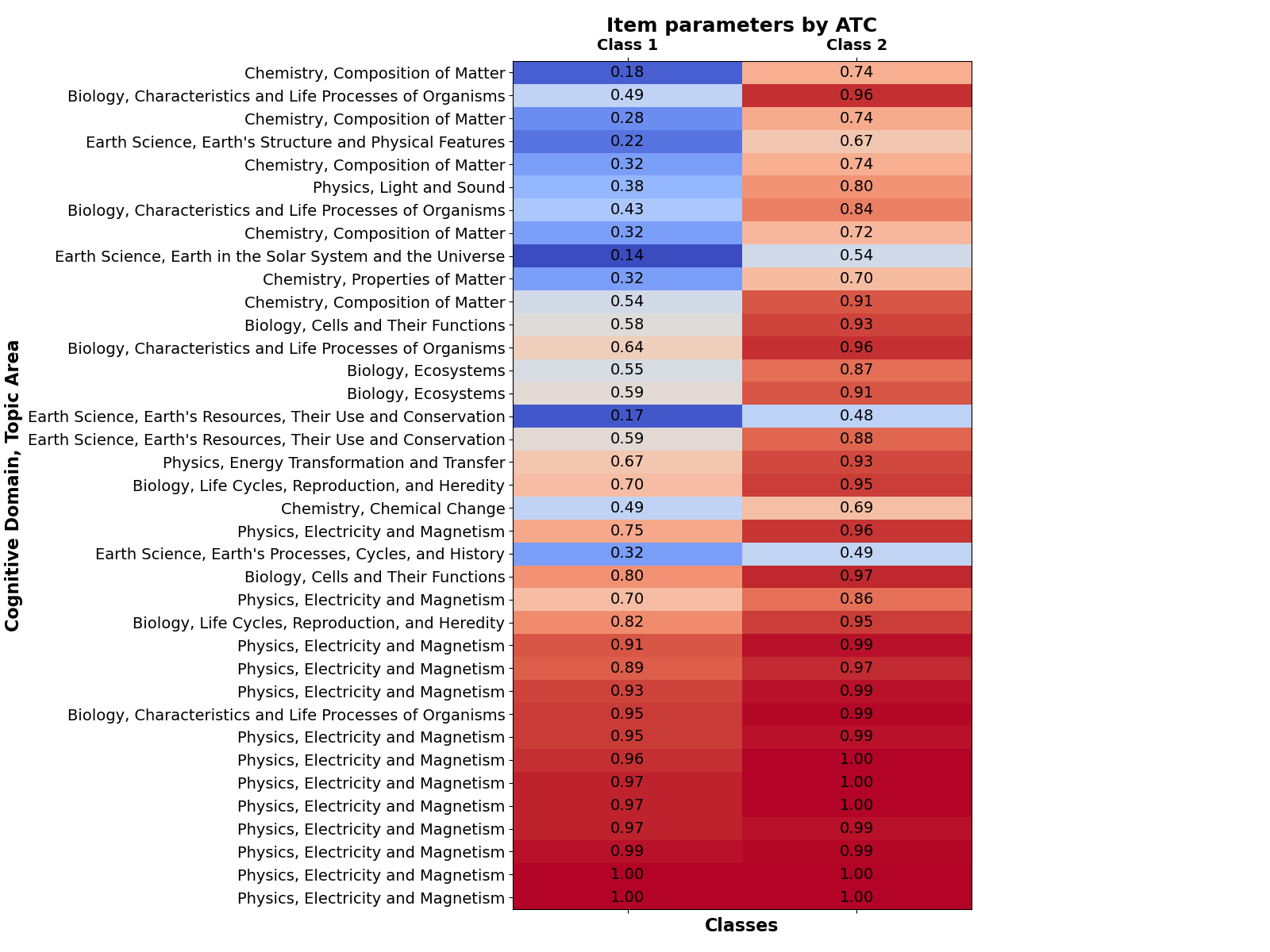}
    \caption{Item parameters  estimated by {\ATC} in the latent class model for the TIMSS data. }
    \label{fig:TIMSS_item_est}
\end{figure}

To facilitate interpretation, we present a heatmap with the the estimated conditional probabilities of  giving the correct response to each item for each latent class, as shown in Figure \ref{fig:TIMSS_item_est}. A closer inspection of the items at the bottom of the figure reveals that these are sub-questions of a single overarching question: ``Components Julia must use to build a circuit.'' The design of these items appears to have lower discriminative power, as students in both classes tend to answer them correctly.

\section{General Theory}\label{sec:theory}

In this section, we conduct a theoretical analysis of the adaptive estimator $\ATC\brac{\bX_0,\bX_1;\Lambda,\hat\bTheta_0,\hat\bTheta_1}$ based on \Cref{alg:GL} and \Cref{alg:bootstrap-quantile}. We start with general results and then derive sharp adaptivity guarantees under the two-component Gaussian mixture model in $\RR^d$.

\subsection{Assumptions and General Theory}
We impose the following assumptions on $\brac{\bZ^*_0,\bZ^*_1}$:
\begin{itemize}[leftmargin=*]
    \item ${Z^*_{0,i}}\overset{i.i.d.}\sim \Pi_0$ and ${Z^*_{1,i}}\overset{i.i.d.}\sim \Pi_1$ for all $i\in[n]$, where $\Pi_0$ and $\Pi_1$ are probability distributions over $[K]$;
    \item $\PP\brac{Z^*_{0,i}\ne Z^*_{1,i}}\le \varepsilon $ for $i\in[n]$.
\end{itemize} 
For technical reasons, we assume the logarithmic posterior densities have separable forms:
\begin{assumption}[Separability]\label{assump:decouple} 
For $m\in\ebrac{0,1}$, there exists $\ebrac{L_{m,i}}_{i=1}^n$ such that $\log L_m\brac{\bZ,\bX;\bTheta}=\sum_{i=1}^n\log L_{m,i}\brac{Z_{i},X_{i};\bTheta}$.
\end{assumption}
\begin{remark}
    \sloppy Note that Assumption \ref{assump:decouple} fails for the posterior of the SBM. In particular, we have $\log L_{\textsf{SBM}}\brac{\bZ,\bX;\brac{\ebrac{\pi_k}_{k=1}^K,\bB}}=\sum_{i=1}^n\log L_{\textsf{SBM},i}\brac{Z_i,X_i;\brac{\ebrac{\pi_k}_{k=1}^K,\bB,\bZ_{-i},\bX_{-i}}}$, where $\bZ_{-i}\in[K]^{N-1}$ is obtained by removing the $i$-th entry of $\bZ$,  $\bX_{-i}\in\RR^{(n-1)\times (n-1)}$ is obtained by removing the $i$-th row and column of $\bX$, and 
\begin{align*}
    \log L_{\textsf{SBM},i}&\brac{Z_i,X_i;\brac{\ebrac{\pi_k}_{k=1}^K,\bB,\bZ_{-i},\bX_{-i}}}\\
    &:=\sum_{j>i}\sqbrac{X_{ij}\log B_{Z_iZ_j}+\brac{1-X_{ij}}\log \brac{1-B_{Z_iZ_j}}}+\log \pi_{Z_i}.
\end{align*}
However, the optimization \eqref{eq:hatZ0Z1-def} can still be decoupled across different $i\in[n]$  by using  $L_{\textsf{SBM},i}\brac{Z_i,X_i;\brac{\ebrac{\pi_k}_{k=1}^K,\bB,\bZ_{-i},\bX_{-i}}}$, in which $\bZ_{-i}$ and $\bX_{-i}$ are viewed as part of parameters. This approach is commonly used  in the network analysis literature \citep{ABH15, abbe2022, gao2017achieving}.
\end{remark}
Consider an imaginary source data $\bar \bX_0=\calG_1\brac{\bZ_0^*,\bTheta^*_1,\bE_1}$, and define
\begin{align*}
    \brac{\bar \bZ_0^{\lambda},\bar \bZ_1^{\lambda}}:=\TC\brac{\bX_0,\bar\bX_0;\lambda,\bTheta^*_0,\bTheta^*_1}.
\end{align*}
We can view $\bar \bX_0$ as the source data generated by the ``clean'' model $\calG_1\brac{\bZ_0^*,\bTheta^*_1,\bE_1}$, as it shares the same latent labels with target data $\bX_0$. Observe that 
\begin{itemize}[leftmargin=*]
    \item $\bar \bZ_0^0$ reduces to the independent task learning estimator using only $\bX_0$.
    \item $\bar \bZ_0^\infty$  reduces to the data-pooling estimator, which can also be regarded as the MAP under the ``null'', i.e.,  $\varepsilon=0$. 
\end{itemize}
It is thereby natural to compare $\frakD\brac{\hat \bZ_0^\lambda,\bZ_0^*}$ to $\frakD\brac{\bar \bZ_0^\infty,\bZ_0^*}$, for which we get
\begin{align*}
    \ab{\frakD\brac{\hat \bZ_0^\lambda,\bZ_0^*}-\frakD\brac{\bar \bZ_0^\infty,\bZ_0^*}}\le \frakD\brac{\hat \bZ_0^\lambda,\bar \bZ_0^\infty}\le \underbrace{\frakD\brac{\bar \bZ_0^\lambda,\bar \bZ_0^\infty}}_{\textsf{``variance"}}+\underbrace{\frakD\brac{\hat \bZ_0^\lambda,\bar \bZ_0^\lambda}}_{\textsf{``bias"}}
\end{align*}
Note that the variance term $\frakD\brac{\bar \bZ_0^\lambda,\bar \bZ_0^\infty}$ only involves a  ``clean'' model (i.e., with perfectly matched cluster labels) and can be estimated by parametric bootstrap if $\bTheta^*_0$ and $\bTheta^*_1$ are known. Define $\brac{\check \bZ_0^{\lambda},\check \bZ_1^{\lambda}}:=\TC\brac{\bX_0,\bX_1;\lambda,\bTheta^*_0,\bTheta^*_1}$.
It can be readily seen that the bias term $\frakD\brac{\hat \bZ_0^\lambda,\bar \bZ_0^\lambda}$  can be decomposed  as
\begin{align*}
    \frakD\brac{\hat \bZ_0^\lambda,\bar \bZ_0^\lambda}\le \underbrace{\frakD\brac{\hat \bZ_0^\lambda,\check \bZ_0^\lambda}}_{\text{induced by error in } \hat\bTheta_0\text{~and~}\hat\bTheta_1}+\underbrace{\frakD\brac{\check \bZ_0^\lambda,\bar \bZ_0^\lambda}}_{\text{induced by mismatch of } \bZ_0^*\text{~and~}\bZ_1^*}.
\end{align*}

Ideally, for Algorithm \ref{alg:GL}, we want an oracle inequality of the form
\[
\frakD\brac{\hat \bZ_0^{\hat \lambda},\bar \bZ_0^\infty}
\lesssim \min_{\lambda\in\Lambda} \{  \phi (\lambda) + \psi (\lambda) \}
+ \text{overhead}
.
\]
We present the following lemma of the same flavor. See \Cref{sec-lem:oracle-gl} for the proof.

\begin{lemma}[Oracle inequality for Goldenshluger-Lepski (GL) method]\label{lem:oracle-gl}
Consider any set $\Lambda$ and $\hat\psi$ on $\Lambda$, let $\hat\lambda$ be obtained from Algorithm \ref{alg:GL}. Suppose there is a deterministic function $\phi$ on $\Lambda$ such that
\begin{enumerate}[leftmargin=*]
    \item $\PP\brac{\frakD\brac{\hat \bZ_0^\lambda,\bar \bZ_0^\lambda}\le \phi\brac{\lambda},\forall\lambda\in\Lambda}\ge 1-\zeta$, 
    \item $\phi$ is non-decreasing.
\end{enumerate}
Then the following inequality holds with probability at least $1-\zeta$:
    \begin{align*}
        \frakD\brac{\hat \bZ_0^{\hat \lambda},\bar \bZ_0^\infty}\le 4\min_{\lambda\in\Lambda}\ebrac{\phi\brac{\lambda}+\hat\psi\brac{\lambda}}+3\xi_{\hat\psi},
    \end{align*}
where $\xi_{\hat\psi}:=\max_{\lambda\in\Lambda}\sqbrac{\frakD\brac{\bar \bZ_0^\lambda,\bar \bZ_0^\infty}-\hat\psi\brac{\lambda}}_{+}$.
\end{lemma}
As result of \Cref{lem:oracle-gl}, if $ \frakD\brac{\bar \bZ_0^\lambda,\bar \bZ_0^\infty} \leq \hat\psi\brac{\lambda} $ for all $\lambda \in \Lambda$, then $\xi_{\hat\psi} = 0$ and
\begin{align}
\frakD\brac{\hat \bZ_0^{\hat \lambda},\bar \bZ_0^\infty}\le 4\min_{\lambda\in\Lambda}\ebrac{\phi\brac{\lambda}+\hat\psi\brac{\lambda}}.
\label{eqn-oracle-simple}
\end{align}
In general, we would hope that $\frakD\brac{\bar \bZ_0^\lambda,\bar \bZ_0^\infty} $ is dominated by  $\hat\psi\brac{\lambda} $ plus a small quantity, so that $\xi_{\hat\psi}$ is small. This requires $\hat\psi$ to be sufficiently large. Meanwhile, we do not want it to be too large because that would make the upper bound in \eqref{eqn-oracle-simple} become vacuous. Fortunately, choosing $\hat\psi$ by quantiles of bootstrap samples does a good job in reconciling the two competing forces.

Define $\wt \bX_0=\calG_0\brac{\bZ_0^*,\hat\bTheta_0,\bE_0}$, $\wt \bX_1=\calG_1\brac{\bZ_0^*,\hat\bTheta_1,\bE_1}$ and let 
\begin{align*}
    \brac{\wt \bZ^{\lambda}_{0},\wt \bZ^{\lambda}_{1}}:=\TC\brac{\wt\bX_0,\wt\bX_1;\lambda,\hat\bTheta_0,\hat\bTheta_1}.
\end{align*}
We cast the following assumption on the parameter estimation error.

\begin{assumption}\label{assump:par-est-general}
    Suppose $(\hat\bTheta_1,\hat\bTheta_2)$ is independent of $\brac{\bX_0,\bX_1}$, and  there exists $\gamma\in[0,0.04)$ such that for $\forall i\in[n]$ and $\forall\lambda\in\Lambda\cup\ebrac{\infty}$,
    \begin{align*}
        \PP\brac{\wt Z^\lambda_{0,i}\ne \bar Z_{0,i}^\lambda}\le \gamma\cdot \PP\brac{\bar Z^\lambda_{0,i}\ne  Z_{0,i}^*}.
    \end{align*}
\end{assumption}
\begin{remark}
    To facilitate the analysis, we assume independence of $(\hat\bTheta_1,\hat\bTheta_2)$  and $\brac{\bX_0,\bX_1}$ in Assumption \ref{assump:par-est-general}. In cases where the parameters are known (e.g., warm-up example in \Cref{sec:warm-up}), this assumption holds with $\gamma=0$. In cases where the parameters need to be estimated, such independence can be achieved by either leveraging additional independent source data (e.g., in the context of item parameter calibration in LCMs and related latent variable models), or utilizing a sample splitting and label alignment trick on $\brac{\bX_0,\bX_1}$.
\end{remark}
Under Assumption \ref{assump:par-est-general}, we have the following observations:
\begin{itemize}[leftmargin=*]
    \item $\brac{\wt \bZ^{\lambda}_{0},\wt \bZ^{\lambda}_{1}}$ is coupled with $\brac{\bar \bZ^{\lambda}_{0},\bar \bZ^{\lambda}_{1}}$ through random seeds $\bE_0$ and $\bE_1$.
    \item $\brac{\wt \bZ^{\lambda}_{0},\wt \bZ^{\lambda}_{1}}\overset{d}{=}\brac{\bar \bZ^{\lambda,b}_{0},\bar \bZ^{\lambda,b}_{1}}$. 
\end{itemize}
The following lemma indicates the validity of using $\hat\psi$ in Algorithm \ref{alg:bootstrap-quantile} as our estimator for $\psi$, whose proof is given in Appendix \ref{pf-lem:bootstrap-quantile}.

\begin{lemma}\label{lem:bootstrap-quantile}
    Suppose Assumption \ref{assump:decouple} and \ref{assump:par-est-general} hold and $B\ge C_0\zeta^{-2}\log\brac{\ab{\Lambda}/\zeta}$ for some universal  constant $C_0>0$. For the $\hat\psi$ defined in Algorithm \ref{alg:bootstrap-quantile}, with probability at least $1-\zeta$ we have 
\begin{align*}
    \xi_{\hat\psi}\le C_1\sqbrac{\frac{\log\brac{\ab{\Lambda}/\zeta}}{n}+\gamma\EE\frakD\brac{\bar \bZ^\infty_{0},  \bZ_{0}^*}},
\end{align*}
and
\begin{align*}
    \hat\psi\brac{\lambda}\le C_2\sqbrac{\EE\frakD\brac{\bar \bZ^{\lambda}_{0}, \bar \bZ^\infty_{0}}+\frac{\log\brac{\ab{\Lambda}/\zeta}}{n}+\gamma\EE\frakD\brac{\bar \bZ^\infty_{0},\bZ_{0}^*}},\qquad \forall \lambda\in\Lambda,
\end{align*} 
for some universal constants $C_1,C_2>0$.

\end{lemma}

Combining \Cref{lem:oracle-gl} and \Cref{lem:bootstrap-quantile}, we immediately obtain the following result for our general transfer clustering procedure:

\begin{theorem}\label{thm:gen-bound}
    Suppose Assumption \ref{assump:decouple} and \ref{assump:par-est-general} hold, and there is a deterministic non-decreasing function $\phi$ on $\Lambda$ such that $\PP\brac{\frakD\brac{\hat \bZ_0^\lambda,\bar \bZ_0^\lambda}\le \phi\brac{\lambda},\forall\lambda\in\Lambda}\ge 1-\zeta/2$. As a consequence, we have with probability at least $1-\zeta$ that 
    \begin{align*}
        \frakD\brac{\hat \bZ_0^{\hat \lambda},\bar \bZ_0^\infty}\le C_0\sqbrac{\min_{\lambda\in\Lambda}\ebrac{\phi\brac{\lambda}+\EE\frakD\brac{\bar\bZ_0^\lambda,\bar\bZ_0^\infty}}+\gamma\EE\frakD\brac{\bar \bZ^\infty_{0},  \bZ_{0}^*}+\frac{\log\brac{\ab{\Lambda}/\zeta}}{n}}.
    \end{align*}
    for some universal constant $C_0>0$.
\end{theorem}
In \Cref{thm:gen-bound}, $\phi\brac{\lambda}$ and  $\EE\frakD\brac{\bar\bZ_0^\lambda,\bar\bZ_0^\infty}$ can be viewed as the bias and the variance terms, respectively. The latter acts as a proxy for $\psi\brac{\lambda}$. The additional term $\gamma\EE\frakD\brac{\bar \bZ^\infty_{0},  \bZ_{0}^*}$ accounts for the error introduced by the parameter estimation, which, together with $n^{-1}\log\brac{1/\zeta}$, is typically negligible. In light of this, our adaptive method achieves the bias-variance trade-off over the grid  $\Lambda$.

\subsection{Two-component Symmetric $d$-dimensional GMM}

We now illustrate our general result using two-component symmetric multivariate GMM. Similar to the univariate case,  let $\{ Z_{0, i}^* \}_{i=1}^n $ and $\{ Z_{1, i}^* \}_{i=1}^n$ be i.i.d.~Rademacher latent variables  such that $\PP\bbrac{Z_{0, i}^*\ne Z_{0, i}^*}\le \varepsilon$. In addition, we have 
$\bX_{0, i} | Z_{0, i}^* \sim N_d( Z_{0, i}^* \bmu , \sigma^2\bI_d )$ and $\bX_{1, i} | Z_{1, i}^* \sim N_d( Z_{0, i}^* \bmu , \sigma^2\bI_d )$ for some $\bmu\in\RR^{d}$ and $\sigma>0$, where $d\ge 1$. We investigate the performance of our method for this multivariate GMM with unknown $\brac{\bmu,\sigma,\varepsilon}$ in the regime $\op{\bmu}/\sigma\rightarrow\infty$ and $\op{\bmu}/\sigma<\sqrt{2\log n}$. Suppose we have any estimator $\brac{\hat\bmu,\hat\sigma}$ that satisfies the following assumption: 
\begin{assumption}\label{assump:mul-par-est}
    Suppose $\brac{\hat\bmu,\hat\sigma}$ is independent of $\brac{\bX_0,\bX_1}$, and   there exists  an event $\calE_{\sf par}$ with $\PP\brac{\calE_{\sf par}}\ge 1-n^{-C_{\sf par}}$ such that on $\calE_{\sf par}$,
    \begin{align}\label{eq:mul-assump-event}
        \ab{\frac{\min_{\theta\in\ebrac{\pm 1}}\op{\hat\bmu-\theta\bmu}}{\op{\bmu}}-1}+\ab{\frac{\hat\sigma}{ \sigma}-1}\le \frac{\delta_n}{4},
    \end{align}
    for some $\delta_n=o(1)$ and $C_{\sf par}\ge 3$.
\end{assumption}
Applying $\TC$ to the two-component symmetric multivariate GMM for any $\lambda>0$ gives
\begin{align*}
    \brac{\hat Z_{0,i}^{\lambda},\hat Z_{1,i}^{\lambda}}&=\argmin_{u,v\in\{\pm1\}}\left \{-\inp{\hat \bmu}{u \bX_{0,i}+v\bX_{1,i}}+\lambda\hat \sigma^2\II\brac{u\ne v}\right \},\quad \forall i\in[n].
\end{align*}
For any grid set a grid set $\Lambda\subset[0,\infty)$, the following theorem provides an error bound for $\hat\bZ^{\hat\lambda}_0=\ATC\brac{\bX_0,\bX_1;\Lambda,\hat\bmu,\hat\sigma}$, whose proof can be found in Appendix \ref{pf-thm:gl-mul-gmm-adap}.
\begin{theorem}\label{thm:gl-mul-gmm-adap}
    Let $\zeta\in(0,1)$. Suppose Assumption \ref{assump:mul-par-est} holds with $\delta_n\le c_0\brac{\log n}^{-3}$
     and assume  $B\ge C_{0}\zeta^{-2}\log\brac{M\zeta^{-1}}$ for some   universal constants $c_0,C_0>0$. There exists some  universal constant $C_1>0$ determined  by $c_0$ and $C_0$  such that with probability at least $1-\zeta$,
    \begin{align*}
    \ell\brac{\hat  \bZ_0^{\hat \lambda}, \bZ_0^*}\le C_{1}\min_{\lambda\in\Lambda}\ebrac{\cM \brac{\frac{\op{\bmu}}{\sigma} , \varepsilon, \lambda } +\frac{\log\brac{M\zeta^{-1}}}{n}}.
\end{align*}
\end{theorem}
\begin{remark}
{\color{black} The condition on $\delta_n$ in \Cref{thm:gl-mul-gmm-adap} is mild. In standard parametric models, $\delta_n=\tilde O(\sqrt{d/n})$ via EM or method of moments \citep{balakrishnan2017statistical,daskalakis2017ten,wu2021randomly,wu2020optimal,doss2020optimal}, so $\delta_n=O((\log n)^{-3})$ whenever $d\lesssim n/\mathrm{polylog}(n)$. In sparse high-dimensional settings, EM attains $\delta_n=\tilde O(\sqrt{s\log d/n})$ with $s$ the sparsity \citep{wang2014high,yi2015regularized,cai2019chime}, again implying $\delta_n=O((\log n)^{-3})$ under standard scaling.}
\end{remark}
Analogous to \Cref{col:optimality-gmm}, we have the following result certifying the adaptivity of $\ATC$, whose  proof is almost the same as that of \Cref{col:optimality-gmm} and hence omitted.
\begin{corollary}\label{col:optimality-mul-gmm}
    Suppose the conditions of \Cref{thm:gl-mul-gmm-adap} hold.  Let $M=\lceil \log^2 n \rceil$ and $\lambda_{j}=jM^{-1} \log n$ for $j=0,\cdots,M$, then there exists some universal constant $C_2>0$ determined  by $C_1$ such that  with probability at least $1-\zeta$,
        \begin{align*}
    \ell\brac{\hat  \bZ_0^{\hat \lambda}, \bZ_0^*}\le C_{2}\sqbrac{\cM \brac{\frac{\op{\bmu}}{\sigma} , \varepsilon, \log \brac{\frac{1 - \varepsilon}{\varepsilon}}} +\frac{\log\brac{M\zeta^{-1}}}{n}}.
\end{align*}
Moreover, if  ${\op{\bmu}}^2/\sigma^2 <\brac{1-c}\log n$ for some constant $c\in(0,1)$, then there exists some universal constant $C_3>0$ determined  by $C_2$ such that with probability at least $1-n^{-10}$,
    \begin{align}\label{eq:adap-multi-upper-bound}
    \ell\brac{\hat  \bZ_0^{\hat \lambda}, \bZ_0^*}\le C_{3}\cM \brac{\frac{\op{\bmu}}{\sigma} , \varepsilon, \log \brac{\frac{1 - \varepsilon}{\varepsilon}}}.
\end{align}
\end{corollary}
\begin{remark}
    Our results can also be extended to the two-component symmetric Gaussian mixture model with unequal $\textsf{SNR}$, as considered in \cite{gao2022testing}. While the analysis follows a similar argument with more tedious calculations, the resulting expressions are more complex to interpret, and hence we omit them for the sake of brevity.
\end{remark}

\color{black}
\subsection{Two-class Contextual LCM}
We consider the two-class \emph{contextual LCM} (CLCM), i.e., the target data follow a two-class LCM and source data follow a two-component $d$-dimensional GMM. Let $\{ Z_{0, i}^* \}_{i=1}^n $ and $\{ Z_{1, i}^* \}_{i=1}^n$ be i.i.d.~Rademacher latent variables  such that $\PP\bbrac{Z_{0, i}^*\ne Z_{0, i}^*}\le \varepsilon$.  For target data, each sample is generated as $X_{0,ij}\mid (\bZ_0^*,\bP^*) \sim \text{Bern}\big(P^*_{j(Z^*_{0,i}+1)/2}\big)$, where $ \bP^*=[a\mathbf{1}_{d_1}, b\mathbf{1}_{d_1}]\in \RR^{d_1\times 2}$ is the item parameter matrix for some $a,b\in(0,1)$. 
For source data, each sample is generated as $\bX_{1, i} | Z_{1, i}^* \sim N_{d_2}( Z_{0, i}^* \bmu , \sigma^2\bI_{d_2} )$  for some $\bmu\in\RR^{d_2}$ and $\sigma>0$.  For this model, we assume $(a,b,\bmu,\sigma)$ are known.

Without loss of generality, we  assume $a>b$ and denote $p_{+}:=a$ and $p_{-}:=b$. Then for each $i\in[n]$, our \TC~ estimator is  given by
\begin{align*}
   \brac{\check Z_{0,i}^\lambda,\check Z_{1,i}^\lambda} \in\argmin_{u,v\in\ebrac{\pm 1}}\ebrac{-\sum_{j=1}^d\sqbrac{X_{0,ij}\log p_u+\brac{1-X_{0,ij}}\log \brac{1-p_u}}+\frac{\op{\bX_{1,i}-v\bmu}^2    }{2\sigma^2}+\lambda\II(u\ne v)}.
\end{align*}

In order to obtain a more interpretable result, let us consider the asymptotics 
\begin{align}\label{eq:asymptotics-clcm}
    a,b=o(1),&\quad a/b=1+o(1), \notag\\
    \textsf{SNR}_0:=d(a-b)^2/(8a),\quad \textsf{SNR}_1:=&~\op{\bmu}^2/(2\sigma^2),\quad 1\ll \textsf{SNR}_0+\textsf{SNR}_1\le \log n.
\end{align}
Then we have 
\begin{theorem}\label{thm:gl-contextual-lcm-adap-simple}
    Let $\zeta\in(0,1)$. Suppose  $B\ge C_{0}\zeta^{-2}\log\brac{M\zeta^{-1}}$ and  $\SNR_1+\SNR_2<(1-c)\log n$ for some universal constants $C_0>0$ and $c\in(0,1)$. Let $M=\lceil \log^2 n \rceil$ and $\lambda_{j}=jM^{-1} \log n$ for $j=0,\cdots,M$,  and define $\alpha:=\dfrac{\log(1/\varepsilon)}{2\brac{\SNR_0+\SNR_1}}\in (0,\infty)$, $r:=\SNR_0/\SNR_1\in(0,\infty)$. Under the asymptotics in \eqref{eq:asymptotics-clcm}, there exists a sequence $\rho=o(1)$ such that with probability at least $1-\zeta$,
$$\ell\brac{\check  \bZ_0^{\hat \lambda}, \bZ_0^*}\lesssim\exp\sqbrac{-(1-\rho)\brac{\SNR_0+\SNR_1}},$$
whenever $\alpha>\dfrac{1}{1+\sqrt{1+r}}$.
\end{theorem}
\Cref{thm:gl-contextual-lcm-adap-simple} is a consequence of a more general result (\Cref{thm:gl-contextual-lcm-adap}), which is deferred  to \Cref{sec:gen-clcm} due to space limit.
Notably, when $r=1$, the threshold $\alpha=\sqrt{2}-1$ reduces to the same threshold in the two-component GMM, where our $\ATC$ can achieve the  rate  as if there is no mismatch.

\color{black}
\section{Disscussion}\label{sec:disc}
In this paper, we have developed a unified transfer learning framework for clustering in the presence of unknown discrepancy between latent labels of target and source data, applicable to a wide range of statistical models. The proposed adaptive transfer clustering (\ATC) algorithm effectively balances the bias-variance trade-off  and its theoretical optimality  under two-component symmetric GMMs is well established. Several future directions are worth exploring. It would be of great interest to extend the analysis in \Cref{thm:gen-bound} to  LCMs and SBMs, which  requires more technical efforts. Moreover, our framework can be adapted to other unsupervised learning tasks beyond clustering. It would be intriguing to examine its empirical and theoretical performance in continuous parameter estimation in models such as mixed membership models and continuous latent factor models.

\noindent{\bf Data Availability Statement }
The data that support the findings of this study are openly available at \cite{lazega2001collegial}, \url{https://timssandpirls.bc.edu/timss2019}, and \url{https://data.world/datasyndrome/relato-business-graph-database}.

\if1\anon
{
\section*{Acknowledgement}
Kaizheng Wang's research is supported by an NSF grant DMS-2210907 and a startup grant at Columbia University. 
Yuqi Gu's research is supported by an NSF grant DMS-2210796 and startup funding from the Department of Statistics at Columbia University.
Part of the research was conducted when Zhongyuan Lyu was affiliated with the Data Science Institute at Columbia University.
The authors report there are no competing interests to declare.
} \fi
\if0\anon
{
} \fi


\appendix

The appendix provides general results for two-component CLCM, a discussion on comparison with the testing procedure, simulation studies, additional  real data applications, and all the proofs of theoretical results.

To start with, we provide \Cref{tab:Z-notation-2} that summarizes the label notation used throughout the paper. 
\begin{table}[!ht]
\centering
\renewcommand{\arraystretch}{1.15}
\setlength{\tabcolsep}{6pt}
\begin{tabular}{p{0.15\textwidth} p{0.75\textwidth}}
\hline
\textbf{Symbol} & \textbf{Meaning / construction} \\
\hline
$\bZ_0^*,\ \bZ_1^*$ 
& True latent labels for target/source under the generative model. \\[2pt]

$\hat{\bZ}_0^{\lambda},\ \hat{\bZ}_1^{\lambda}$
& TC labels at tuning $\lambda$ under the estimated parameters; output of $\textsf{TC}(\bX_0,\bX_1;\lambda,\hat\bTheta_0,\hat\bTheta_1)$. \\[2pt]

$\check{\bZ}_0^{\lambda},\ \check{\bZ}_1^{\lambda}$ 
& TC labels at $\lambda$ under the true parameters; output of $\textsf{TC}(\bX_0,\bX_1;\lambda,\bTheta_0^*,\bTheta_1^*)$. \\[2pt]

$\bar{\bZ}_0^{\lambda},\ \bar{\bZ}_1^{\lambda}$
& TC labels at $\lambda$ under $\varepsilon=0$ and the true parameters; output of $\textsf{TC}(\bar\bX_0,\bar\bX_0;\lambda,\bTheta_0^*,\bTheta_1^*)$. \\[2pt]

$\wt{\bZ}_0^{\lambda},\ \wt{\bZ}_1^{\lambda}$
& TC labels at $\lambda$ on pseudo-data generated under $\varepsilon=0$ and the estimated parameters; output of $\textsf{TC}(\wt\bX_0,\wt\bX_1;\lambda,\hat\bTheta_0,\hat\bTheta_1)$. \\[2pt]

$\bar{\bZ}_{0,\mathrm{inf}}^{\lambda,q},\ \bar{\bZ}_{1,\mathrm{inf}}^{\lambda,q}$
& TC labels from the {inference bootstrap} draw $q$ at $\lambda$, generated under $\varepsilon=0$ and the true parameters;  output of $\textsf{TC}(\bX_{0,\rm inf}^q,\bX_{1,\rm inf}^q;\lambda,\bTheta_0^*,\bTheta_1^*)$. \\[4pt]
\hline
\end{tabular}
\caption{Quick reference for label notation.}
\label{tab:Z-notation-2}
\end{table}
\color{black}
\section{General Results for Two-class CLCM}\label{sec:gen-clcm}
In this section, we present two general results for CLCM, whose proof can be found in \Cref{pf-thm:gl-contextual-lcm-adap} and \Cref{pf-thm:gl-contextual-lcm-adap-asymp}. We start with the following bound:
\begin{theorem}\label{thm:gl-contextual-lcm-adap}
    Let $\zeta\in(0,1)$. Suppose  $B\ge C_{0}\zeta^{-2}\log\brac{M\zeta^{-1}}$ for some   universal constant $C_0>0$, then there exists a universal constant $C_1>0$ depending only on $C_0$ such that with probability at least $1-\zeta$,
    \begin{align*}
    \ell\brac{\check  \bZ_0^{\hat \lambda}, \bZ_0^*}&\le C_1\Bigg[ \exp\brac{-{\frac{d }{2}D_{1/2}\brac{a\mid\mid b}-\frac{\mu^2}{2\sigma^2}}}\\
    &+\min_{\lambda\in\Lambda}\Bigg\{\Phi\brac{\frac{\mu}{
    \sigma}-\frac{\lambda \sigma}{2
    \mu}}\exp\Big(-d \min\ebrac{D_{\rm KL}\brac{p^*-\delta_\lambda\mid\mid a},D_{\rm KL}\brac{p^*+\delta_\lambda\mid\mid b}}\Big)\\
    &+\Phi\brac{-\frac{\mu}{
    \sigma}-\frac{\lambda \sigma}{2
    \mu}}\exp\Big(-d \min\ebrac{D_{\rm KL}\brac{p^*+\delta_\lambda\mid\mid a},D_{\rm KL}\brac{p^*-\delta_\lambda\mid\mid b}}\II\brac{\lambda\ge \lambda_{\max}}\Big])\\
    &+\varepsilon\exp\Big(-d\min\ebrac{D_{\rm KL}\brac{p^*+\delta_\lambda\mid\mid  a},D_{\rm KL}\brac{p^*-\delta_\lambda\mid\mid  b}}\II\brac{\lambda\le \lambda_{\min}}\Big)\Bigg\}+\frac{\log\brac{M/\zeta}}{n}\Bigg],
\end{align*}
where $D_{1/2}(a\mid\mid b):=-2\log\brac{\sqrt{ab}+\sqrt{(1-a)(1-b)}}$ is the Renyi divergence of order $1/2$ between $\text{Bern}(a)$ and $\text{Bern}(b)$, and $D_{\rm KL}(a\mid\mid b):=a\log(a/b)+(1-a)\log[(1-a)/(1-b)]$ is the KullbackÃ¢â‚¬â€œLeibler divergence between $\text{Bern}(a)$ and $\text{Bern}(b)$, $\lambda_{\min}:=d\min\ebrac{D_{\rm KL}\brac{a\mid\mid  b},D_{\rm KL}\brac{b\mid\mid  a}}$ and 
$\lambda_{\max}:=d\max\ebrac{D_{\rm KL}\brac{a\mid\mid  b},D_{\rm KL}\brac{b\mid\mid  a}}$.
\end{theorem}
Under the asymptotics in \eqref{eq:asymptotics-clcm}, we can get a more explicit bound as follows: 
\begin{theorem}\label{thm:gl-contextual-lcm-adap-asymp}
    Suppose the conditions of \Cref{thm:gl-contextual-lcm-adap} hold. Let $M=\lceil \log^2 n \rceil$ and $\lambda_{j}=jM^{-1} \log n$ for $j=0,\cdots,M$,  Under the asymptotics in \eqref{eq:asymptotics-clcm}, there exists a sequence $\rho=o(1)$ such that with probability at least $1-\zeta$,
\begin{align*}
  & \ell\brac{\check  \bZ_0^{\hat \lambda}, \bZ_0^*}\lesssim \exp\sqbrac{-\brac{1-\rho}\brac{\textsf{SNR}_0+\textsf{SNR}_1}}+\exp\brac{-(1-\rho)\brac{\SNR_0+\SNR_1}\calT_r(\alpha)}+\frac{\log\brac{M/\zeta}}{n},
\end{align*}
where $\alpha$ and $r$ are defined in \Cref{thm:gl-contextual-lcm-adap-simple}, and  
\begin{align*}
  \calT_r&(\alpha):=\begin{cases}
      2\alpha+\dfrac{r}{1+r}, & \alpha\in\Big[0,\dfrac{1-r}{2\brac{1+r}}\Big)\\
      \alpha+\dfrac{1}{2}+\dfrac{U^2_r(\alpha)}{8r\brac{1+\sqrt{1-c_rU_r(\alpha) }}^2}, & \alpha\in\Big[\dfrac{1-r}{2\brac{1+r}},\dfrac{1+r}{2}\Big)\\
      2\alpha, & \alpha\in\Big[\dfrac{1+r}{2},\infty\Big)
    \end{cases},
\end{align*}
with $U_r(\alpha):=2\alpha(1+r)-(1-r)$ and $c_r:=(1-r)/(4r)$.
Moreover, if  $\SNR_1+\SNR_2<(1-c)\log n$ for some constant $c\in(0,1)$, we have 
$$\ell\brac{\check  \bZ_0^{\hat \lambda}, \bZ_0^*}\lesssim\exp\sqbrac{-(1-\rho)\brac{\SNR_0+\SNR_1}},$$
with probability at least $1-\zeta$ whenever $\alpha>\dfrac{1}{1+\sqrt{1+r}}$.
\end{theorem}

\color{black}
\section{Comparison with Testing Procedure}\label{sec:testing}
\begin{figure}[!tb]
    \centering
    \includegraphics[width=0.7\linewidth]{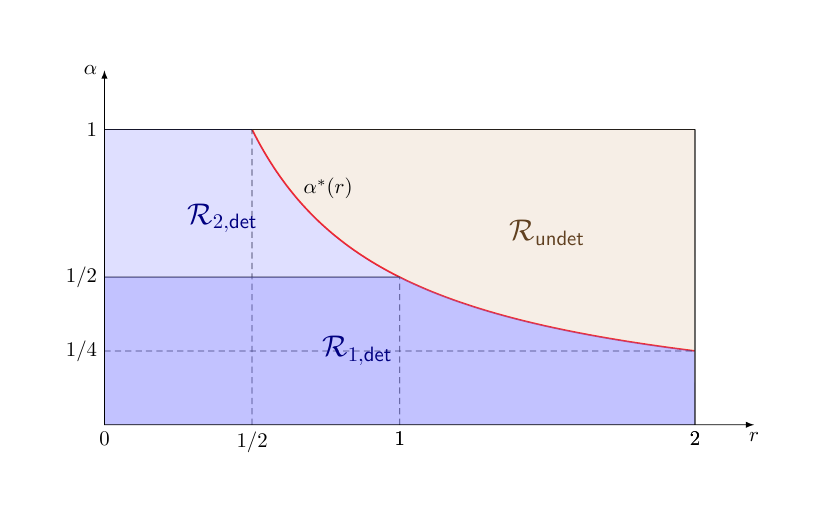}
    \caption{Comparison with the hypothesis test's detection boundary. Below the red curve $\alpha=\alpha^*(r)$, i.e., Region $\calR_{1,\sf det}\bigcup\calR_{2,\sf det}$, consistent test exists for \eqref{eq:testing}; above the red curve, i.e., Region $\calR_{\sf undet}$, there is no consistent test for \eqref{eq:testing}. Region $\calR_{\sf undet}$: transfer clustering performs as well as data pooling when no consistent test can distinguish $\bZ_0^*$ and $\bZ_1^*$. Region $\calR_{1,\sf det}$: transfer clustering beats data pooling while we  can have a consistent test to distinguish $\bZ_0^*$ and $\bZ_1^*$. Region $\calR_{2,\sf det}$: transfer clustering performs as well as data pooling while we  can have a consistent test to distinguish $\bZ_0^*$ and $\bZ_1^*$.}
    \label{fig:Comparison}
\end{figure}

\cite{gao2022testing} proposed a testing procedure for the following hypothesis testing problem:
\begin{align}\label{eq:testing}
    H_0: \ell\brac{\bZ_0^*,\bZ_1^*}=0 \quad \text{vs.} \quad H_1: \ell\brac{\bZ_0^*,\bZ_1^*}> \varepsilon
\end{align} 
for some $\varepsilon\ge 0$, under the regime $\varepsilon=n^{-\beta}$ and $\mu/\sigma=\sqrt{r\log n}$ with some constants $\beta,r>0$. In particular (adapted to our notation), they derived a sharp boundary for \eqref{eq:testing} in the sense that the likelihood ratio test and a Higher-Criticism-type test \citep{donoho2004higher} is consistent when $\alpha<\alpha^*\brac{r}$, and no test is consistent when $\alpha>\alpha^*(r)$, where 
\begin{align*}
\alpha^*(r):=\begin{cases}
        \frac{1}{4}\brac{3+\frac{1}{r}},& 0<r\le 1/5\\
        \frac{1}{2r}\sqrt{1-\brac{1-2r}^2_+}, & r>1/5.
    \end{cases}
\end{align*}
\begin{itemize}[leftmargin=*]
    \item When $0<\alpha<\alpha^*(r)\wedge 1/2$ (Region $\calR_{1,\sf det}$ in \Cref{fig:Comparison}), $\varepsilon$ is sufficiently large so that we can tell the the difference between $\bZ_0^*$ and $\bZ_1^*$.  A natural strategy is to use independent task learning estimator solely based on $\bX_0$, leading to an error rate of  $\exp\brac{-\textsf{SNR}\brac{1+o(1)}}$. On the other hand, applying data pooling leads to an error rate of $\exp\brac{-4\alpha\cdot \textsf{SNR}\brac{1+o(1)}}$. In contrast, the transfer learning with optimal $\lambda$ leads to a vanishingly smaller error rate $\exp\brac{-(1+\alpha)^2\cdot \textsf{SNR}\brac{1+o(1)}}$.
    \item When $1/2<\alpha<\alpha^*(r)\wedge 1$ (Region $\calR_{2,\sf det}$ in \Cref{fig:Comparison}), $\varepsilon$ remains  large enough to be detectable. Here, independent task learning estimator is still sub-optimal. The transfer learning with optimal $\lambda$ achieves the ideal error rate $\exp\brac{-2\cdot \textsf{SNR}\brac{1+o(1)}}$, which is also attainable by data pooling. This regime reveals a  subtle difference between clustering and testing: even when the discrepancy is detectable, the optimal estimator can be constructed by ignoring it.
    \item When  $\alpha^*(r)<\alpha<1$ (See Region $\calR_{\sf undet}$ in \Cref{fig:Comparison}),  $\varepsilon$ is so small that no test can reliably tell the difference between $\bZ_0^*$ and $\bZ_1^*$. By definition of $\alpha$ and $\alpha^*\brac{r}$, we have $\beta>1$ in this regime. Both the transfer learning with optimal $\lambda$ and data pooling result in the error rates  of $\exp\brac{-\min\ebrac{2,{2}/{r}}\cdot \textsf{SNR}\brac{1+o(1)}}$.
\end{itemize}

\color{black}
\section{Extension of $\ATC$ to Hypothesis Testing}
The framework in \cite{gao2022testing} requires both the target and source data to follow two-component GMMs. Here we show that our methodology enables testing for general parametric families. Formally, we are interested in testing the following hypothesis:
\begin{align}\label{eq:test-extention}
	H_0:\varepsilon=0\qquad \text{v.s.}\qquad H_1:\varepsilon>0.
\end{align}
For illustration, we consider the case when $(\bTheta_0^*,\bTheta_1^*)$ is given. We proceed as follows.
\begin{itemize}
	\item Compute $\hat\psi(\lambda)$ for $\lambda\in\Lambda$ by parametric bootstrap (\Cref{alg:bootstrap-quantile} with $(\bTheta_0^*,\bTheta_1^*)$), $\hat\phi(\lambda):=\max_{\lambda^\prime\in\Lambda,\lambda^\prime<\lambda}\sqbrac{\frakD\brac{\check \bZ_0^{\lambda},\check \bZ_0^{\lambda^\prime}}-\hat\psi\brac{\lambda^\prime}}_+$ for $\lambda\in\Lambda$. Then, compute test statistics $$T:=\min_{\lambda\in\Lambda}\ebrac{\hat\psi(\lambda)+\hat\phi(\lambda)}.$$
	\item For $B_{\rm inf}>0$, generate bootstrap samples $\ebrac{(\bX_{0,\rm inf}^q,\bX_{1,\rm inf}^q)}_{q=1}^{B_{\rm inf}}$ with $(\bTheta_0^*,\bTheta_1^*)$ and $\varepsilon=0$.  For $q=1,\cdots,B_{\rm inf}$, compute $\brac{\bar \bZ^{\lambda,q}_{0,\rm inf},\bar \bZ^{\lambda,q}_{1,\rm inf}}=\textsf{TC}\brac{\bX^q_{0,\rm inf},\bX^q_{1,\rm inf};\lambda,\bTheta_0^*,\bTheta_1^*}$, $\hat\phi_q(\lambda):=\max_{\lambda^\prime\in\Lambda,\lambda^\prime<\lambda}\sqbrac{\frakD\brac{\bar \bZ_{0, \rm inf}^{\lambda,q},\bar \bZ_{0,\rm inf}^{\lambda^\prime,q}}-\hat\psi\brac{\lambda^\prime}}_+$ for $\lambda\in\Lambda$, and 
	\begin{align*}
		T_q:=\min_{\lambda\in\Lambda}\ebrac{\hat\psi(\lambda)+\hat\phi_q(\lambda)}.
	\end{align*}
	\item Reject $H_0$ if $T>Q_{1-\alpha}^{\rm inf}:=(1-\alpha)$ quantile of $\ebrac{T_1, ..., T_{B_{\rm inf}}}$.
\end{itemize}
\begin{proposition}
    For the testing problem in \eqref{eq:test-extention}, we have $\PP\brac{T\ge Q_{1-\alpha}^{\rm inf}\mid H_0}\le \alpha$.
\end{proposition}
\textit{Proof}: Let $\PP_0$ denote the joint law of $(\bX_0,\bX_1)$ under $H_0:\varepsilon=0$. By construction, $\ebrac{(\bX_{0,\rm inf}^q,\bX_{1,\rm inf}^q)}_{q=1}^{B_{\rm inf}}$ are i.i.d. from $\PP_0$ and independent of $(\bX_0,\bX_1)$. By definition, $T,T_1,\cdots,T_{B_{\rm inf}}$ are i.i.d. as each is the same measurable functional  $F(\cdot\mid \hat\psi)$ applied to an independent draw from $\PP_0$. The result follows immediately by the definition of $1-\alpha$ quantile.

\color{black}

\section{Simulation Studies}\label{sec:numerical}
We conduct  numerical simulations to validate the effectiveness of our proposed estimator $\hat\bZ$  compared to $\hat\bZ^{\ITL}$ and $\hat\bZ^{\DP}$. 
\if1\anon
{
Our code for reproducing results in this section is available at \url{https://github.com/ZhongyuanLyu/ATC}. 
} \fi
\if0\anon
{
Our code for reproducing results in this section is available on GitHub (link omitted for blind peer review).
} \fi

\begin{figure}[h!]
    \centering
    \includegraphics[width=0.45\linewidth]{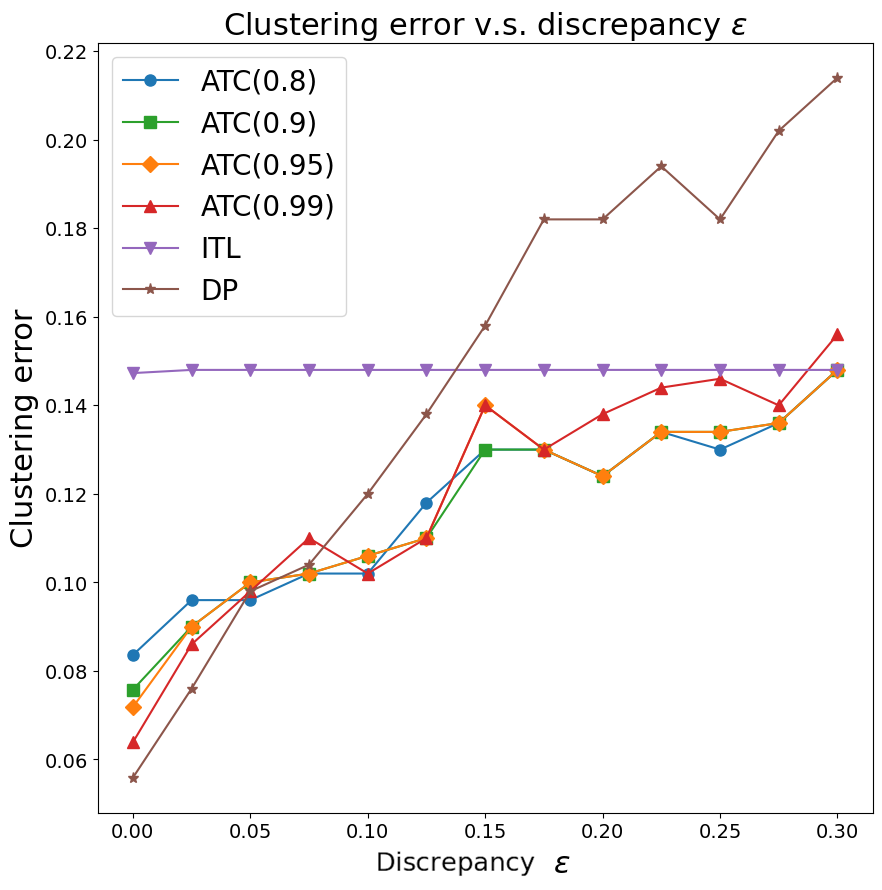}
     \includegraphics[width=0.45\linewidth]{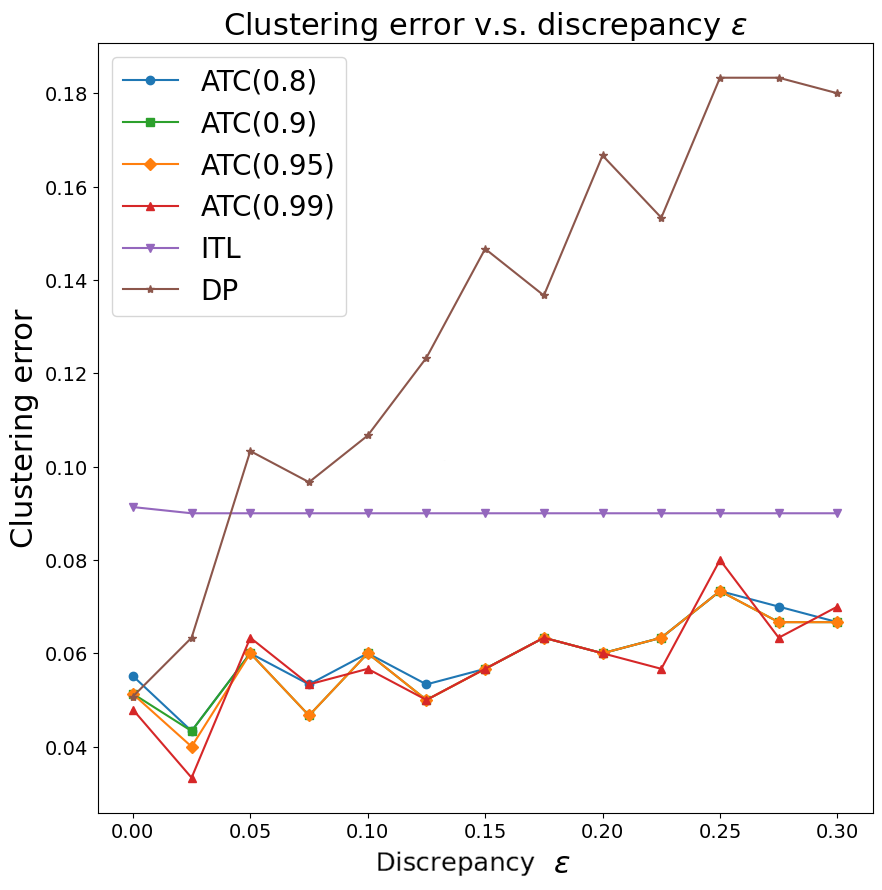}
    \caption{Clustering error v.s. discrepancy $\varepsilon$.  (Left) Target: GMM; Source: GMM. Parameters are set to be $(n,K,d,\upsilon)=(500, 2, 10, 0.54)$. (Right)  Target: SBM; Source: GMM. Parameters are set to be $(n,K,d,p,q,\upsilon)=(300, 2, 10, 0.4, 0.3, 0.79)$.  Each error point represents the average of 50 replications.}
    \label{fig:gmm_err_epsilon}
\end{figure}

\begin{figure}[h!]
    \centering
    \includegraphics[width=0.45\linewidth]{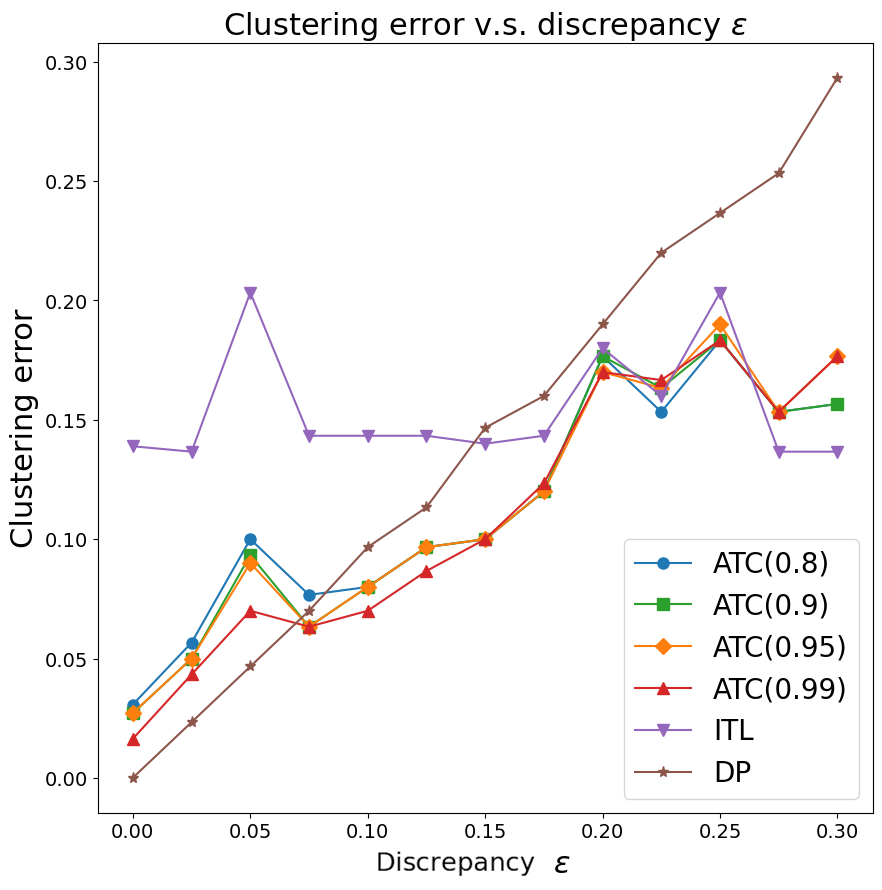}
     \includegraphics[width=0.45\linewidth]{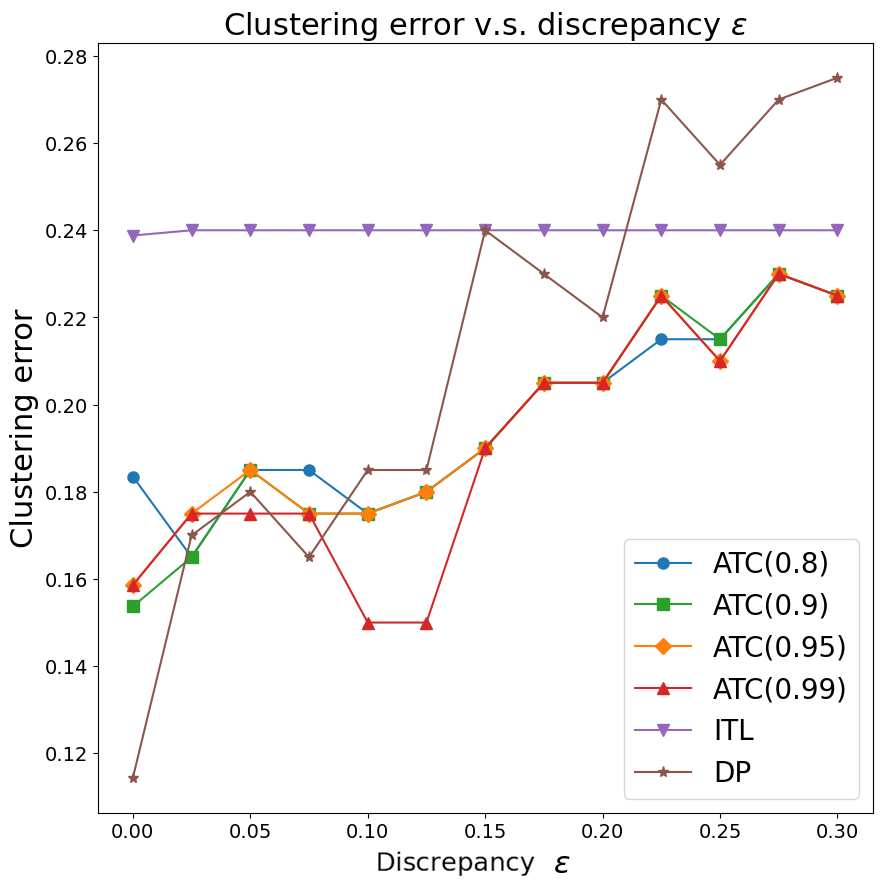}
    \caption{Clustering error v.s. discrepancy $\varepsilon$. (Left) Target: GMM; Source: SBM. Parameters are set to be $(n,K,p,q,d,\upsilon)=(300, 2, 0.5, 0.3, 10, 0.75)$. (Right) Target: LCM; Source: GMM. Parameters are set to be $(n,K,d,\Delta, \upsilon)=(200, 2,  10, 0.1, 0.76)$. Each error point represents the average of 50 replications.}
    \label{fig:csbm_targetgmm_err_epsilon}
\end{figure}

\begin{figure}[!htp]
    \centering
    \includegraphics[width=0.45\linewidth]{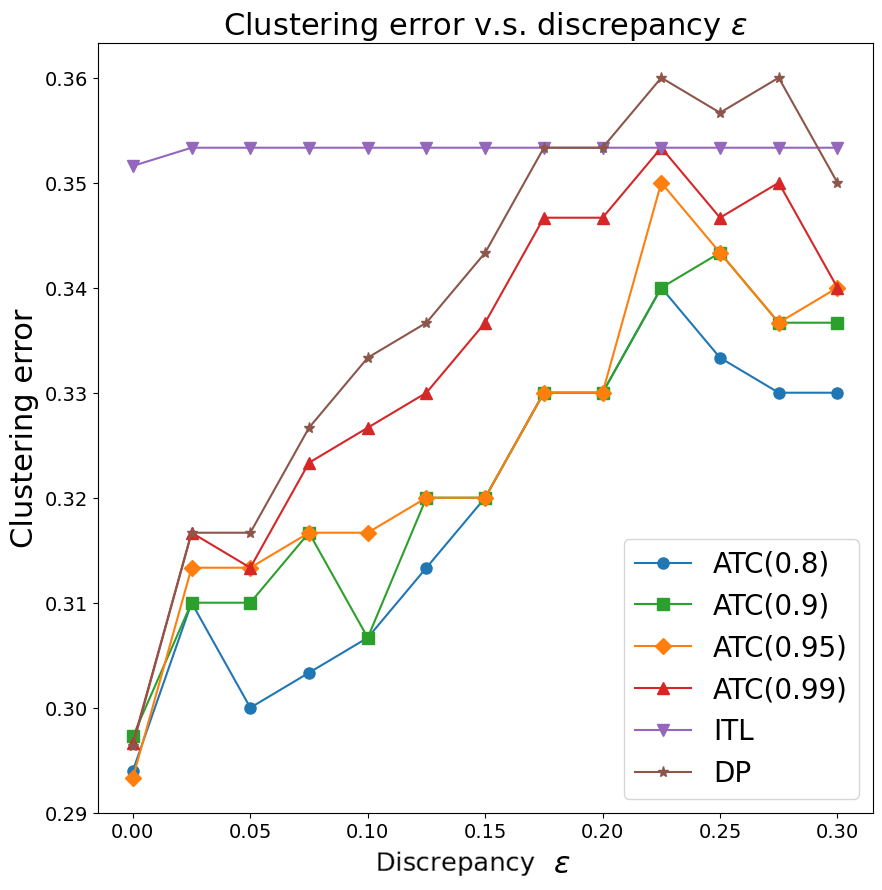}
    \caption{Clustering error v.s. discrepancy $\varepsilon$.  Target: LCM; Source: LCM. Parameters are set to be $(n,K,d,\Delta)=(300, 2,  15, 0.1)$. Each error point represents the average of 50 replications.}
    \label{fig:lcm_err_epsilon}
\end{figure}

Given $\varepsilon\in[0,1/2]$, the random labels $\bZ_0$ and $\bZ_1$ are generated as follows:
\begin{itemize}[leftmargin=*]
    \item Generate $Z_{0,i}\overset{i.i.d.}\sim \textsf{Unif}\ebrac{1,2}$ for $i\in[n]$.
    \item Generate $Z_{1,i}\in\ebrac{1,2}$ by $\PP\brac{Z_{1,i}=Z_{0,i}}=1-\varepsilon$ and $\PP\brac{Z_{1,i}\ne Z_{0,i}}=\varepsilon$ for $i\in[n]$.
\end{itemize}
Given the label $\bZ$ (either $\bZ_0$ or $\bZ_1$), we use the following procedure to generate our random samples:
\begin{itemize}[leftmargin=*]
    \item (GMM) Randomly generate $\a\sim N_d\brac{0,\bI_d}$ and set $\bmu=\frac{\a }{\op{\a}}\cdot \upsilon$, then generate $\bX_i\sim N_d\brac{\bmu,\bI_d}$ for $i\in[n]$.
    \item (SBM) Set $\bB=q\mathbf{1}_K\mathbf{1}_K^\top +(p-q)\bI_K$ and generate  $X_{ij}\sim \textsf{Bernoulli}(B_{Z_i,Z_j})$ for $i,j\in[n]$.
    \item (LCM) Set $\bTheta=[(0.5-\Delta)\mathbf{1}_{d},(0.5+\Delta)\mathbf{1}_{d}]$ and generate  $X_{ij}\sim \textsf{Bernoulli}(\Theta_{j,Z_i})$ for $i\in[n]$ and $j\in[d]$.
\end{itemize}
In all simulations, we set $K=2$ and consider four different scenarios:
\begin{itemize}[leftmargin=*]
    \item In the first setting (left panel in Figure \ref{fig:gmm_err_epsilon}), we use GMM as both target and source distribution with $(n,d,\upsilon)=(500, 10, 0.54)$.
    \item In the second setting (right panel in Figure \ref{fig:gmm_err_epsilon}), we use SBM as target  distribution and GMM as source distribution with $(n,d,p,q,\upsilon)=(300, 10, 0.4, 0.3, 0.79)$.
    \item In the third setting (left panel in Figure \ref{fig:csbm_targetgmm_err_epsilon}), we use GMM as target  distribution and SBM as source distribution with $(n,d,p,q,\upsilon)=(300, 10, 0.5, 0.3, 0.75)$.
    \item In the fourth setting (right panel in Figure \ref{fig:csbm_targetgmm_err_epsilon}), we use LCM as target  distribution and GMM as source distribution with $(n,d,\Delta,\upsilon)=(200, 10, 0.1, 0.76)$.
    \item In the fifth setting (Figure \ref{fig:lcm_err_epsilon}), we use LCM as both target  and  source distribution with $(n,d,\Delta,\upsilon)=(200, 10, 0.1, 0.76)$.
\end{itemize}
In each setting, we fix $\upsilon$, $(p,q)$, or $\Delta$ (representing the signal-to-noise ratio in GMM, SBM and LCM, respectively), hence $\varepsilon$ controls the signal-to-noise ratio in transfer learning, analogous to $\alpha$ defined in Section \ref{subsec:tl-procedure}. We apply our method with different choices of quantile parameter $1-\zeta\in\ebrac{0.8,0.9,0.95,0.99}$. The numerical results are consistent with our theoretical findings.  As shown in  Figure \ref{fig:gmm_err_epsilon}-\ref{fig:lcm_err_epsilon}, our estimator $\hat\bZ$ (\ATC) outperforms, or at least matches, $\hat\bZ^{\ITL}$ and $\hat\bZ^{\DP}$ (denoted by \ITL~and \DP, respectively), which indicates the benefit of our proposed method. In addition, it can be seen that our method is robust to the choice of $\zeta$.


\begin{figure}[!htp]
    \centering
    \includegraphics[width=0.45\linewidth]{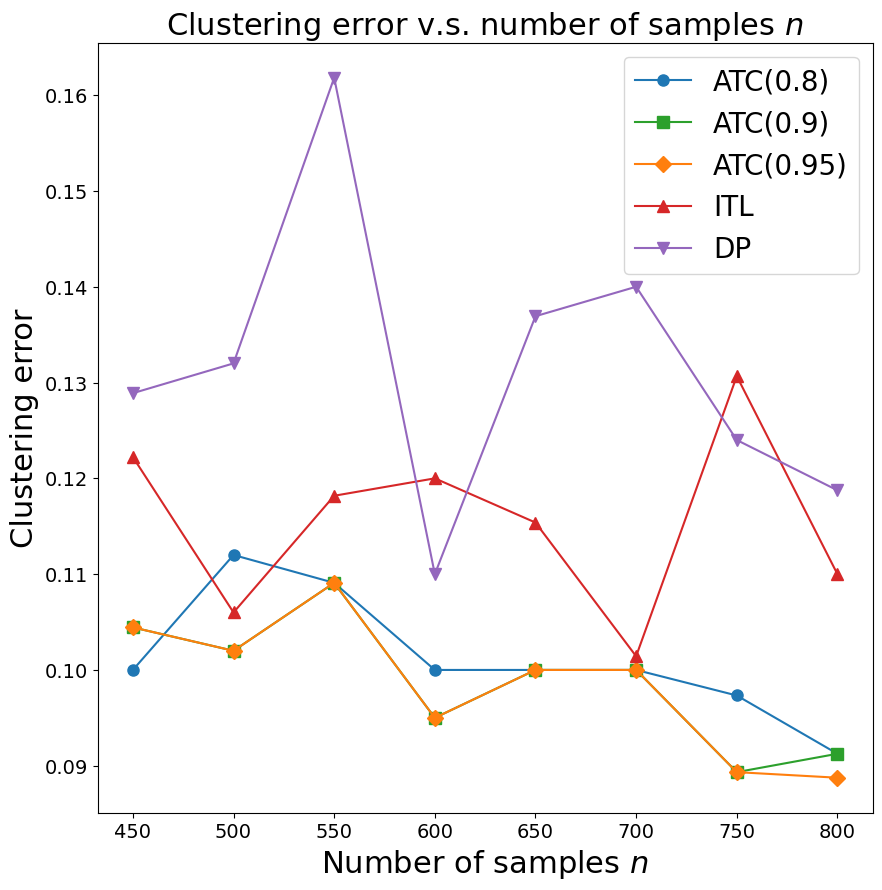}
    \includegraphics[width=0.45\linewidth]{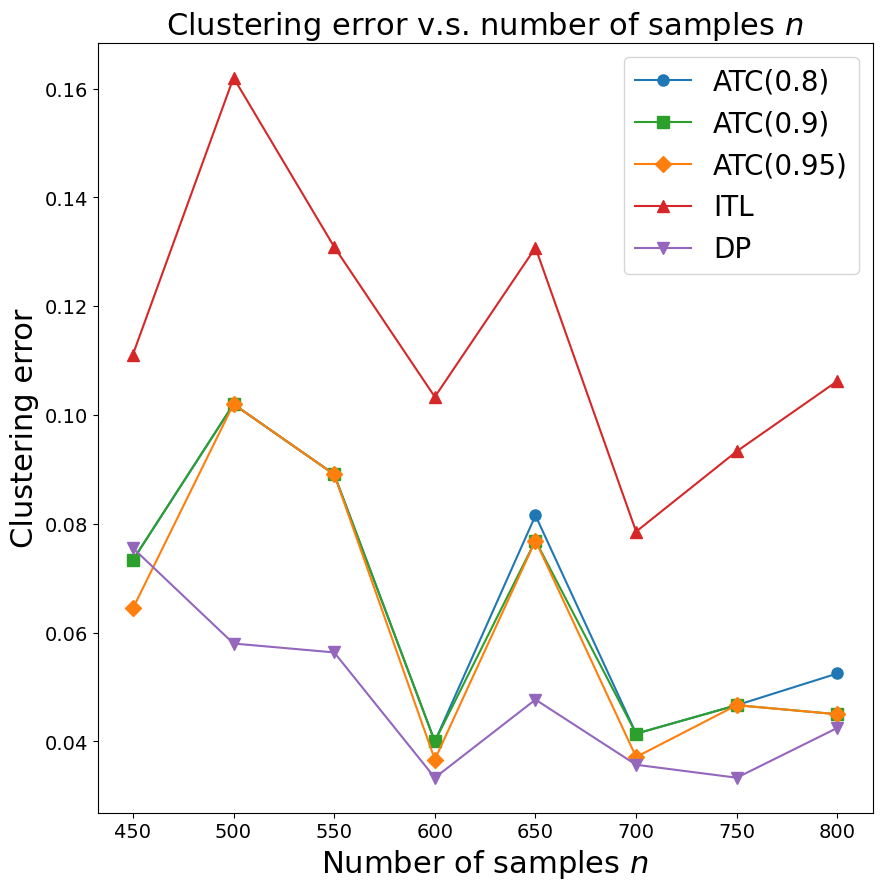}
    \caption{Clustering error v.s. number of samples $n$.  (Left) Target: GMM; Source: GMM. Parameters are set to be $(\varepsilon,K,d)=(0.2, 2,  10)$ and $n\in\ebrac{100,150,\cdots,500}$.  (Right) Target: GMM; Source: GMM. Parameters are set to be $(\varepsilon,K,d)=(0, 2,  10)$ and $n\in\ebrac{100,150,\cdots,500}$.  Each error point represents the average of 50 replications.}
    \label{fig:gmm_err_n}
\end{figure}

\begin{figure}[!htp]
    \centering
    \includegraphics[width=0.45\linewidth]{figs/gmm_err_vs_eps_mu0.2_n_varying.png}
    \includegraphics[width=0.45\linewidth]{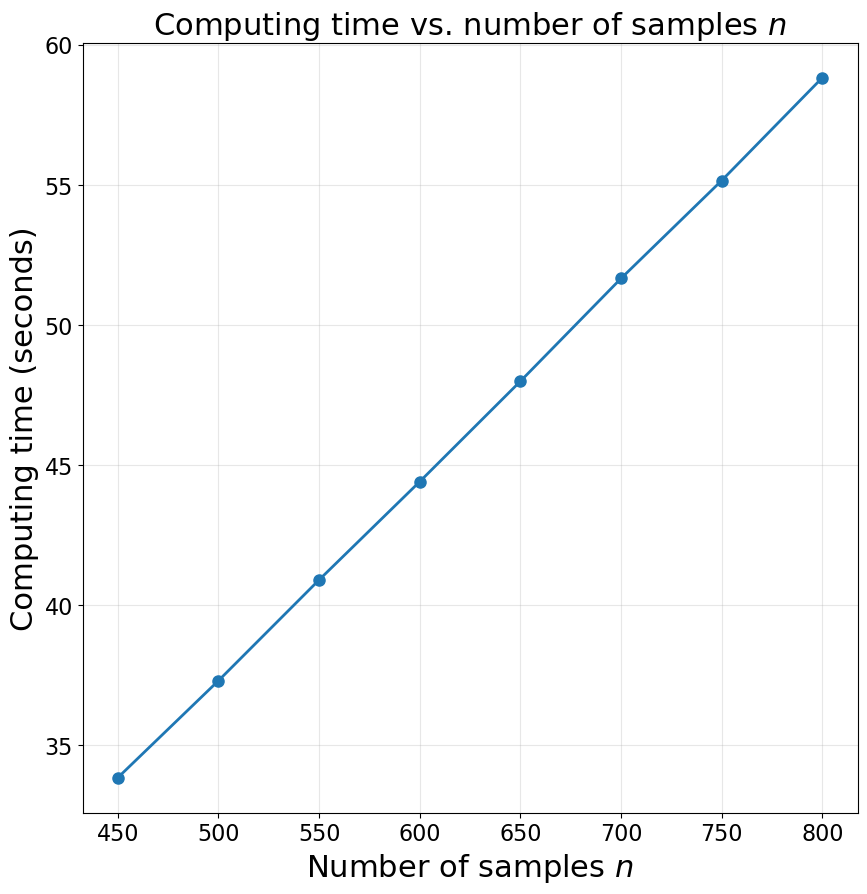}
    \caption{Robustness and scalability of \ATC~with $n$ varying. (Left) Clustering error v.s. number of samples $n$.  (Right) Computing time (seconds) v.s. number of samples $n$.  Target: GMM; Source: GMM. Parameters are set to be $(\varepsilon, K,d)=(0.2, 2,  10)$ and $n\in\ebrac{450,\cdots,800}$. Each point represents the average of 50 replications.}
    \label{fig:gmm_robust_n}
\end{figure}

\begin{figure}[!htp]
    \centering
    \includegraphics[width=0.45\linewidth]{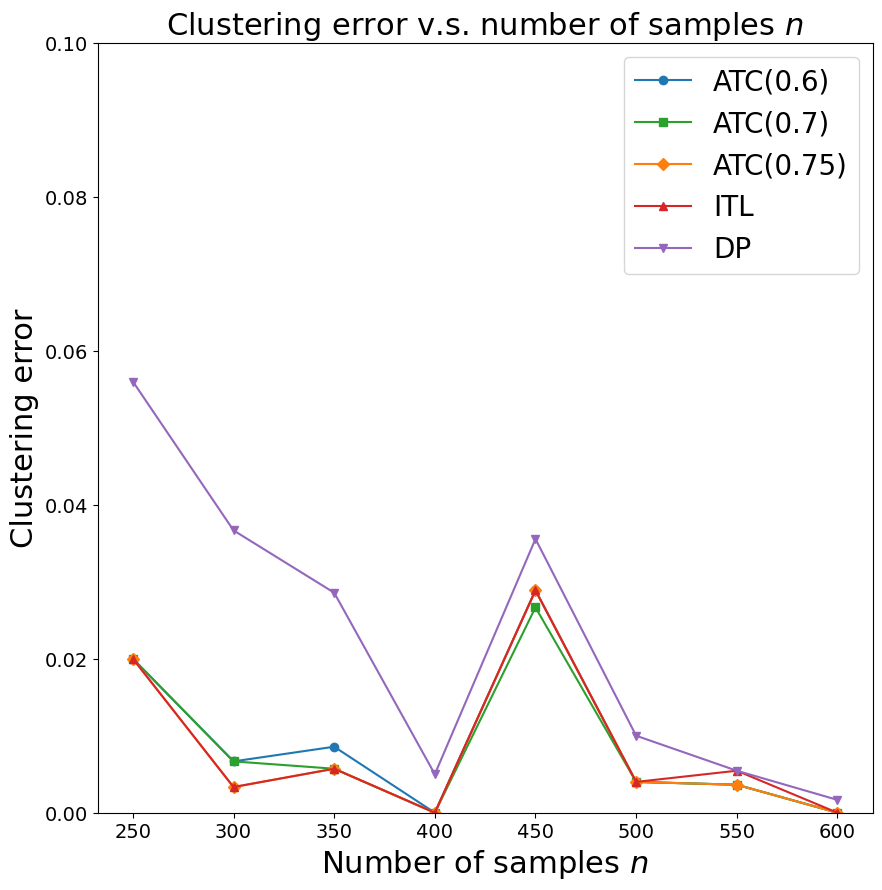}
    \includegraphics[width=0.45\linewidth]{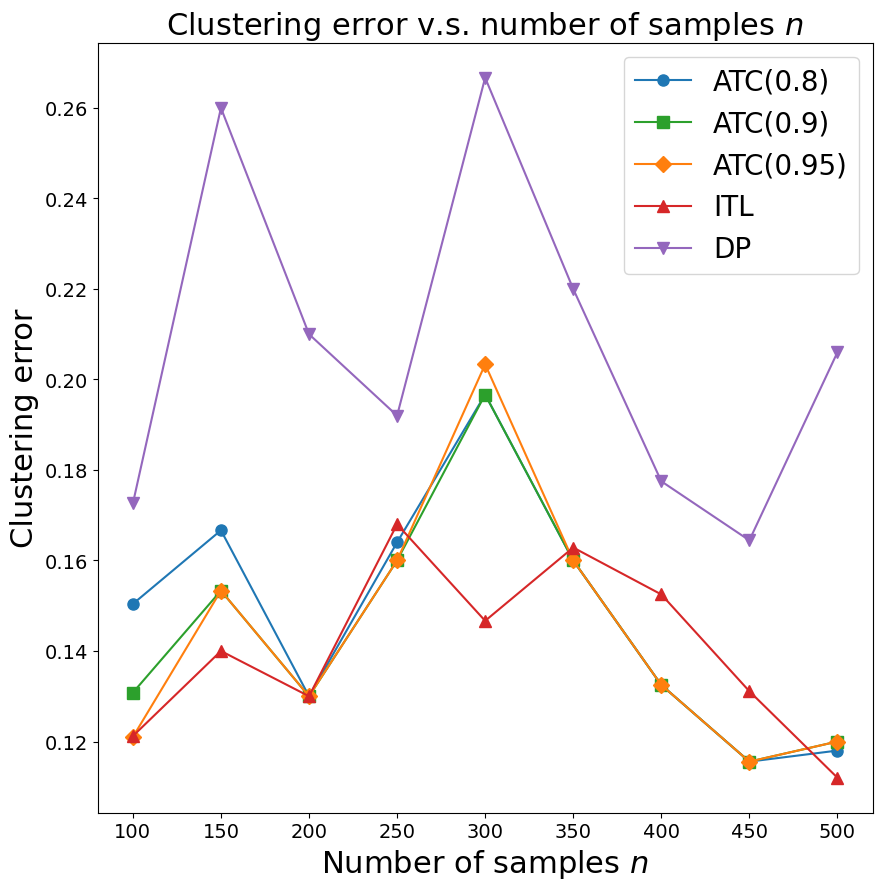}
    \caption{Clustering error v.s. number of samples $n$.  (Left) Target: SBM; Source: GMM. Parameters are set to be $(\varepsilon,K,d,p,q)=(0.2, 2,  10, 0.4, 0.25)$ and $n\in\ebrac{250,300,\cdots,600}$. (Right) Target: LCM; Source: GMM. Parameters are set to be $(\varepsilon,K,d,\Delta)=(0.2, 2,  10, 0.1)$ and $n\in\ebrac{100,150,\cdots,500}$. Each error point represents the average of 50 replications.}
    \label{fig:csbm_clcm_err_n}
\end{figure}

\color{black}
We also conduct experiments with $n$ varying with $(K,d)=(2,10)$ and $\varepsilon\in\ebrac{0,0.2}$.
\begin{itemize}[leftmargin=*]
    \item In the seventh setting (\Cref{fig:gmm_err_n} and \ref{fig:gmm_robust_n}), we use GMM as both target and source distribution with $n\in\ebrac{100,150,\cdots,500}$,  $\upsilon=0.2\log n$ under $\varepsilon=0.2$  (left) and $\varepsilon=0$  (right).
    \item In the eighth setting (left panel in \Cref{fig:csbm_clcm_err_n}), we use SBM as target  distribution and GMM as source distribution with $n\in\ebrac{200,250,\cdots,600}$, $\upsilon=0.25\log n$ and $(p,q)=(0.4, 0.25)$.
    \item In the ninth setting (right panel in \Cref{fig:csbm_clcm_err_n}), we use LCM as target  distribution and GMM as source distribution with $n\in\ebrac{100,150,\cdots,500}$, $\upsilon=0.25\log n$ and $\Delta=0.1$.
\end{itemize}

\color{black}
\section{Additional Real Data Applications}
\subsection{Business Relation Network}\label{subsec:BRN}
We  test our method on the Business Relation Network with contextual information, 
publicly available at \url{https://data.world/datasyndrome/relato-business-graph-database}. We use the data pre-processed by \cite{xu2023covariate}, from which we consider the supplier network among $n=312$ companies as the target data, and covariate vectors of dimension $d=102$ being the
standardized closing prices of these companies in the stock market  from 2021-01-01 to 2021-06-01 as the source data. The companies we consider here belong to $K=5$  
 sectors, including ``health'', ``information technology'',
``financials'', ``consumer discretionary'', and ``industrials'', which are treated as their true community labels $\bZ_0^*$ for target data.  {\color{black}True labels $\bZ_1^*$ were not available for the source data (closing prices). However, sectors provide a reasonable proxy for the target data (supplier network), as they are more accurately reflected by inter-company networks; thus, we treat them as $\bZ_0^*$.}

\begin{figure}[!tb]
    \centering
    \includegraphics[width=0.45\linewidth]{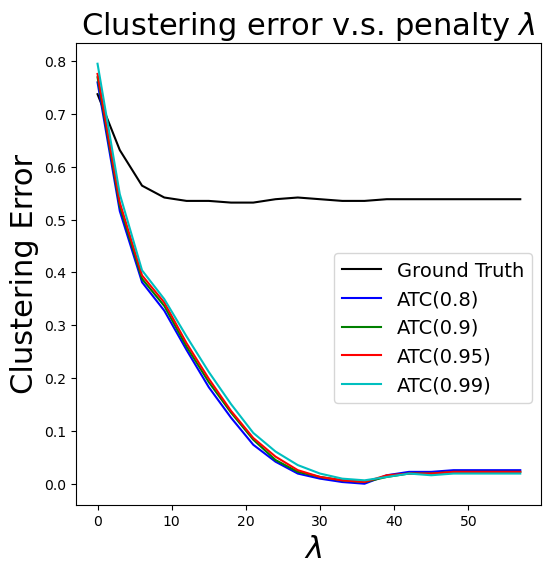}
    \caption{True clustering error (Ground Truth) and estimated excess error ({\sf ATC}) v.s. penalty $\lambda$ for Business Relation Network.  Target: supplier network (SBM); Source: standardized closing prices (GMM).}
    \label{fig:BRN_lam}
\end{figure}
\begin{table}[!htbp]
    \centering
    \resizebox{\linewidth}{!}{  
    \begin{tabular}{ccccccccccccc}
        \textbf{Type} & \textbf{CASC} & \textbf{SDP} & 
        \textbf{NAC} & \textbf{DP} & \textbf{Target} & 
        \textbf{Oracle} & \textbf{ATC(0.8)} & \textbf{ATC(0.9)} & \textbf{ATC(0.95)} & \textbf{ATC(0.99)} \\
        \hline
        Misclassification & 0.692 & 0.663& 0.625 & 0.551 & 0.734 & 
        {0.532} & {\bf 0.535} & {\bf 0.535} & {\bf 0.535} & {\bf 0.535} \\
        ARI &0.018 &0.032 &0.071 & 0.129  & 0.003  & 0.138 & {\bf 0.134} & {\bf 0.134} & {\bf 0.134} & {\bf 0.134}
    \end{tabular}
    }
    \caption{Error rates for business relation network }
    \label{tab:BRN_error}
\end{table}

Figure \ref{fig:BRN_lam} shows the performance of \ATC~with different  $1-\zeta\in\ebrac{0.8,0.9,0.95,0.99}$. It is suggested that we shall choose a large $\lambda$, which has a similar performance as data pooling. The misclassification error is $0.734$ by using the target data,  
which indicates the extremely low SNR in this dataset. By our adaptive choice of $\lambda$, we are able to achieve a lower misclassification error of $0.535$ for all choices of  $\zeta$. It's worth mentioning that the best misclassification error reported in \cite{xu2023covariate} is $0.599$, which utilizes $5$ layers of network. 

Similar to the lawyers network, we also  apply CASC, SDP, and  NAC to this dataset and the result is given in \Cref{tab:BRN_error}.  Our approach $\sf ATC$ achieves the best performance in terms of misclassification error and ARI. Note that the ATC improvement over DP in \Cref{tab:BRN_error} is $0.016$, i.e., about $312\times 0.016\approx 5$ companies.
While this margin over DP is modest, ATC yields a substantial gain over target-only learning ($0.734\to 0.535$,
about $62$ companies) and remains close to the oracle ($0.532$), suggesting that in this low-SNR setting aggressive borrowing is beneficial and data pooling is already near-optimal. 

\section{Auxiliary Lemmas}
\begin{lemma}\label{lem:bernoulli-bernstein}
    Let $\ebrac{X_{0,i}}_{i=1}^n$ be i.i.d. \textsf{Ber}$\brac{p}$. For any $\vartheta\in(0,1],\delta\in(0,1)$ we have
\begin{align*}
    &\PP\brac{\frac{1}{n}\sum_{i=1}^nX_{0,i}\le \brac{1+\vartheta}p+\frac{7\log\brac{1/\delta}}{6\vartheta n}}\ge 1-\delta,\\
    &\PP\brac{\frac{1}{n}\sum_{i=1}^nX_{0,i}\ge \brac{1-\vartheta}p-\frac{7\log\brac{1/\delta}}{6\vartheta n}}\ge 1-\delta.
\end{align*}
\end{lemma}
\textit{Proof}: see Section \ref{sec-lem:bernoulli-bernstein}.
\begin{lemma}\label{lem:approx-cdf-normal} 
For any $x>0$, we have
    \begin{align*}
        \frac{\sqrt{2/\pi}}{x+\sqrt{x^2+4 }}\exp\brac{-\frac{x^2}{2}}\le \Phi\brac{-x}\le \frac{\sqrt{2/\pi}}{x+\sqrt{x^2+8/\pi }}\exp\brac{-\frac{x^2}{2}}.
    \end{align*}
\end{lemma}
\begin{lemma}\label{lem:perturb-cdf-normal} 
For any $x\in\RR$ and $\Delta>0$, we have 
    \begin{align*}
        \Phi\brac{x+\Delta}-\Phi\brac{x}\le \frac{\Delta}{\sqrt{2\pi}}\exp\brac{-\frac{\min\ebrac{x^2,\brac{x+\Delta}^2}}{2}},
    \end{align*}
    and
    \begin{align*}
        \Phi\brac{x+\Delta}-\Phi\brac{x}\ge \frac{\Delta}{\sqrt{2\pi}}\exp\brac{-\frac{\max\ebrac{x^2,\brac{x+\Delta}^2}}{2}}. 
    \end{align*}
\end{lemma}
\textit{Proof}: The result immediately follows by definition of $\Phi$.

We also present the following assumption and lemma which are needed in our main proofs.

\begin{assumption}\label{assump:par-est}
    Consider $\brac{\bX_0,\bX_1}$ which follows the two-component symmetric univariate GMM model with  parameter $\brac{\mu,\sigma,\varepsilon}$. Suppose $\brac{\hat\mu,\hat\sigma}$ is independent of $\brac{\bX_0,\bX_1}$, and   there exists  an event $\calE_{\sf par}$ with $\PP\brac{\calE_{\sf par}}\ge 1-n^{-C_{\sf par}}$ such that on $\calE_{\sf par}$,
    \begin{align}\label{eq:assump-event}
        \max\ebrac{\frac{\ab{{\hat\mu}/{\hat\sigma}-{\mu}/{\sigma}}}{\mu/\sigma}, \frac{\ab{\hat \sigma/{\hat \mu}-\sigma/{\mu}}}{\sigma/{\mu}}}\le \delta_n,
    \end{align}
    for some $\delta_n=o(1)$ and $C_{\sf par}\ge 3$.
\end{assumption}
\begin{lemma}\label{lem:lower-bound-mis-error-lam-adap}
    Suppose Assumption \ref{assump:par-est} holds. For each $i\in[n]$,
\begin{align*}
    \Prob\brac{\bar Z^\lambda_{0,i}\ne Z^*_{0,i}}\ge &\max\ebrac{\Phi\brac{-\frac{\sqrt{2}\mu}{\sigma}}\brac{1-2\Phi\brac{-\frac{\lambda}{\sqrt{2}\mu/\sigma}}}, \sqbrac{\Phi\brac{-\frac{\mu}{\sigma}}}^2}\\
    &+\Phi\brac{-\frac{\mu}{\sigma}-\frac{\lambda}{2\mu/\sigma}}\Phi\brac{\frac{\mu}{\sigma}-\frac{\lambda}{2\mu/\sigma}}.
\end{align*}
\end{lemma}
\textit{Proof}: see Section \ref{sec-lem:lower-bound-mis-error-lam-adap}.

\section{Proofs of Main Results}
\sloppy
Throughout the proof,  we write $\PP_{a,b}\brac{\cdot}:=\PP\brac{\cdot\mid (Z^*_{0,i},Z^*_{1,i})=(a,b)}$ for brevity.

\subsection{Proof of Thereom \ref{thm:mis-error-lam} }\label{sec-prop:mis-error-lam-proof}

Define
\[
L(u, v) = \mu \brac{u X_{0,i}+vX_{1,i}}+\lambda\sigma^2\II\brac{u\ne v}, \qquad u,v \in \{ \pm 1 \}.
\]
If $\check Z^{\lambda}_{0,i}=-1$, then
\begin{align*}
    L(-1, 1) &\leq \min \{  L(1, 1), L(1, -1), L(-1, -1) \}
\quad\\
\text{or}\qquad L(-1, -1) &\leq \min \{ L(1, 1), L(1, -1), L(-1, 1) \}
\end{align*}
should happen. Let $K=\frac{\lambda \sigma^2}{2\mu} $. The aforementioned two events translate to 
\begin{align}\label{eq:-1+1-check-gmm}
     X_{0,i} \leq  X_{1,i}, \quad X_{0,i} \leq -K,\quad  X_{1,i}\geq K
\end{align}
and
\begin{align}\label{eq:-1-1-check-gmm}
        X_{0,i}+X_{1,i} \leq 0,\quad X_{0,i} \leq K,\quad X_{1,i} \leq K ,
\end{align}
respectively. We start with the case $(Z^*_{0,i},Z^*_{1,i})=(1,1)$. Note that
\begin{align*}
    &\PP_{1,1} \brac{ \text{Event }\eqref{eq:-1+1-check-gmm} } = \PP_{1,1}\brac{X_{0,i} \leq -K, X_{1,i} \geq K}
    = \Phi \bigg(
    \frac{-K-\mu}{\sigma} 
\bigg)
\bigg[
    1-\Phi\brac{\frac{K-\mu}{\sigma}}
\bigg]
\end{align*}
and
\begin{align*}
&\PP_{1,1}
\brac{
\text{Event }\eqref{eq:-1-1-check-gmm} 
}\le \PP_{1,1}\brac{X_{0,i}+X_{1,i} \leq 0}= \Phi \bigg(
-\frac{\sqrt{2}\mu}{\sigma}
\bigg).
\end{align*}
Hence,
\begin{align}
    \Prob_{1,1} ( \check  Z^\lambda_{0,i}\ne Z^*_{0,i} )\le \Phi\brac{-\frac{\sqrt{2}\mu}{\sigma}}+
 \Phi \bigg(
\frac{-K-\mu}{\sigma} 
\bigg)
\Phi
\bigg(
\frac{-K+\mu}{\sigma}
\bigg)
.
\label{eqn-prop-TL-1}
\end{align}
We then consider the case $(Z^*_{0,i},Z^*_{1,i})=(1,-1)$. Note that
\begin{align*}
&\PP_{1,-1} \brac{ \text{Event }\eqref{eq:-1+1-check-gmm}  } =  \PP_{1,-1}\brac{X_{0,i}\leq -K, X_{1,i} \geq 
K }=\Phi\brac{\frac{-K-\mu}{\sigma} }
\bigg[
1-\Phi\brac{\frac{K+\mu}{\sigma} }
\bigg]
.
\end{align*}
and
\begin{align*}
    &\PP_{1,-1} \brac{ \text{Event }\eqref{eq:-1-1-check-gmm}   }\le \PP_{1,-1}\brac{X_{0,i}<K }= \Phi\brac{\frac{K-\mu}{\sigma} }.
\end{align*}
Thus we can conclude that 
\begin{align}
    \Prob_{1,-1} ( \check Z^\lambda_{0,i}\ne Z^*_{0,i} ) \le  \Phi\brac{\frac{K-\mu}{\sigma} }+\Phi\brac{\frac{-K-\mu}{\sigma} }.
\label{eqn-prop-TL-2}
\end{align}
By symmetry, the right-hand sides of \eqref{eqn-prop-TL-1} and \eqref{eqn-prop-TL-2} can upper bound $\PP (  \check Z^{\lambda}_{0,i}\ne  Z^*_{0,i} | Z^*_{0,i} = Z^*_{1,i}  )$ and $\PP (  \check Z^{\lambda}_{0,i}\ne  Z^*_{0,i} | Z^*_{0,i} \neq Z^*_{1,i}  ) $, respectively. Therefore,
\begin{align*}
& \PP ( \check Z^{\lambda}_{0,i}\ne  Z^*_{0,i} )\\
&= \PP (  \check Z^{\lambda}_{0,i}\ne  Z^*_{0,i} | Z^*_{0,i} = Z^*_{1,i}  ) \PP ( Z^*_{0,i} = Z^*_{1,i} )
    + \PP (  \check Z^{\lambda}_{0,i}\ne  Z^*_{0,i} | Z^*_{0,i} \neq Z^*_{1,i}  ) \PP ( Z^*_{0,i} \neq Z^*_{1,i}  ) \\
    & \leq 
\bigg[
 \Phi\brac{-\frac{\sqrt{2}\mu}{\sigma}}+
\Phi \bigg(
\frac{-K-\mu}{\sigma} 
\bigg)
\Phi
\bigg(
\frac{-K+\mu}{\sigma}
\bigg)
\bigg]
 \cdot [1 -  \PP ( Z^*_{0,i} \neq Z^*_{1,i}  ) ] \\
&\quad + 
\bigg[
\Phi\brac{\frac{K-\mu}{\sigma} }+\Phi\brac{\frac{-K-\mu}{\sigma} }
\bigg]
\cdot  \PP ( Z^*_{0,i} \neq Z^*_{1,i}  ) \\
& =  \Phi\brac{-\frac{\sqrt{2}\mu}{\sigma}}+
\Phi \bigg(
\frac{-K-\mu}{\sigma} 
\bigg)
\Phi
\bigg(
\frac{-K+\mu}{\sigma}
\bigg)
\\
& \quad + 
\PP ( Z^*_{0,i} \neq Z^*_{1,i}  ) \cdot
\bigg\{
\Phi \bigg(
\frac{K-\mu}{\sigma} 
\bigg) \bigg[
1 + \Phi \bigg(
\frac{-K-\mu}{\sigma} 
\bigg)
\bigg]
- \Phi \bigg( -\frac{\sqrt{2}\mu}{\sigma} \bigg)
\bigg\} .
\end{align*}
We conclude that
\[
\PP ( \check Z^{\lambda}_{0,i}\ne  Z^*_{0,i} )
\leq 
 \Phi\brac{-\frac{\sqrt{2}\mu}{\sigma}}+
\Phi \bigg(
-\frac{\mu}{\sigma}-\frac{\lambda}{2\mu/\sigma}
\bigg)
\Phi\brac{\frac{\mu}{\sigma} - \frac{\lambda}{2\mu/\sigma}}
+ 2 \varepsilon \Phi \bigg(
-\frac{\mu}{\sigma}+\frac{\lambda}{2\mu/\sigma}
\bigg).
\]

\subsection{Proof of Theorem \ref{thm:oracle-gmm}}\label{pf-thm:oracle-gmm}
Without loss of generality we assume $\sigma=1$,  otherwise we can always proceed our analysis on standardized data $\ebrac{X_{0,i}/\sigma,i\in[n]}$. We take $\lambda^*=\log\brac{1/\varepsilon}$. By Proposition \ref{thm:mis-error-lam}, we obtain that 
\begin{align*}
    &\Prob\brac{\bar Z^\lambda_{0,i}\ne Z^*_{0,i}}\\
    &\le \brac{1-\varepsilon}\sqbrac{\Phi\brac{-\sqrt{2}\mu}+\Phi\brac{-\brac{\frac{\log\brac{1/\varepsilon}}{2\mu}-\mu} }\Phi\brac{-\mu-\frac{\log\brac{1/\varepsilon}}{2\mu }}}\\
    &+\varepsilon\sqbrac{2\Phi\brac{-\brac{\mu-\frac{\log\brac{1/\varepsilon}}{2\mu}} }\wedge1}.
\end{align*}
We first consider the case when $0\le \alpha<\sqrt{2}-1$, for which we obtain that 
\begin{align*}
    &\Prob\brac{\bar Z^\lambda_{0,i}\ne Z^*_{0,i}}\\
    &\le \Phi\brac{-\sqrt{2}\mu}+\Phi\brac{-\mu-\frac{\log\brac{1/\varepsilon}}{2\mu }}+2\exp\brac{-2\mu^2\cdot \alpha}\Phi\brac{-\brac{\mu-\frac{\log\brac{1/\varepsilon}}{2\mu}}}\\
    &\le\frac{1}{2\sqrt{\pi }\mu }\exp\brac{-\mu^2}+ \frac{1}{\sqrt{2\pi }(1+\alpha)\mu}\exp\brac{-\frac{\mu^2}{2}\brac{1+\alpha}^2}\\
    &+\frac{2}{\sqrt{2\pi}(1-\alpha)\mu }\exp\brac{-\frac{\mu^2}{2}(1+\alpha)^2}\\
    &\le \frac{1}{2\sqrt{\pi }\mu }\exp\brac{-\mu^2}+ \frac{3}{\sqrt{2\pi}(1-\alpha)\mu }\exp\brac{-\frac{\mu^2}{2}(1+\alpha)^2}\\
    &\le \frac{C_1}{\mu }\exp\brac{-\frac{\mu^2}{2}(1+\alpha)^2},
\end{align*}
for some universal constant $C_1>0$. where the last inequality holds since $(1+\alpha)^2\le 2$.

It remains to consider $\alpha>\sqrt{2}-1$. 
If $\sqrt{2}-1<\alpha<1/2$, then 
\begin{align*}
    &\Prob\brac{\bar Z^\lambda_{0,i}\ne Z^*_{0,i}}\\
        &\le \Phi\brac{-\sqrt{2}\mu}+\Phi\brac{-\mu-\frac{\log\brac{1/\varepsilon}}{2\mu }}+2\exp\brac{-2\mu^2\cdot \alpha}\Phi\brac{-\brac{\mu-\frac{\log\brac{1/\varepsilon}}{2\mu }}}\\
    &\le\frac{1}{2\sqrt{\pi }\mu }\exp\brac{-\mu^2}+ \frac{1}{\sqrt{2\pi }(1+\alpha)\mu}\exp\brac{-\frac{\mu^2}{2}\brac{1+\alpha}^2}\\
    &+\frac{2}{\sqrt{2\pi}(1-\alpha)\mu }\exp\brac{-\frac{\mu^2}{2}(1+\alpha)^2}\\
    &\le \frac{1}{2\sqrt{\pi }\mu }\exp\brac{-\mu^2}+ \frac{3}{\sqrt{2\pi}(1-\alpha)\mu }\exp\brac{-\frac{\mu^2}{2}(1+\alpha)^2}\\
    &\le \frac{\wt C_2}{\mu }\exp\brac{-{\mu^2}}.
\end{align*}
for some universal constant $\wt C_2>0$. If $\alpha\ge 1/2$  then 
\begin{align*}
    &\Prob\brac{\bar Z^\lambda_{0,i}\ne Z^*_{0,i}}\\
        &\le \Phi\brac{-\sqrt{2}\mu}+\Phi\brac{-\mu-\frac{\log\brac{1/\varepsilon}}{2\mu }}+\exp\brac{-2\mu^2\cdot \alpha}\\
    &\le\frac{1}{2\sqrt{\pi }\mu }\exp\brac{-\mu^2}+ \frac{1}{\sqrt{2\pi }(1+\alpha)\mu}\exp\brac{-\frac{\mu^2}{2}\brac{1+\alpha}^2}+\exp\brac{-2\mu^2\cdot \alpha}\\
    &\le \frac{1}{\sqrt{\pi}\mu}\exp\brac{-{\mu^2}}+\exp\brac{-\mu^2\cdot 2\alpha}.
\end{align*}
The proof is completed by collecting all bounds for three cases.

\subsection{Proof of Theorem \ref{thm:gmm-lb}}\label{sec-thm:gmm-lb-proof}
Fix $i\in[n]$, denote the posterior log-likelihood function for $( Z^*_{0,i},Z^*_{1,i} )$ as
\begin{align*}
    g\brac{z_0,z_1\mid X_0,X_1} =\log\PP\brac{Z^*_{0,i}=z_0,Z^*_{1,i}=z_1\mid X_{0,i}=x_0,X_{1,i}=x_1}.
\end{align*}
It is easy to see that 
\begin{align*}
        g\brac{z_0,z_1\mid x_0,x_1}
        &=\frac{\mu }{\sigma^2}\brac{z_0x_0+z_1x_1}-\log\brac{\frac{1-\varepsilon}{\varepsilon}}\II\brac{z_0\ne z_1}+C\brac{x_0,x_1}
\end{align*}
where $C\brac{x_0,x_1}$ is a function independent of $(z_0,z_1)$. For any $\bZ,\bZ^\prime\in\ebrac{\pm1}^n$, define $s  ( \bZ,\bZ^\prime ) =\argmin_{u\in\ebrac{\pm1}}\sum_{i=1}^n\II\brac{Z_i\ne uZ_i^\prime}$.
Notice that for any estimator $( \hat \bZ_{0},\hat \bZ_{1} )$ and any $i\in[n]$, we have 
\begin{align}\label{eq:lb-decomp}
    &\PP\brac{\brac{\hat Z_{0,i},\hat Z_{1,i}}\ne \brac{Z^*_{0,i}, Z^*_{1,i}}}\notag\\
    &\ge \PP\brac{g\brac{ Z^*_{0,i}, Z^*_{1,i}\mid X_{0,i},X_{1,i}}<\max_{\brac{z_0,z_1}\ne \brac{Z^*_{0,i}, Z^*_{1,i}}}g\brac{z_0,z_1\mid X_{0,i},X_{1,i}}}.
\end{align}
where the inequality holds due to the Bayes optimal estimator under $0$-$1$ loss is given by the posterior mode (MAP). 
It suffices to lower bound the event inside the RHS of \eqref{eq:lb-decomp}, denoted by $\calE_{\sf lb}$. 

Define $K=\frac{\sigma^2}{2\mu}\log\brac{\frac{1-\varepsilon}{\varepsilon}}$. When $\brac{ Z^*_{0,i}, Z^*_{1,i}}=\brac{1,1}$, $\calE_{\sf lb}$ is equivalent to 
\begin{align*}
    \ebrac{X_{0,i}<-K}\cup\ebrac{X_{1,i}<-K}\cup \ebrac{X_{0,i}+X_{1,i}<0}.
\end{align*}
Hence we have 
\begin{align*}
&\PP_{1,1} \brac{\calE_{\sf lb} } \\
&=\PP_{1,1}\brac{X_{0,i}<-K}+\PP_{1,1}\brac{X_{1,i}<-K}+\PP_{1,1}\brac{X_{0,i}+X_{1,i}<0}\\
&\quad -\PP_{1,1}\brac{\ebrac{X_{0,i}<-K}\bigcap\ebrac{X_{0,i}+X_{1,i}<0}}
    -\PP_{1,1}\brac{\ebrac{X_{1,i}<-K}\bigcap\ebrac{X_{0,i}+X_{1,i}<0}}\\
    &=\PP_{1,1}\brac{\ebrac{X_{0,i}<-K}\bigcap\ebrac{X_{0,i}+X_{1,i}>0}}
    +\PP_{1,1}\brac{\ebrac{X_{1,i}<-K}\bigcap\ebrac{X_{0,i}+X_{1,i}>0}} 
\\&\quad    +\PP_{1,1}\brac{X_{0,i}+X_{1,i}<0}\\
    &= \Phi\brac{-\frac{\sqrt{2}\mu}{\sigma}}+2\PP_{1,1}\brac{\ebrac{X_{0,i}<-K}\bigcap\ebrac{X_{0,i}+X_{1,i}>0}}.
\end{align*}
where the last equality holds due to $X_{0,i}\mid\{Z^*_{0,i}=Z^*_{1,i}\}\overset{d}{=}X_{1,i}\mid\{Z^*_{0,i}=Z^*_{1,i}\}$. We claim that
\begin{align}\label{eq:lb-claim}
    \PP_{1,1}\brac{\ebrac{X_{0,i}<-K}\cap\ebrac{X_{0,i}+X_{1,i}>0}}\ge \frac{1}{2}\PP_{1,1}\brac{X_{0,i}<-K}\PP_{1,1}\brac{X_{1,i}>K}.
\end{align}
To see it, let $\calS_0:=\ebrac{\brac{x,y}\in\RR^2:x<-K,x+y>0}$, $\calS_1:=\ebrac{\brac{x,y}\in\RR^2:y>K,x+y<0}$ and $p\brac{x}:=\frac{1}{\sqrt{2\pi}\sigma}\exp\brac{-\frac{\brac{x-\mu}^2}{2\sigma^2}}$. Define   $I_0:=\int_{\calS_0}p(x)p(y) \rd x \rd y$ and $I_1:=\int_{\calS_1}p(x)p(y) \rd x \rd y$. By definition we have $I_0=\PP_{1,1}\brac{\ebrac{X_{0,i}<-K}\bigcap\ebrac{X_{0,i}+X_{1,i}>0}}$ and $$I_0+I_1=\PP_{1,1}\brac{X_{0,i}<-K}\PP_{1,1}\brac{X_{1,i}>K}.$$ To show \eqref{eq:lb-claim}, it boils down  to show $I_0\ge I_1$. Notice that 
\begin{align*}
    I_1&=\int_{\calS_1}p(x)p(y) \rd x \rd y=\int_{\calS_0}p(-y)p(-x) \rd x \rd y&& \text{(By change of variable)}\\
    &=\int_{\calS_0}\frac{1}{2\pi\sigma^2}\exp\brac{-\frac{\brac{x+\mu}^2} {2\sigma^2}}\exp\brac{-\frac{\brac{y+\mu}^2}{2\sigma^2}} \rd x \rd y
    \\&
    =\int_{\calS_0}p(x)p(y)e^{-\frac{\mu}{\sigma^2}\brac{x+y}} \rd x \rd y\\
    &\le \int_{\calS_0}p(x)p(y) \rd x \rd y =I_0.&& (\text{Since~} e^{-\frac{\mu}{\sigma^2}\brac{x+y}}<1\text{~in~}\calS_0)
\end{align*}
Hence we can conclude that \eqref{eq:lb-claim} holds. Then,
\begin{align}
    &\PP_{1,1} \brac{\calE_{\sf lb}  }\ge
    \Phi\brac{-\frac{\sqrt{2}\mu}{\sigma}}+\Phi\brac{\frac{-\mu-K}{\sigma}}\Phi\brac{\frac{\mu-K}{\sigma}}.
\label{eqn-lb-11}
\end{align}

When $\brac{ Z^*_{0,i}, Z^*_{1,i}}=\brac{1,-1}$, $\calE_{\sf lb}$ is equivalent to 
\begin{align*}
    \ebrac{X_{0,i}<K}\cup\ebrac{X_{1,i}>-K}\cup \ebrac{X_{0,i}<X_{1,i}}.
\end{align*}
Hence we have 
\begin{align}
    &\PP_{1,-1}\brac{ \calE_{\sf lb}  }
=1-\PP_{1,-1}\brac{\ebrac{X_{0,i}>K}\cap\ebrac{X_{1,i}<-K}\cap \ebrac{X_{0,i}>X_{1,i}}} \notag\\
    &=1-\PP_{1,-1}\brac{X_{0,i}>K}\PP_{1,-1}\brac{X_{1,i}<-K} \notag\\
    &=1 - \sqbrac{\PP_{1,-1}\brac{X_{0,i}>K}}^2 \notag\\
& = 1 - 
\Phi^2 \bigg(
\frac{\mu - K}{\sigma}
\bigg)
\ge \PP_{1,-1}\brac{X_{0,i}<K}=\Phi\brac{\frac{K - \mu}{\sigma} 
},
\label{eqn-lb-1-1}
\end{align}
where the penultimate equality holds due to $X_{0,i}\mid\{Z^*_{0,i}=-Z^*_{1,i}\}\overset{d}{=}-X_{1,i}\mid\{Z^*_{0,i}=-Z^*_{1,i}\}$. By symmetry, the investigations of $\brac{ Z^*_{0,i}, Z^*_{1,i}}=\brac{-1,-1}$ and $\brac{ Z^*_{0,i}, Z^*_{1,i}}=\brac{-1,1}$ are almost the same as the above and hence omitted. Combining \eqref{eq:lb-decomp}, \eqref{eqn-lb-11} and \eqref{eqn-lb-1-1}, we arrive at
\begin{align*}
& \PP\brac{\brac{\hat Z_{0,i},\hat Z_{1,i}}\ne \brac{ Z^*_{0,i}, Z^*_{1,i}}} 
\\& 
\ge ( 1-\varepsilon ) \sqbrac{
    \Phi\brac{-\frac{\sqrt{2}\mu}{\sigma}}+\Phi\brac{\frac{-\mu-K}{\sigma}}\Phi\brac{\frac{\mu-K}{\sigma}}  
}+ \varepsilon 
\Phi\brac{\frac{K - \mu}{\sigma} 
}.
\end{align*}
We finish the proof by using the fact that
\begin{align*}
\PP\brac{ (\hat Z_{0,i},\hat Z_{1,i}) \ne \brac{ Z^*_{0,i}, Z^*_{1,i}}} 
\leq 
\PP\brac{ \hat Z_{0,i} \ne  Z^*_{0,i} }
+
\PP\brac{ \hat Z_{1,i} \ne  Z^*_{1,i} }
= 2 \PP\brac{ \hat Z_{0,i} \ne  Z^*_{0,i} } .
\end{align*}

\subsection{Proofs of Theorems \ref{thm:gl-gmm}  and  \ref{thm:gl-gmm-adap}}\label{sec-thm:gl-gmm-proof}
Note that Theorem \ref{thm:gl-gmm} is a special case of Theorem \ref{thm:gl-gmm-adap}, Assumption \ref{assump:par-est} holds with  $\delta_n=0$. Hence it suffices to prove Theorem \ref{thm:gl-gmm-adap}.
 
We will invoke Lemma \ref{lem:oracle-gl}. To that end, we first need to show that 
\begin{align*}
    \PP\brac{\ell\brac{\hat \bZ_0^\lambda,\bar \bZ_0^\lambda}\le \phi\brac{\lambda},\forall\lambda\in\Lambda}\ge 1-\zeta
\end{align*}
with some $\zeta\in(0,1)$ and a non-decreasing $\phi$. Notice that,
\begin{align*}
    \ell\brac{\hat \bZ_0^\lambda,\bar \bZ_0^\lambda}\le \ell\brac{\check  \bZ_0^\lambda,\bar \bZ_0^\lambda}+\ell\brac{\hat \bZ_0^\lambda,\check \bZ_0^\lambda},\quad \lambda\in\Lambda.
\end{align*}

We need the following two lemmas whose proofs are deferred to \Cref{sec-lem:checkbar-prob-proof} and \Cref{sec-lem:checkhat-prob-proof}. 
\begin{lemma}\label{lem:checkbar-prob}
For each $i\in[n]$ and $\lambda\ge 0$, we have
\begin{align*}
\PP\brac{\bar Z^\infty_{0,i}\ne \bar Z^\lambda_{0,i}}&\le 4
\Phi\brac{-\frac{\mu}{\sigma}- \frac{\lambda \sigma}{2\mu}}\Phi\brac{\frac{\mu}{\sigma}-\frac{\lambda \sigma}{2\mu}}
,\\
    \PP\brac{\check  Z_{0,i}^\lambda\ne \bar Z_{0,i}^\lambda}&\le \varepsilon\sqbrac{2\Phi\brac{-\frac{\mu}{\sigma}+\frac{\lambda}{2\mu/ \sigma}}\wedge1 }.
\end{align*}
\end{lemma}

\begin{lemma}\label{lem:checkhat-prob}
        Suppose  Assumption \ref{assump:par-est} holds with  $\delta_n=o\brac{\brac{\mu/\sigma}^{-2}}$, then for each $i\in[n]$ and $\lambda\ge 0$ we have
    \begin{align*}
    \PP\brac{\hat Z_{0,i}^\lambda\ne \check Z_{0,i}^\lambda}&\le \frac{C\delta_n\lambda\brac{1+\delta_n\lambda}}{\mu/\sigma} \sqbrac{\Phi\brac{-\frac{\mu}{\sigma}-\frac{\lambda}{2\mu/\sigma}}+\varepsilon\cdot \Phi\brac{-\frac{\mu}{\sigma}+\frac{\lambda}{2\mu/\sigma}}}+n^{-C_{\sf par}}.
\end{align*}
where $C>0$ is an universal  constant.
\end{lemma}

By Lemma \ref{lem:checkbar-prob} and Lemma \ref{lem:checkhat-prob}, we get that 
\begin{align*}
    \EE\ell\brac{\hat \bZ_0^\lambda,\bar \bZ_0^\lambda}\le \EE\ell\brac{\hat \bZ_0^\lambda,\check  \bZ_0^\lambda}+\EE\ell\brac{\check \bZ_0^\lambda,\bar \bZ_0^\lambda}\le 3\varepsilon \Phi\brac{-\frac{\mu}{\sigma}+\frac{\lambda}{2\mu/ \sigma}}.
\end{align*}
provided that $\lambda\lesssim\log n$, $\delta_n=o\brac{\brac{\log n}^{-1}}$ due to $\mu/\sigma\lesssim\sqrt{\log n}$ and $\delta_n=o\brac{\brac{\frac{\mu/\sigma}{\log n}}^2}$, and $C_{\sf par}\ge 3$. Combined with Lemma \ref{lem:bernoulli-bernstein}, with probability at least $1-\zeta/2$ that for any $\lambda\in\Lambda$,
\begin{align*}
    &\ell\brac{\hat \bZ_0^\lambda,\bar \bZ_0^\lambda}\le 4\varepsilon\Phi\brac{-\frac{\mu}{\sigma}+\frac{\lambda}{2\mu/ \sigma}}+\frac{7\log\brac{2M/\zeta}}{2n}=:\phi\brac{\lambda}
\end{align*}
It is readily seen that $\phi\brac{\lambda}$ is non-decreasing. 
 By Lemma \ref{lem:oracle-gl}, we obtain that with probability at least $1-\zeta/2$, 
\begin{align}\label{eq:oracle-gl-gmm}
        \ell\brac{\hat \bZ_0^{\hat \lambda},\bar \bZ_0^\infty}\le 4\min_{\lambda\in\Lambda}\ebrac{\phi\brac{\lambda}+\hat\psi\brac{\lambda}}+3\xi_{\hat\psi}.
\end{align}
Next we will invoke Lemma \ref{lem:bootstrap-quantile}. To that end, we need the following lemma to verify Assumption \ref{assump:par-est-general} holds under Assumption \ref{assump:par-est}, whose proof can be found in \Cref{sec-slem:assump-1-gmm}.
\begin{lemma}\label{lem:assump-1-gmm}
    Suppose  Assumption \ref{assump:par-est} holds with $\delta_n=o\brac{\frac{1}{\brac{\mu/\sigma}^{3}}\wedge \brac{\frac{\mu/\sigma}{\log n}}^2}$ and $\max_{\lambda\in\Lambda}\lambda\lesssim \log n$, then Assumption \ref{assump:par-est-general} is satisfied for two-component symmetric GMM with $\gamma=o(1)$.
\end{lemma}

By Lemma \ref{lem:bootstrap-quantile} and Lemma \ref{lem:assump-1-gmm} with $\gamma=0.004$, we can conclude that with probability at least $1-\zeta/2$,
\begin{align*}
    \xi_{\hat\psi}\le \frac{2105\log\brac{16/\zeta}}{n}+2\EE\ell\brac{\bar \bZ^\infty_{0},  \bZ_{0}^*},
\end{align*}
and
\begin{align*}
    \hat\psi\brac{\lambda}\le 8\EE\ell\brac{\bar \bZ^{\lambda}_{0}, \bar \bZ^\infty_{0}}+\frac{42\log\brac{32/\zeta}}{n}+0.05\EE\ell\brac{\bar \bZ^\infty_{0},\bZ_{0}^*}.
\end{align*}  
Combined with \eqref{eq:oracle-gl-gmm}, we can conclude that with probability at least $1-\zeta$,
\begin{align*}
    \ell\brac{\hat \bZ_0^{\hat \lambda},\bar \bZ_0^\infty}\le C\min_{\lambda\in\Lambda}\sqbrac{\varepsilon\Phi\brac{-\frac{\mu}{\sigma}+\frac{\lambda}{2\mu/ \sigma}}+\EE\ell\brac{\bar \bZ^{\lambda}_{0}, \bar \bZ^\infty_{0}}+\EE\ell\brac{\bar \bZ^\infty_{0},\bZ_{0}^*}+\frac{\log\brac{M/\zeta}}{n}}.
\end{align*}
for some universal constant $C>0$. By using $\EE\ell\brac{\bar \bZ^\infty_{0},\bZ_{0}^*}=\Phi\brac{-\sqrt{2}\mu/\sigma}$ and Lemma \ref{lem:checkbar-prob}, we obtain that  with probability at least $1-\zeta$,
\begin{align}\label{eq:hatbar-bound}
    \ell&\brac{\hat \bZ_0^{\hat \lambda},\bar \bZ_0^\infty}\\
    &\le \wt C\min_{\lambda\in\Lambda}\sqbrac{\Phi\brac{-\frac{\sqrt{2}\mu}{\sigma}}+\Phi\brac{-\frac{\mu}{\sigma}- \frac{\lambda \sigma}{2\mu}}\Phi\brac{\frac{\mu}{\sigma}-\frac{\lambda \sigma}{2\mu}}+\varepsilon\Phi\brac{-\frac{\mu}{\sigma}+\frac{\lambda}{2\mu/ \sigma}}+\frac{\log\brac{M/\zeta}}{n}}\notag.
\end{align}
On the other hand, by Lemma \ref{lem:bernoulli-bernstein} we have 
\begin{align}\label{eq:barinfstar-bound}
    &\PP\brac{\ell\brac{\bar \bZ_0^\infty,\bZ_0^*}\le 2\Phi\brac{-\frac{\sqrt{2}\mu}{\sigma}}+\frac{7\log\brac{2/\zeta}}{6 n}}\ge 1-\zeta/2.
\end{align}
The proof is completed by using triangle inequality and combining \eqref{eq:hatbar-bound} and \eqref{eq:barinfstar-bound}.

\subsection{Proof of Theorem \ref{thm:gl-mul-gmm-adap}}\label{pf-thm:gl-mul-gmm-adap}
We first prove the result for $d=1$ using  the original samples $\brac{\bX_0,\bX_1}$ with unknown $\brac{\mu,\sigma,\varepsilon}$ under \Cref{assump:par-est}, as the proof for $d\ge 2$ can be reduced to univariate case.

The following theorem can be regarded as a generalized version of \Cref{thm:gl-gmm} fully adaptive to $\brac{\mu,\sigma,\varepsilon}$ in the univariate case, whose proof can be found in \Cref{sec-thm:gl-gmm-proof}.
\begin{theorem}\label{thm:gl-gmm-adap}
    \sloppy Let $\hat  \bZ_0^{\hat \lambda}:=\ATC\brac{\bX_0,\bX_1;\Lambda,\hat\mu,\hat\sigma}$. Suppose Assumption \ref{assump:par-est} holds with $\delta_n=o\brac{\frac{1}{\brac{\mu/\sigma}^{3}}\wedge\brac{\frac{\mu/\sigma}{\log n}}^2}$. 
    Let $\zeta\in(0,1)$, and assume that $B\ge C_{0}\zeta^{-2}\log\brac{M\zeta^{-1}}$ for some universal constant $C_{0}>0$, then with probability at least $1-\zeta$, we have
    \begin{align*}
    \ell\brac{\hat  \bZ_0^{\hat \lambda}, \bZ_0^*}\le C_{1}\min_{\lambda\in\Lambda}\ebrac{\cM \brac{\frac{\mu}{\sigma} , \varepsilon, \lambda } +\frac{\log\brac{M\zeta^{-1}}}{n}}.
\end{align*}
for some universal constant $C_1>0$.
\end{theorem}
\sloppy
Now we turn to the proof for $d\ge 2$. By definition, for $\forall i\in[n]$ we have
\begin{align}\label{eq:adaptive-multi-gmm-procedure}
    \brac{\hat Z_{0,i}^{\lambda},\hat Z_{1,i}^{\lambda}}&=\argmin_{u,v\in\{\pm1\}}\left \{-\inp{\hat \bmu}{u \bX_{0,i}+v\bX_{1,i}}+\lambda\hat \sigma^2\II\brac{u\ne v}\right \}\\
    &= \argmin_{u,v\in\{\pm1\}}\left \{-\op{\hat\bmu}\brac{\frac{\hat\bmu^\top \bX_{0,i}}{\op{\hat\bmu}}+\frac{\hat\bmu^\top \bX_{1,i}}{\op{\hat\bmu}}}+\lambda\hat \sigma^2\II\brac{u\ne v}\right \}\notag.
\end{align}
In view of \eqref{eq:adaptive-multi-gmm-procedure}, we can construct projected samples $\big\{\wt X_{0,i},\wt X_{1,i}\big\}_{i=1}^n$ using the estimated parameters   $\brac{\hat\bmu,\hat\sigma}$:
\begin{align*}
    \wt X_{0,i}:=\frac{\hat\bmu^\top \bX_{0,i}}{\op{\hat\bmu}},\quad \wt X_{1,i}:=\frac{\hat\bmu^\top \bX_{1,i}}{\op{\hat\bmu}},\quad \forall i\in[n].
\end{align*}
We thus have $\TC\brac{\bX_0,\bX_1;\lambda,{\hat\bmu},\hat\sigma}=\TC\bbrac{\wt\bX_0,\wt\bX_1;\lambda,\op{\hat\bmu},\hat\sigma}$ for any $\lambda>0$. Note that we can view  $\brac{\op{\hat\bmu},\hat \sigma}$ as fixed parameters, which serves as a good approximation of $\brac{\op{\bmu},\sigma}$ by \Cref{assump:mul-par-est}.
Denote $\mu:=\op{\bmu}$ and $\wt \mu:=\op{\hat\bmu}^{-1}{\hat\bmu^\top\bmu}$.  Due to independence, we have 
$$\wt X_{0,i}| Z_{0,i}^*\sim N\bbrac{Z_{0,i}^*\wt\mu,\sigma^2},\quad\wt X_{1,i}| Z_{1,i}^*\sim N\bbrac{Z_{1,i}^*\wt\mu,\sigma^2},\quad \forall i\in[n].$$ 
In addition, let 
\begin{align*}
    \wt X^b_{0,i}:=\frac{\hat\bmu^\top \bX^b_{0,i}}{\op{\hat\bmu}},\quad \wt X_{1,i}:=\frac{\hat\bmu^\top \bX^b_{1,i}}{\op{\hat\bmu}},\quad \forall i\in[n],\quad \forall b\in[B].
\end{align*}
We thus have 
\begin{align*}
    \wt X_{0,i}^b| Z_{0,i}^*\sim N\bbrac{Z_{0,i}^*\op{\hat\bmu},\sigma^2},\quad\wt X^b_{1,i}| Z_{1,i}^*\sim N\bbrac{Z_{1,i}^*\op{\hat\bmu},\sigma^2},\quad \forall i\in[n],\quad \forall b\in[B].
\end{align*}
Similar to \eqref{eq:adaptive-multi-gmm-procedure}, we have $\TC\brac{\bX_0^b,\bX_1^b;\lambda,{\hat\bmu},\hat\sigma}=\TC\bbrac{\wt\bX_0^b,\wt\bX_1^b;\lambda,\op{\hat\bmu},\hat\sigma}$ for any $\lambda>0$. Therefore, $\hat  \bZ_0^{\hat \lambda}$ given by $\ATC$ using $\bbrac{ \bX_{0}, \bX_{1}}$ and  $\brac{\hat\bmu,\hat\sigma}$ is exactly the same as the output of $\ATC$ using the projected samples $\bbrac{\wt \bX_{0},\wt \bX_{1}}$ and $\brac{\op{\hat\bmu},\hat\sigma}$,  i.e., $\hat \bZ_0^{\hat \lambda}=\ATC\brac{\bX_0,\bX_1;\Lambda,{\hat\bmu},\hat\sigma}=\ATC\bbrac{\wt\bX_0,\wt\bX_1;\Lambda,\op{\hat\bmu},\hat\sigma}$. The problem is then reduced to the univariate case with samples $\big\{\wt X_{0,i},\wt X_{1,i}\big\}_{i=1}^n$ and unknown $\brac{\wt \mu,\sigma,\varepsilon}$. 

We start by noting that $\delta_n=o\brac{\frac{1}{\brac{\op{\bmu}/\sigma}^{3}}\wedge\brac{\frac{\op{\bmu}/\sigma}{\log n}}^2}$ implies that $\delta_n=o\brac{\frac{1}{\brac{\wt\mu/\sigma}^{3}}\wedge\brac{\frac{\wt\mu/\sigma}{\log n}}^2}$ by \Cref{assump:mul-par-est}, as 
\begin{align*}
     \ab{\frac{\wt\mu}{\mu}-1}=\frac{\ab{{\hat\bmu^\top\bmu}-\op{\hat\bmu}\op{\bmu }}}{\op{\hat\bmu}\op{\bmu }}\le \frac{2\op{\hat\bmu-\bmu}\op{\bmu} }{\brac{1-\delta_n}\op{\bmu }^2}=O\brac{\delta_n}.
\end{align*}
We have the following lemma whose proof is deferred to \Cref{pf:par-est}.
\begin{lemma}\label{lem:par-est} 
Under \Cref{assump:mul-par-est}, \Cref{assump:par-est} is satisfied for $\bbrac{\wt \bX_{0},\wt \bX_{1}}$ with parameter $\brac{\wt \mu,\sigma,\varepsilon}$. 
\end{lemma}
By \Cref{lem:par-est} and \Cref{thm:gl-gmm-adap}, we immediately obtain that with probability at least $1-\zeta$,
    \begin{align*}
    \ell\brac{\hat  \bZ_0^{\hat \lambda}, \bZ_0^*}\le C_{1}\min_{\lambda\in\Lambda}\ebrac{\cM \brac{\frac{\wt \mu}{\sigma} , \varepsilon, \lambda } +\frac{\log\brac{M\zeta^{-1}}}{n}}.
\end{align*}
Recall that 
\begin{align*}
    \cM ( s , \varepsilon, \lambda )= \Phi ( - \sqrt{2} s ) +\Phi \bigg(-s-\frac{\lambda}{2s}\bigg)\Phi\bigg(s - \frac{\lambda}{2 s }\bigg)+ 2 \varepsilon \Phi \bigg(- s +\frac{\lambda}{2 s }\bigg).
\end{align*}
 It suffices to show that $\cM \brac{\wt \mu/\sigma , \varepsilon, \lambda }\lesssim \cM \brac{\mu/\sigma ,\varepsilon, \lambda }$. By Lemma \ref{lem:approx-cdf-normal}, we have 
\begin{align*}
    \Phi\brac{-\frac{\sqrt{2}\wt \mu }{\sigma}}\lesssim\frac{\wt\mu}{\mu}\exp\brac{\frac{\ab{\mu^2-\wt\mu^2}}{\sigma^2}}\Phi\brac{-\frac{\sqrt{2} \mu }{\sigma}}.
\end{align*}

Moreover, using $\delta_n=O\brac{\brac{\log n}^{-1}}$ and $\mu^2/\sigma^2\lesssim \log n$, we have
\begin{align*}
    \exp\brac{\frac{\mu^2-\wt\mu^2}{\sigma^2}}=\exp\brac{\frac{\brac{\mu-\wt\mu}\brac{\mu+\wt\mu}}{\sigma^2}}\le \exp\brac{\frac{C\delta_n\mu^2}{\sigma^2}}\lesssim 1.
\end{align*}
Hence we obtain that $\Phi\brac{-\sqrt{2}\wt\mu/\sigma}\lesssim \Phi\brac{-\sqrt{2}\mu/\sigma}$. Then define
\begin{align*}
    \eta_0:={{\frac{\mu}{\sigma}-\frac{\lambda}{\mu/\sigma}}},\quad \Delta:=\ab{- \frac{\wt \mu}{\sigma} +\frac{\lambda}{\wt \mu/\sigma }-\brac{-\frac{ \mu}{\sigma} +\frac{\lambda}{ \mu/\sigma }}}
\end{align*}
Notice that \Cref{assump:mul-par-est} implies that $\Delta=O\brac{\delta_n\ab{\eta_0}}$. Hence  by \Cref{lem:perturb-cdf-normal} we  have
\begin{align*}
    \Phi \brac{- \frac{\wt \mu}{\sigma} +\frac{\lambda}{\wt \mu/\sigma }}&\le \Phi\brac{- \eta_0}+\frac{C\delta_n\ab{\eta_0}}{\sqrt{2\pi}}\exp\brac{-\frac{\eta_0^2}{2}}\exp\brac{C\delta_n\eta_0^2}\\
    &\lesssim \Phi\brac{- \eta_0}+\frac{\delta_n\ab{\eta_0}}{\sqrt{2\pi}}\exp\brac{-\frac{\eta_0^2}{2}}.
\end{align*}
where the second inequality holds due to $\delta_n=O\brac{\brac{\log n}^{-1}}$ and $\ab{\eta_0}\lesssim \log n$. For $\eta_0>0$,  we have
\begin{align*}
    \Phi\brac{- \eta_0}\ge\frac{\sqrt{2/\pi}}{\eta_0+\sqrt{\eta_0^2+4 }}\exp\brac{-\frac{\eta_0^2}{2}}\gtrsim \frac{\delta_n\ab{\eta_0}}{\sqrt{2\pi}}\exp\brac{-\frac{\eta_0^2}{2}}.
\end{align*}
For $\eta_0<0$,  we have
\begin{align*}
    \frac{\delta_n\ab{\eta_0}}{\sqrt{2\pi}}\exp\brac{-\frac{\eta_0^2}{2}}<\frac{1}{2}<\Phi\brac{- \eta_0}.
\end{align*}
Thus we can conclude that 
\begin{align*}
    \Phi \brac{- \frac{\wt \mu}{\sigma} +\frac{\lambda}{\wt \mu/\sigma }}\lesssim \Phi \brac{-\eta_0}=\Phi \brac{- \frac{ \mu}{\sigma} +\frac{\lambda}{ \mu/\sigma }}.
\end{align*}
Similarly, we could obtain that 
\begin{align*}
    \Phi \brac{\frac{\wt \mu}{\sigma} -\frac{\lambda}{\wt \mu/\sigma }}\lesssim \Phi \brac{\frac{ \mu}{\sigma} -\frac{\lambda}{ \mu/\sigma }},\quad \Phi \brac{-\frac{\wt \mu}{\sigma} -\frac{\lambda}{\wt \mu/\sigma }}\lesssim \Phi \brac{-\frac{ \mu}{\sigma} -\frac{\lambda}{ \mu/\sigma }}.
\end{align*}
We thus conclude the proof by using the definition of $\cM$.

\subsection{Proof of Theorem \ref{thm:gl-contextual-lcm-adap}}\label{pf-thm:gl-contextual-lcm-adap}
It suffices to prove the result for $d_2=1$ using  the original samples $\brac{\bX_0,\bX_1}$, as the proof for $d_2\ge 2$ can be reduced to univariate case similar to the proof of \Cref{thm:gl-mul-gmm-adap}. Then our estimator reduces to 
\begin{align*}
   \brac{\check Z_{0,i}^\lambda,\check Z_{1,i}^\lambda} \in\argmin_{u,v\in\ebrac{\pm 1}}\ebrac{-\sum_{j=1}^d\sqbrac{X_{0,ij}\log p_u+\brac{1-X_{0,ij}}\log \brac{1-p_u}}+\frac{\brac{X_{1,i}-v\mu}^2    }{2\sigma^2}+\lambda\II(u\ne v)}.
\end{align*}
Note that we can write $X_{0,ij}=\II\big(E_{0,ij}\le P_{j(Z^*_{0,i}+1)/2}\big)$ where $\bE_0\in\RR^{n\times d_1}$ has i.i.d. Unif$[0,1]$ entries,  and $\bX_{1}=\mu\bZ_1^*+\bE_1$ where $\bE_1\in\RR^{n}$ has i.i.d $N(0,\sigma^2)$ entries. By construction, the imaginary source data is generated as $\bar \bX_{0}=\mu\bZ_0^\star+\bE_1$, and
\begin{align*}
   \brac{\bar Z_{0,i}^\lambda,\bar Z_{1,i}^\lambda} \in\argmin_{u,v\in\ebrac{\pm 1}}\ebrac{-\sum_{j=1}^d\sqbrac{X_{0,ij}\log p_u+\brac{1-X_{0,ij}}\log \brac{1-p_u}}+\frac{(\bar X_{0,i}-v\mu)^2    }{2\sigma^2}+\lambda\II(u\ne v)}.
\end{align*}
We will invoke Lemma \ref{lem:oracle-gl}. To that end, we first need to show that 
\begin{align*}
    \PP\brac{\ell\brac{\check \bZ_0^\lambda,\bar \bZ_0^\lambda}\le \phi\brac{\lambda},\forall\lambda\in\Lambda}\ge 1-\zeta
\end{align*}
with some $\zeta\in(0,1)$ and a non-decreasing $\phi$. We need the following two lemmas whose proofs are deferred to \Cref{pf-lem:checkbar-prob-lcm} and \Cref{pf-lem:barinflam-prob-lcm} . 
\begin{lemma}\label{lem:checkbar-prob-lcm}
    For each $i\in[n]$ and $\lambda\ge 0$, we have
    \begin{align*}
          \PP\brac{\check Z_{0,i}^\lambda\ne \bar Z_{0,i}^\lambda}\le \varepsilon\exp\Big[-d\min\ebrac{D_{\rm KL}\brac{p^*+\delta_\lambda\mid\mid  a},D_{\rm KL}\brac{p^*-\delta_\lambda\mid\mid  b}}\II\brac{\lambda\le \lambda_{\min}}\Big],
    \end{align*}
    where we define $\lambda_{\min}:=d\min\ebrac{D_{\rm KL}\brac{a\mid\mid  b},D_{\rm KL}\brac{b\mid\mid  a}}$,
    \begin{align*}
    p^*:=\brac{\log\frac{1-b}{1-a}}/\sqbrac{{\log \frac{a(1-b)}{b(1-a)} }},\qquad \delta_{\lambda}:={\lambda}/\sqbrac{d\log \frac{a(1-b)}{b(1-a)} }.
\end{align*}
\end{lemma}
\begin{lemma}\label{lem:barinflam-prob-lcm}
    For each $i\in[n]$ and $\lambda\ge 0$, we have
    \begin{align*}
          &\PP\brac{\bar Z_{0,i}^\infty\ne \bar Z_{0,i}^\lambda}\le 2\Phi\brac{\frac{\mu}{
    \sigma}-\frac{\lambda \sigma}{2
    \mu}}\exp\Big[-d \min\ebrac{D_{\rm KL}\brac{p^*-\delta_\lambda\mid\mid a},D_{\rm KL}\brac{p^*+\delta_\lambda\mid\mid b}}\Big]\\
    &+\Phi\brac{-\frac{\mu}{
    \sigma}-\frac{\lambda \sigma}{2
    \mu}}\exp\Big[-d \min\ebrac{D_{\rm KL}\brac{p^*+\delta_\lambda\mid\mid a},D_{\rm KL}\brac{p^*-\delta_\lambda\mid\mid b}}\II\brac{\lambda\ge \lambda_{\max}}\Big],
    \end{align*}
    where $\lambda_{\max}:=d\max\ebrac{D_{\rm KL}\brac{a\mid\mid  b},D_{\rm KL}\brac{b\mid\mid  a}}$.
\end{lemma}
By Lemma \ref{lem:checkbar-prob-lcm}, we get that 
\begin{align*}
    \EE\ell\brac{\check \bZ_0^\lambda,\bar \bZ_0^\lambda}\le \varepsilon\exp\Big[-d\min\ebrac{D_{\rm KL}\brac{p^*+\delta_\lambda\mid\mid  a},D_{\rm KL}\brac{p^*-\delta_\lambda\mid\mid  b}}\II\brac{\lambda\le \lambda_{\min}}\Big].
\end{align*}
Combined with Lemma \ref{lem:bernoulli-bernstein}, with probability at least $1-\zeta/2$ that for any $\lambda\in\Lambda$,
\begin{align*}
    &\ell\brac{\check \bZ_0^\lambda,\bar \bZ_0^\lambda}\le \underbrace{2\varepsilon\exp\Big[-d\min\ebrac{D_{\rm KL}\brac{p^*+\delta_\lambda\mid\mid  a},D_{\rm KL}\brac{p^*-\delta_\lambda\mid\mid  b}}\II\brac{\lambda\le \lambda_{\min}}\Big]+\frac{7\log\brac{2M/\zeta}}{2n}}_{=:\phi\brac{\lambda}}.
\end{align*}
It is readily seen that $\phi\brac{\lambda}$ is non-decreasing. 
 By Lemma \ref{lem:oracle-gl}, we obtain that with probability at least $1-\zeta/2$, 
\begin{align}\label{eq:oracle-gl-clcm}
        \ell\brac{\check \bZ_0^{\hat \lambda},\bar \bZ_0^\infty}\le 4\min_{\lambda\in\Lambda}\ebrac{\phi\brac{\lambda}+\hat\psi\brac{\lambda}}+3\xi_{\hat\psi}.
\end{align}
Then we invoke Lemma \ref{lem:bootstrap-quantile} to conclude that with probability at least $1-\zeta/2$,
\begin{align*}
    \xi_{\hat\psi}\lesssim  \frac{\log\brac{16/\zeta}}{n},\qquad \hat\psi\brac{\lambda}\lesssim  \EE\ell\brac{\bar \bZ^{\lambda}_{0}, \bar \bZ^\infty_{0}}+\frac{\log\brac{32/\zeta}}{n}.
\end{align*}
Combined with \eqref{eq:oracle-gl-clcm}, we can conclude that with probability at least $1-\zeta$,
\begin{align*}
    \ell\brac{\hat \bZ_0^{\hat \lambda},\bar \bZ_0^\infty}\le C\min_{\lambda\in\Lambda}\sqbrac{\phi(\lambda)+\EE\ell\brac{\bar \bZ^{\lambda}_{0}, \bar \bZ^\infty_{0}}}.
\end{align*}
for some universal constant $C>0$. 
 We thus get the desired result by combining the above bound with \Cref{lem:barinflam-prob-lcm} and the following lemma, whose proof is in \Cref{pf-lem:barstar-prob-lcm}:
\begin{lemma}\label{lem:barstar-prob-lcm}
    For each $i\in[n]$, we have
    \begin{align*}
          &\PP\brac{\bar Z_{0,i}^\infty\ne  Z_{0,i}^*}\le \exp\ebrac{-\sqbrac{\frac{d }{2}D_{1/2}\brac{a\mid\mid b}+\frac{\mu^2}{2\sigma^2}}}.
    \end{align*}
\end{lemma}

\subsection{Proof of Theorem \ref{thm:gl-contextual-lcm-adap-asymp}}\label{pf-thm:gl-contextual-lcm-adap-asymp}
We start by noticing that under our asymptotics,
\begin{align*}
    p^*=b+D_{\rm KL}(b\mid\mid a)\sqbrac{\log\frac{a(1-b)}{b(1-a)}}^{-1}=\frac{a+b}{2}+o(a-b)
\end{align*}
is asymptotically the midpoint of $(b,a)$. Moreover, for $\delta_\lambda=O(a-b)$, 
\begin{align*}
    D_{\rm KL}(p^*-\delta_\lambda\mid\mid a)&=\frac{(a-b)^2}{8a}\brac{1+\frac{2\delta_\lambda}{a-b}}^2[1+o(1)],\\
    D_{\rm KL}(p^*+\delta_\lambda\mid\mid a)&=\frac{(a-b)^2}{8a}\brac{1-\frac{2\delta_\lambda}{a-b}}^2[1+o(1)],\\
    D_{\rm KL}(p^*-\delta_\lambda\mid\mid b)&=\frac{(a-b)^2}{8a}\brac{1-\frac{2\delta_\lambda}{a-b}}^2[1+o(1)],\\
    D_{\rm KL}(p^*+\delta_\lambda\mid\mid b)&=\frac{(a-b)^2}{8a}\brac{1+\frac{2\delta_\lambda}{a-b}}^2[1+o(1)],\\
    D_{1/2}(a\mid\mid b)&=\frac{(a-b)^2}{4a}[1+o(1)].
\end{align*}
Also, we have
\begin{align*}
    D_{\rm KL}(a\mid\mid b)=\frac{(a-b)^2}{2a}\brac{1+\frac{a-b}{3a}+o\brac{\frac{a-b}{a}}},\\
     D_{\rm KL}(b\mid\mid a)=\frac{(a-b)^2}{2a}\brac{1-\frac{a-b}{3a}+o\brac{\frac{a-b}{a}}},\\
     \lambda_{\min}=dD_{\rm KL}\brac{b\mid\mid a},\qquad  \lambda_{\max}=dD_{\rm KL}\brac{a\mid\mid b}.
\end{align*}
Collecting the above results and \Cref{thm:gl-contextual-lcm-adap} gives us the following simplified bound:
\begin{align}\label{eq:asymp-exp}
  & \ell\brac{\hat  \bZ_0^{\hat \lambda}, \bZ_0^*}\notag\\
  &\lesssim \exp\ebrac{-\brac{\textsf{SNR}_0+\textsf{SNR}_1}\sqbrac{1+o(1)}}\notag\\
    &+\min_{\lambda\in\Lambda}\Bigg\{\exp\brac{-\textsf{SNR}_1\brac{1-\frac{\lambda}{4\textsf{SNR}_1}}^2[1+o(1)]}\exp\brac{-\textsf{SNR}_0\brac{1+\frac{\lambda}{4\textsf{SNR}_0}}^2[1+o(1)]}\notag\\
    &+\exp\brac{-\textsf{SNR}_1\brac{1+\frac{\lambda}{4\textsf{SNR}_1}}^2[1+o(1)]}\exp\brac{-\textsf{SNR}_0\brac{1-\frac{\lambda}{4\textsf{SNR}_0}}^2\II\brac{\lambda\ge \lambda_{\max}}[1+o(1)]}\notag\\
    &+\varepsilon\exp\brac{-\textsf{SNR}_0\brac{1-\frac{\lambda}{4\textsf{SNR}_0}}^2\II\brac{\lambda\le \lambda_{\min}}[1+o(1)]}\Bigg\}+\frac{\log\brac{M/\zeta}}{n},
\end{align}
which holds with probability at least $1-\zeta$.

It remains to prove the second claim. Denote $s_k:=\textsf{SNR}_k$ for $k\in\ebrac{0,1}$. Under the current asymptotics, we can further upper bound \eqref{eq:asymp-exp} by replacing $1+o(1)$ term with $1-\rho$ for some $\rho=o(1)$. Hence we only need to focus on the following two terms based on the range of $\lambda$:
\begin{align}\label{eq:large-lam-exp}
    e^{-(1-\rho)\log(1/\varepsilon)}+e^{-(1-\rho)\sqbrac{s_0\brac{\frac{\lambda}{4s_0}-1}^2+s_1\brac{\frac{\lambda}{4s_1}+1}^2}}+e^{-(1-\rho)\sqbrac{s_1\brac{\frac{\lambda}{4s_1}-1}^2+s_0\brac{\frac{\lambda}{4s_0}+1}^2}},
    \end{align}
  for $\lambda\ge dD_{\rm KL}(a\mid\mid b)$, and
\begin{align}\label{eq:small-lam-exp}
e^{-(1-\rho)\sqbrac{\log(1/\varepsilon)+s_0\brac{1-\frac{\lambda}{4s_0}}^2}}+e^{-(1-\rho)s_1\brac{1+\frac{\lambda}{4s_1}}^2}+e^{-(1-\rho)\sqbrac{s_1\brac{1-\frac{\lambda}{4s_1}}^2+s_0\brac{1+\frac{\lambda}{4s_0}}^2}},
\end{align}
for $\lambda\le dD_{\rm KL}(b\mid\mid a)$.
We first consider $\lambda\ge dD_{\rm KL}(a\mid\mid b)$. Define
\begin{align*}
    G_{>,1}(\lambda):=-s_0\brac{\frac{\lambda}{4s_0}-1}^2-s_1\brac{\frac{\lambda}{4s_1}+1}^2,\quad G_{>,2}(\lambda):=-s_0\brac{\frac{\lambda}{4s_0}+1}^2-s_1\brac{\frac{\lambda}{4s_1}-1}^2.
\end{align*}
Some direct calculations lead to 
\begin{align*}
   G_{>,1}(\lambda)=  G_{>,2}(\lambda)=-\brac{\frac{\lambda^2}{16s_0}+\frac{\lambda^2}{16s_1}+s_0+s_1}.
\end{align*}
Therefore, we get
\begin{align*}
    \min_{\lambda>dD_{\rm KL}(a\mid\mid b)}\ebrac{\varepsilon+e^{G_{>,1}(\lambda)}+e^{G_{>,2}(\lambda)}}=\min_{\lambda>dD_{\rm KL}(a\mid\mid b)}\ebrac{\varepsilon+2\exp\sqbrac{-\brac{\frac{\lambda^2}{16s_0}+\frac{\lambda^2}{16s_1}+s_0+s_1}}}.
\end{align*}
The balanced choice of $\lambda$ is to satisfy 
\begin{align}\label{eq:large-lambda-choice}
\lambda>4\sqrt{\frac{s_0s_1}{s_0+s_1}\sqbrac{\log\brac{\frac{1}{\varepsilon}}-\brac{s_0+s_1}}_+}.
\end{align}
Thus  we have $ \min_{\lambda>dD_{\rm KL}(a\mid\mid b)}\ebrac{\varepsilon+e^{G_{>,1}(\lambda)}+e^{G_{>,2}(\lambda)}}\asymp \varepsilon$.

Then we consider $\lambda\le dD_{\rm KL}(b\mid\mid a)$. Define
\begin{align*}
    G_{<,1}(\lambda):=-s_1\brac{1+\frac{\lambda}{4s_1}}^2,\quad G_{<,2}(\lambda):=-s_0\brac{1+\frac{\lambda}{4s_0}}^2-s_1\brac{1-\frac{\lambda}{4s_1}}^2.
\end{align*}
Notice that 
\begin{align*}
    G_{<,1}(\lambda)-G_{<,2}(\lambda)&=-2s_1\brac{\frac{\lambda}{2s_1}}+s_0\brac{1+\frac{\lambda}{4s_0}}^2=s_0\brac{1-\frac{\lambda}{4s_0}}^2>0.
\end{align*}
Since $\lambda\le dD_{\rm KL}(b\mid\mid a)\le 4s_0$, we can conclude that 
\begin{align}\label{eq:small-lambda-exp}
    &\min_{\lambda<dD_{\rm KL}(b\mid\mid a)}\ebrac{ \varepsilon e^{-s_0\brac{1-\frac{\lambda}{4s_0}}^2}+e^{G_{<,1}(\lambda)}+e^{G_{<,2}(\lambda)}}\notag\\
    &\asymp \min_{\lambda<dD_{\rm KL}(b\mid\mid a)}\ebrac{\varepsilon\exp\sqbrac{-s_0\brac{1-\frac{\lambda}{4s_0}}^2}+\exp\sqbrac{-s_1\brac{1+\frac{\lambda}{4s_1}}^2}}.
\end{align}
In light of \eqref{eq:small-lambda-exp}, the balanced choice of $\lambda$ is to satisfy 
\begin{align}\label{eq:lam-eq-clcm}
\Gamma(\lambda)=\log\brac{\frac{1}{\varepsilon}}.
\end{align}
where we define $\Gamma(\lambda):=A\lambda^2+\lambda+s_1-s_0$  and $A:=\dfrac{1}{16}\brac{\dfrac{1}{s_1}-\dfrac{1}{s_0}}=\dfrac{s_0-s_1}{16s_0s_1}$. Notice that 
\begin{align*}
    \Gamma^\prime(\lambda)=1+\lambda\dfrac{s_0-s_1}{8s_0s_1}.
\end{align*}
When $s_0\ge s_1$, $\Gamma^\prime(\lambda)>0$. When $s_0< s_1$, $\Gamma^\prime(\lambda)$ decreases with $\lambda$ and  $\Gamma^\prime(4s_0)=\dfrac{1}{2}\brac{1+\dfrac{s_0}{s_1}}>0$. Hence we can conclude that $\Gamma(\lambda)$ in increasing on  $[0,4s_0]$.

We then consider the case when $s_1>s_0$. Moreover, when $0\le \log\brac{1/\varepsilon}<\Gamma(0)=s_1-s_0$, the first term in \eqref{eq:small-lambda-exp} always dominates and  we shall choose $\lambda=0$. When $\log\brac{1/\varepsilon}\in[\Gamma(0),\Gamma(4s_0)]=\sqbrac{s_1-s_0,\dfrac{\brac{s_0+s_1}^2}{s_1}}$, we can get the solution to \eqref{eq:lam-eq-clcm} as 
\begin{align*}
    \lambda_*=\frac{1- \sqrt{1-4A\sqbrac{s_0-s_1+\log\brac{1/\varepsilon}}}}{2A}=\frac{2\sqbrac{\log\brac{1/\varepsilon}+s_0-s_1}}{1+ \sqrt{1-4A\sqbrac{\log\brac{1/\varepsilon}+s_0-s_1}}}.
\end{align*}
where we have discarded the other root due to the range of $\lambda$. For $\log\brac{1/\varepsilon}>\Gamma(4s_0)=\dfrac{\brac{s_0+s_1}^2}{s_1}$, then the second term in \eqref{eq:small-lambda-exp} always dominates and  we shall choose $\lambda=dD_{\rm KL}(b\mid\mid a)$. To conclude, we choose
\begin{align*}
    \lambda_{\rm bal}:=\begin{cases}
       0, & \log\brac{1/\varepsilon}\in[0,s_1-s_0)\\
        \dfrac{2\sqbrac{\log\brac{1/\varepsilon}+s_0-s_1}}{1+ \sqrt{1-4A\sqbrac{\log\brac{1/\varepsilon}+s_0-s_1}}}, & \log\brac{1/\varepsilon}\in\Big[s_1-s_0,\dfrac{\brac{s_0+s_1}^2}{s_1}\Big] \\
        dD_{\rm KL}(b\mid\mid a), & \log\brac{1/\varepsilon}\in\Big[\dfrac{\brac{s_0+s_1}^2}{s_1},\infty \Big)
    \end{cases}.
\end{align*}
Similarly, when $s_1<s_0$, we choose
\begin{align*}
    \lambda_{\rm bal}:=\begin{cases}
        \dfrac{2\sqbrac{\log\brac{1/\varepsilon}+s_0-s_1}}{1+ \sqrt{1-4A\sqbrac{\log\brac{1/\varepsilon}+s_0-s_1}}}, & \log\brac{1/\varepsilon}\in\Big[0,\dfrac{\brac{s_0+s_1}^2}{s_1}\Big] \\
        dD_{\rm KL}(b\mid\mid a), & \log\brac{1/\varepsilon}\in\Big[\dfrac{\brac{s_0+s_1}^2}{s_1},\infty \Big)
    \end{cases}.
\end{align*}
Combined with \eqref{eq:small-lambda-exp}, we get
\begin{align*}
    &\min_{\lambda<dD_{\rm KL}(b\mid\mid a)}\ebrac{ \varepsilon e^{-s_0\brac{1-\frac{\lambda}{4s_0}}^2}+e^{G_{<,1}(\lambda)}+e^{G_{<,2}(\lambda)}}\notag\\
    &\lesssim  \begin{cases}
        \varepsilon \exp\brac{-s_0}, & \log\brac{1/\varepsilon}\in[0,s_1-s_0)\\
       {\varepsilon}^{1/2}\exp\brac{-\dfrac{s_0+s_1}{2}-\dfrac{\lambda_{*}^2}{32}\brac{\dfrac{1}{s_0}+\dfrac{1}{s_1}}}, & \log\brac{1/\varepsilon}\in\Big[s_1-s_0,\dfrac{\brac{s_0+s_1}^2}{s_1}\Big] \\
        \varepsilon, & \log\brac{1/\varepsilon}\in\Big[\dfrac{\brac{s_0+s_1}^2}{s_1},\infty \Big)
    \end{cases}.
\end{align*}
Combined with the case $\lambda>dD_{\rm KL}(a\mid\mid b)$ and \eqref{eq:large-lambda-choice}, we obtain the same bound holds up to constant over $\lambda\in[0,\infty)$, i.e.,
\begin{align}\label{eq:T-bound}
    &\min_{\lambda\in[0,\infty)}\Bigg\{\exp\brac{-s_1\brac{1-\frac{\lambda}{4s_1}}^2(1-\rho)}\exp\brac{-s_0\brac{1+\frac{\lambda}{4s_0}}^2(1-\rho)}\notag\\
    &\hspace{1cm}+\exp\brac{-s_1\brac{1+\frac{\lambda}{4s_1}}^2(1-\rho)}\exp\brac{-s_0\brac{1-\frac{\lambda}{4s_0}}^2\II\brac{\lambda\ge \lambda_{\max}}(1-\rho)}\notag\\
    &\hspace{1cm}+\varepsilon\exp\brac{-s_0\brac{1-\frac{\lambda}{4s_0}}^2\II\brac{\lambda\le \lambda_{\min}}(1-\rho)}\Bigg\}\notag\\
    &\lesssim  \begin{cases}
        \varepsilon^{1-\rho} \exp\brac{-\brac{1-\rho}s_0}, & \log\brac{1/\varepsilon}\in[0,s_1-s_0)\\
       {\varepsilon}^{(1-\rho)/2}\exp\ebrac{-\brac{1-\rho}\sqbrac{\dfrac{s_0+s_1}{2}+\dfrac{\lambda_{*}^2}{32}\brac{\dfrac{1}{s_0}+\dfrac{1}{s_1}}}}, & \log\brac{1/\varepsilon}\in\Big[s_1-s_0,\dfrac{\brac{s_0+s_1}^2}{s_1}\Big] \\
        \varepsilon^{1-\rho}, & \log\brac{1/\varepsilon}\in\Big[\dfrac{\brac{s_0+s_1}^2}{s_1},\infty \Big)
    \end{cases}.
\end{align}
Next, we claim the same upper bound holds for $\lambda\in\Lambda$ up to  constant. To see this, note that by construction, 
\begin{align*}
    \frac{\lambda_{j}-\lambda_{j-1}}{4s_k}=o\brac{\frac{1}{s_k}},\qquad j\in[M].
\end{align*}
Suppose $\lambda_{\rm bal}\in[\lambda_{j^*},\lambda_{j^*+1})$ for some $j^*=0,\cdots,M-1$. It is readily seen that 
\begin{align}\label{eq:exp-plus-approx}
   \exp\brac{-s_k\brac{1+ \frac{\lambda_{j^*}}{4s_k}}^2}&\le \exp\brac{-s_k\brac{1+ \frac{\lambda_{\rm bal}}{4s_k}}^2}\exp\brac{\frac{\lambda_{\rm bal}-\lambda_{j^*}}{2}}\notag\\
   &\lesssim\exp\brac{-s_k\brac{1+ \frac{\lambda_{\rm bal}}{4s_k}}^2}.
\end{align}
Similarly, we have 
\begin{align}\label{eq:exp-minus-approx}
   \exp\brac{-s_k\brac{1-\frac{\lambda_{j^*}}{4s_k}}^2} &\le \exp\brac{-s_k\brac{1-\frac{\lambda_{\rm bal}}{4s_k}}^2}\exp\brac{\frac{\lambda_{\rm bal}-\lambda_{j^*}}{2}}\notag\\
   &\lesssim\exp\brac{-s_k\brac{1-\frac{\lambda_{\rm bal}}{4s_k}}^2}.
\end{align}
By combining \eqref{eq:large-lam-exp}, \eqref{eq:small-lam-exp}, \eqref{eq:T-bound}, \eqref{eq:exp-plus-approx}, \eqref{eq:exp-minus-approx},  and the assumption on $\Lambda$, we have with probability at least $1-\zeta$,
\begin{align*}
     \ell\brac{\hat  \bZ_0^{\hat \lambda}, \bZ_0^*}\lesssim \sqbrac{\exp\brac{-\brac{s_0+s_1}}}^{1-\rho}+  \sqbrac{\calT(s_0,s_1,\varepsilon)}^{1-\rho}+\frac{\log (M/\zeta)}{n},
\end{align*}
where we define
\begin{align*}
  \calT&(s_0,s_1,\varepsilon):=\begin{cases}
        \varepsilon \exp\brac{-s_0}, & \log\brac{1/\varepsilon}\in[0,s_1-s_0)\\
       \varepsilon^{1/2}\exp\ebrac{-\dfrac{1}{2}\sqbrac{s_0+s_1+\dfrac{\lambda_{*}^2}{16}\brac{\dfrac{1}{s_0}+\dfrac{1}{s_1}}}}, & \log\brac{1/\varepsilon}\in\Big[s_1-s_0,\dfrac{(s_0+s_1)^2}{s_1}\Big] \\
         \varepsilon, & \log\brac{1/\varepsilon}\in\Big[\dfrac{(s_0+s_1)^2}{s_1},\infty \Big)
    \end{cases}.
\end{align*}
Observe that $\calT$ is piecewise in 
$L:=\log(1/\varepsilon)$, decreasing on each piece, and the endpoint values of adjacent pieces agree in order (hence no order-level discontinuity). Moreover, when $L\in\Big[\dfrac{(s_0+s_1)^2}{s_1},\infty \Big)$, we have $\calT(s_0,s_1,\varepsilon)\le \exp\brac{-\dfrac{(s_0+s_1)^2}{s_1}}=o\brac{\exp\brac{-(s_0+s_1)}}$. When $L\in[0,s_1-s_0)$, we have $\calT(s_0,s_1,\varepsilon)\ge  \exp\brac{-s_1}=\omega\brac{\exp\brac{-(s_0+s_1)}}$. Thus it suffices to consider $L\in\Big[s_1-s_0,\dfrac{(s_0+s_1)^2}{s_1}\Big]$. Define $ L^\dagger$ such that 
\begin{align}\label{eq:def-L-dag}
   \frac{ L^\dagger+s_0+s_1}{2}+\frac{\lambda(L^\dagger)^2}{32}\brac{\frac{1}{s_0}+\frac{1}{s_1}}=s_0+s_1,
\end{align}
where $\lambda(L)$ is the solution to the followin equation:
\begin{align}\label{eq:def-lam-L}
    \dfrac{s_0-s_1}{16s_0s_1}\lambda^2+\lambda+s_1-s_0=L.
\end{align}
Combining \eqref{eq:def-L-dag} and \eqref{eq:def-lam-L}, we get $  \lambda(L^\dagger)^2+16s_1\lambda(L^\dagger)-16s_0s_1=0$. Therefore, we get 
\begin{align*}
   \lambda(L^\dagger)=4s_1\brac{\sqrt{1+\frac{s_0}{s_1}}-1}.
\end{align*}
Plugging it back to \eqref{eq:def-L-dag}, we get
\begin{align*}
   L^\dagger=\frac{2\brac{s_0+s_1}}{1+\sqrt{1+s_0/s_1}}.
\end{align*}
It is not difficult to check that $L^\dagger\in \Big[s_1-s_0,\dfrac{(s_0+s_1)^2}{s_1}\Big]$. Then when $L\ge L^\dagger$, we have $\calT(s_0,s_1,\varepsilon)\le \exp\brac{-(s_0+s_1)}$ due to the definition of $L^\dagger$ and monotonicity of $\calT$ in $L$. It remains to show equivalence of $\calT(s_0,s_1,\varepsilon)$ and $\calT_r(\alpha)$ when $\log\brac{1/\varepsilon}\in\Big[s_1-s_0,\dfrac{(s_0+s_1)^2}{s_1}\Big]$, as the the case for $\log\brac{1/\varepsilon}\in [0,s_1-s_0)$ is straightforward. By definition, we have
\begin{align*}
    \lambda_*:=2s_1\cdot \frac{U_r(\alpha)}{1+\sqrt{1-c_rU_r(\alpha) }},
\end{align*}
and hence 
\begin{align*}
    \calT&(s_0,s_1,\varepsilon)=\exp\sqbrac{-(s_0+s_1)\brac{\alpha+\frac{1}{2}+\frac{\lambda_*^2}{32\brac{s_0+s_1}}\brac{\frac{1}{s_0}+\frac{1}{s_1}}}}.
\end{align*}
The proof for the first claim is thereby completed by noticing 
\begin{align*}
    \frac{\lambda_*^2}{32\brac{s_0+s_1}}\brac{\frac{1}{s_0}+\frac{1}{s_1}}=\frac{U^2_r(\alpha)}{8r\brac{1+\sqrt{1-c_rU_r(\alpha) }}^2}.
\end{align*}
For the second claim, it suffices to notice that the data pooling error rate $\exp(-(1-\rho)(s_0+s_1))$ dominates whenever
\begin{align*}
   \frac{\log(1/\varepsilon)}{2(s_0+s_1)}> \frac{L^\dagger}{2(s_0+s_1)}=\frac{1}{1+\sqrt{1+r}}.
\end{align*}

\section{Other Proofs}

\subsection{Proof of Corollary \ref{col:optimality-gmm}}\label{sec-col:optimality-gmm-proof}
We first show the following holds:
\begin{align}\label{eq:minimum-over-inf}
        \min_{\lambda\in\Lambda}\cM \brac{\frac{\mu}{\sigma} , \varepsilon, \lambda}\lesssim \cM \brac{\frac{\mu}{\sigma} , \varepsilon, \log\brac{\frac{1-\varepsilon}{\varepsilon}} }.
\end{align}
Suppose that $\log\brac{\varepsilon^{-1}\brac{1-\varepsilon}}\in[\lambda_{j^*},\lambda_{j^*+1})$ for some $j^*\in[M]$. Denote 
\begin{align*}
    \Delta:=\frac{\log n}{2M\mu/\sigma},\quad\eta_0:={{\frac{\mu}{\sigma}-\frac{\log\brac{\varepsilon^{-1}\brac{1-\varepsilon}}}{2\mu/\sigma}}}.
\end{align*}
If  $\ab{\eta_0}=o\brac{\mu/\sigma}$ we  have  $\log\brac{\varepsilon^{-1}\brac{1-\varepsilon}}=\brac{2\mu^2/\sigma^2}\brac{1+o(1)}$ and hence 
\begin{align*}
    \varepsilon\Phi \brac{- \frac{\mu}{\sigma} +\frac{\log\brac{\varepsilon^{-1}\brac{1-\varepsilon}}}{2 \mu/\sigma }}\le   \exp\brac{-\frac{2\mu^2}{\sigma^2}\brac{1+o(1)}}=o\brac{\Phi\brac{-\sqrt{2}\mu/{\sigma}}}.
\end{align*}
Similarly, we have
\begin{align*}
    &\Phi\brac{-\frac{\mu}{\sigma}-\frac{\log\brac{\varepsilon^{-1}\brac{1-\varepsilon}}}{2\mu/\sigma}}\Phi\brac{\frac{\mu}{\sigma}-\frac{\log\brac{\varepsilon^{-1}\brac{1-\varepsilon}}}{2\mu/\sigma}}\\
    &\le \Phi\brac{-\frac{2\mu}{\sigma}\brac{1+o(1)}}=o\brac{\Phi\brac{-\sqrt{2}\mu/{\sigma}}}.
\end{align*}
Since $\Delta=o\brac{\mu/\sigma}$, we can deduce that 
\begin{align*}
    &\varepsilon\Phi\brac{-\frac{\mu}{\sigma}+\frac{\lambda_{j^*}}{2\mu/\sigma}}\le \varepsilon\Phi\brac{-\frac{\mu}{\sigma}+\frac{\log\brac{\varepsilon^{-1}\brac{1-\varepsilon}}}{2\mu/\sigma}+\Delta}=o\brac{\Phi\brac{-\sqrt{2}\mu/{\sigma}}},\\
    &\Phi \brac{-\frac{\mu}{\sigma}-\frac{\lambda_{j^*}}{2\mu/\sigma}}\Phi\brac{\frac{\mu}{\sigma} - \frac{\lambda_{j^*}}{2\mu/\sigma }}=o\brac{\Phi\brac{-\sqrt{2}\mu/{\sigma}}}.
\end{align*}
Thus we have 
\begin{align*}
    &\min_{\lambda\in\Lambda}\ebrac{\Phi \brac{-\frac{\mu}{\sigma}-\frac{\lambda}{2\mu/\sigma}}\Phi\brac{\frac{\mu}{\sigma} - \frac{\lambda}{2\mu/\sigma }}+ 2 \varepsilon \Phi \brac{- \frac{\mu}{\sigma} +\frac{\lambda}{2 \mu/\sigma }}}\\
    &\le\Phi \brac{-\frac{\mu}{\sigma}-\frac{\lambda_{j^*}}{2\mu/\sigma}}\Phi\brac{\frac{\mu}{\sigma} - \frac{\lambda_{j^*}}{2\mu/\sigma }}+ 2 \varepsilon \Phi \brac{- \frac{\mu}{\sigma} +\frac{\lambda_{j^*}}{2 \mu/\sigma }}=o\brac{\Phi\brac{-\sqrt{2}\mu/{\sigma}}}.
\end{align*}
As a result of definition of $\cM$, we obtain that 
\begin{align*}
    \min_{\lambda\in\Lambda}\cM \brac{\frac{\mu}{\sigma} , \varepsilon, \lambda}\le \cM \brac{\frac{\mu}{\sigma} , \varepsilon, \log\brac{\frac{1-\varepsilon}{\varepsilon}} }.
\end{align*}
Notice that the above argument also holds for any  $\eta_0<0$, in which we have  $\log\brac{\varepsilon^{-1}\brac{1-\varepsilon}}>2\mu^2/\sigma^2$, and $\Phi\brac{-\sqrt{2}\mu/\sigma}$ is the dominant term. It suffices to consider $\eta_0\asymp\mu/\sigma$ ($\eta_0\le \mu/\sigma$ by definition). Notice that in this regime, we have
\begin{align*}
    \cM \brac{\frac{\mu}{\sigma} , \varepsilon, \log\brac{\frac{1-\varepsilon}{\varepsilon}} }\ge \Phi\brac{-\sqrt{2}\mu/\sigma}+\frac{1}{2}\Phi \brac{-\frac{\mu}{\sigma}-\frac{\log\brac{\varepsilon^{-1}\brac{1-\varepsilon}}}{2\mu/\sigma}}+ 2 \varepsilon \Phi \brac{- \eta_0}.
\end{align*}
On the other hand, by Lemma \ref{lem:perturb-cdf-normal}  we have 
\begin{align*}
    \Phi \brac{- \frac{\mu}{\sigma} +\frac{\lambda_{j^*+1}}{2 \mu/\sigma }}\le \Phi \brac{- \eta_0+\Delta}\le \Phi\brac{-\eta_0}+\frac{\Delta}{\sqrt{2\pi}}\exp\brac{-\frac{\brac{\eta_0-\Delta}^2}{2}}.
\end{align*}
Notice that  $\Delta\exp\brac{-{\brac{\eta_0-\Delta}^2}/{2}}\asymp\Delta\exp\brac{-{\eta_0^2}/{2}}$ and  $\Phi\brac{-\eta_0}\asymp{\eta_0}^{-1}\exp\brac{-\frac{\eta_0^2}{2}}$, hence we have 
\begin{align*}
    \Phi \brac{- \frac{\mu}{\sigma} +\frac{\lambda_{j^*+1}}{2 \mu/\sigma }}\lesssim\Phi\brac{-\eta_0},
\end{align*}
provided that $\Delta=o\brac{\brac{\mu/\sigma}^{-1}}$, which is further implied by $M=\omega\brac{\log n}$. Thus we have 
\begin{align*}
    &\min_{\lambda\in\Lambda}\ebrac{\Phi \brac{-\frac{\mu}{\sigma}-\frac{\lambda}{2\mu/\sigma}}\Phi\brac{\frac{\mu}{\sigma} - \frac{\lambda}{2\mu/\sigma }}+ 2 \varepsilon \Phi \brac{- \frac{\mu}{\sigma} +\frac{\lambda}{2 \mu/\sigma }}}\\
    &\le\Phi \brac{-\frac{\mu}{\sigma}-\frac{\lambda_{j^*+1}}{2\mu/\sigma}}\Phi\brac{\frac{\mu}{\sigma} - \frac{\lambda_{j^*+1}}{2\mu/\sigma }}+ 2 \varepsilon \Phi \brac{- \frac{\mu}{\sigma} +\frac{\lambda_{j^*+1}}{2 \mu/\sigma }}\\
    &\lesssim \Phi\brac{-\frac{\mu}{\sigma}-\frac{\log\brac{\varepsilon^{-1}\brac{1-\varepsilon}}}{2\mu/\sigma}}+\varepsilon\Phi\brac{-\eta_0}\lesssim  \cM \brac{\frac{\mu}{\sigma} , \varepsilon, \log\brac{\frac{1-\varepsilon}{\varepsilon}} }.
\end{align*}
This concludes the proof for \eqref{eq:minimum-over-inf}. It suffices to combine the \eqref{eq:minimum-over-inf}, \Cref{thm:gl-gmm} and \Cref{thm:oracle-gmm} to conclude that the additive term $\log (M\zeta^{-1})/n$ is ignorable provided that $\log\brac{\zeta^{-1}}=O\brac{\log n}$ and 
\begin{align*}
    \textsf{SNR}\cdot \min\ebrac{\brac{1+\alpha}^2,2}<\log n.
\end{align*}

\subsection{Proof of Proposition \ref{prop:ind-dp-gmm}}\label{sec-prop:ind-dp-gmm-proof}
Fix $i\in[n]$. We have $\{ X_{0,i} > 0 \} \subseteq \{ \check  Z^{\ITL}_{0,i}=1 \}
\subseteq \{ X_{0,i} \geq  0  \}$. Then,
\begin{align*}
    &\PP ( \check Z^{\ITL}_{0,i}\ne  Z^*_{0,i} \mid Z^*_{0,i}=1 ) \\
    &=\Prob\brac{X_{0,i}< 0 \mid Z^*_{0,i}=1 }=\Prob \bigg( \frac{X_{0,i}-\mu}{\sigma} < -\frac{\mu}{\sigma} 
\bigg| Z^*_{0,i}=1  
\bigg)
=\Phi\brac{-\frac{\mu}{\sigma} }.
\end{align*}
By symmetry, we can obtain the desired result for  $\PP ( \check Z^{\ITL}_{0,i}\ne Z^*_{0,i} )$. 

It remains to study $ \check Z^{\DP}_{0,i}$. From $\{ X_{0,i}+X_{1,i} > 0 \} \subseteq \{ \check  Z^{\DP}_{0,i}=1 \}
\subseteq \{ X_{0,i}+X_{1,i} \geq 0  \}$ we obtain that
\begin{align}
\PP_{1, 1} \brac{\check Z^{\DP}_{0,i}\ne  Z^*_{0,i}  } 
& =\Prob_{1, 1} \brac{X_{0,i}+X_{1,i}<0
} 
\notag\\&
 =\Prob_{1, 1} \bigg( \frac{X_{0,i}+X_{1,i}-2\mu}{\sqrt{2}\sigma}< -\sqrt{2}\cdot \frac{\mu}{\sigma} \bigg)
 =\Phi\bigg( -\frac{\sqrt{2}\mu}{\sigma}\bigg).
 \label{eqn-prop-baselines-1}
\end{align}
On the other hand,
\begin{align}
    \PP_{1, -1} ( \check Z^{\DP}_{0,i} \ne  Z^*_{0,i} )
    & =\Prob_{1,-1} \brac{X_{0,i}+X_{1,i}<0
    }  =\Prob_{1,-1} \brac{\frac{X_{0,i}+X_{1,i}}{\sqrt{2}\sigma}< 0 }=\frac{1}{2}.
 \label{eqn-prop-baselines-2}
\end{align}
By symmetry, the right-hand sides of \eqref{eqn-prop-baselines-1} and \eqref{eqn-prop-baselines-2} can upper bound $\PP (  \check Z^{\DP}_{0,i}\ne  Z^*_{0,i} | Z^*_{0,i} = Z^*_{1,i}  )$ and $\PP (  \check Z^{\DP}_{0,i}\ne  Z^*_{0,i} | Z^*_{0,i} \neq Z^*_{1,i}  ) $, respectively.
Therefore,
\begin{align*}
&\PP ( \check Z^{\DP}_{0,i}\ne  Z^*_{0,i} )\\
&= \PP (  \check Z^{\DP}_{0,i}\ne  Z^*_{0,i} | Z^*_{0,i} = Z^*_{1,i}  ) \PP ( Z^*_{0,i} = Z^*_{1,i} )
+ \PP (  \check Z^{\DP}_{0,i}\ne  Z^*_{0,i} | Z^*_{0,i} \neq Z^*_{1,i}  ) \PP ( Z^*_{0,i} \neq Z^*_{1,i}  ) \\
& = \Phi\bigg( -\frac{\sqrt{2}\mu}{\sigma}\bigg) \cdot [1 -  \PP ( Z^*_{0,i} \neq Z^*_{1,i}  ) ] + \frac{1}{2} \cdot  \PP ( Z^*_{0,i} \neq Z^*_{1,i}  ) \\
& = \Phi\bigg( -\frac{\sqrt{2}\mu}{\sigma}\bigg) + \bigg[ \frac{1}{2} -  \Phi\bigg( -\frac{\sqrt{2}\mu}{\sigma}\bigg) \bigg] \PP ( Z^*_{0,i} \neq Z^*_{1,i}  ).
\end{align*}
We have
\[
0 \leq \PP ( \check Z^{\DP}_{0,i}\ne  Z^*_{0,i} ) - \Phi\bigg( -\frac{\sqrt{2}\mu}{\sigma}\bigg) \leq \frac{\varepsilon}{2}.
\]

\section{Proofs of Lemmas}
  \subsection{Proof of Lemma \ref{lem:naive-eps}}
Without loss of generality, we assume $\sigma=1$. Notice that for any $i\in[n]$, we have
\begin{align*}
\II\brac{\check Z^{\ITL}_{0,i}\ne  \check Z^{\ITL}_{1,i}}&=\II\brac{ Z^*_{0,i}\ne  Z^*_{1,i},\check Z^{\ITL}_{0,i}=Z^*_{0,i}, \check Z^{\ITL}_{1,i}=Z^*_{1,i}}\\
&+\II\brac{ Z^*_{0,i}\ne  Z^*_{1,i},\check Z^{\ITL}_{0,i}\ne Z^*_{0,i}, \check Z^{\ITL}_{1,i}\ne Z^*_{1,i}}\\
&+\II\brac{ Z^*_{0,i}=  Z^*_{1,i},\check Z^{\ITL}_{0,i}=Z^*_{0,i}, \check Z^{\ITL}_{1,i}\ne Z^*_{1,i}}\\
&+\II\brac{ Z^*_{0,i}=  Z^*_{1,i},\check Z^{\ITL}_{0,i}\ne Z^*_{0,i}, \check Z^{\ITL}_{1,i}= Z^*_{1,i}}.
\end{align*}
Averaging  over $i\in[n]$, taking expectations on both sides and using \Cref{prop:ind-dp-gmm}, we obtain that 
  \begin{align*}
\EE\brac{\hat \varepsilon}&=\PP\brac{\check Z^{\ITL}_{0,i}\ne  \check Z^{\ITL}_{1,i}}\\
&=\varepsilon\cdot \sqbrac{\sqbrac{1-\Phi \brac{-\frac{\mu}{\sigma}}}^2+\sqbrac{\Phi \brac{-\frac{\mu}{\sigma}}}^2}+2\brac{1-\varepsilon}\Phi \brac{-\frac{\mu}{\sigma}}\sqbrac{1-\Phi \brac{-\frac{\mu}{\sigma}}}.
  \end{align*}
  We thus get 
  \begin{align*}
    \ab{\EE\brac{\hat \varepsilon}-\varepsilon}&=2\varepsilon\cdot \sqbrac{1+\Phi \brac{-\frac{\mu}{\sigma}}}\Phi \brac{-\frac{\mu}{\sigma}}+2\brac{1-\varepsilon}\Phi \brac{-\frac{\mu}{\sigma}}\sqbrac{1-\Phi \brac{-\frac{\mu}{\sigma}}}\\
    &=2\sqbrac{1-\brac{1-2\varepsilon}\Phi \brac{-\frac{\mu}{\sigma}}}\Phi \brac{-\frac{\mu}{\sigma}}\ge \Phi \brac{-\frac{\mu}{\sigma}}.
  \end{align*}
  Moreover, since $\Prob\brac{\check Z^{\ITL}_{0,i}\ne  \check Z^{\ITL}_{1,i}}<1/2$ and $\Phi\brac{-\mu/\sigma}<1/4$ for $\mu/\sigma\rightarrow\infty$, we have 
\begin{align*}
    \text{Var}\brac{\hat\varepsilon}&=\frac{1}{n^2}\sum_{i=1}^n\Prob\brac{\check Z^{\ITL}_{0,i}\ne  \check Z^{\ITL}_{1,i}}\brac{1-\Prob\brac{\check Z^{\ITL}_{0,i}\ne \check Z^{\ITL}_{1,i}}}\\
    &\ge  \frac{1}{2n}\Prob\brac{\check Z^{\ITL}_{0,i}\ne  \check Z^{\ITL}_{1,i}}\ge \frac{1}{4n}\sqbrac{\varepsilon+\Phi\brac{-\frac{\mu}{\sigma}}}.
\end{align*}  
Therefore, we have
\begin{align*}
    \EE\brac{\hat \varepsilon-\varepsilon}^2&=\text{Var}\brac{\hat\varepsilon}+\brac{\EE\brac{\hat \varepsilon}- \varepsilon}^2\\
    &\ge \frac{1}{4n}\sqbrac{\varepsilon+\Phi\brac{-\frac{\mu}{\sigma}}}+\sqbrac{\Phi \brac{-\frac{\mu}{\sigma}}}^2.  
\end{align*}
We thus arrive at
\begin{align*}
    &\EE\brac{\frac{\hat \varepsilon}{\varepsilon}-1}^2\\
    &\ge \frac{1/\varepsilon}{4n}+\frac{1/\varepsilon^2}{4n}\cdot \Phi\brac{-\frac{\mu}{\sigma}}+\sqbrac{\frac{1}{\varepsilon}\cdot \Phi \brac{-\frac{\mu}{\sigma}}}^2\\
    &\gtrsim    \frac{1}{n^{1-\beta}}+ \frac{1}{\mu/\sigma}\exp\brac{-\frac{\mu^2}{2\sigma^2}+\brac{2\beta-1}\log n}+\sqbrac{ \frac{1}{\mu/\sigma}\exp\brac{-\frac{\mu^2}{2\sigma^2}+\beta\log n}}^2.
\end{align*}
The desired result follows directly.

\subsection{Proof of Lemma \ref{lem:oracle-gl}}\label{sec-lem:oracle-gl}
First note that for any $\lambda\in\Lambda$ and $\lambda^\prime<\lambda$ we have
\begin{align*}
    &\frakD\brac{\hat \bZ_0^{\lambda},\hat \bZ_0^{\lambda^\prime}}-\hat\psi\brac{\lambda^\prime}\\
    &\le \frakD\brac{\hat \bZ_0^{\lambda},\bar \bZ_0^{\infty}}+\frakD\brac{\hat \bZ_0^{\lambda^\prime},\bar \bZ_0^{\infty}}-\hat\psi\brac{\lambda^\prime}\\
    &\le \frakD\brac{\hat \bZ_0^{\lambda},\bar \bZ_0^{\lambda}}+ \frakD\brac{\bar \bZ_0^{\lambda},\bar \bZ_0^{\infty}}+\frakD\brac{\hat \bZ_0^{\lambda^\prime},\bar \bZ_0^{\lambda^\prime}}+\frakD\brac{\bar \bZ_0^{\lambda^\prime},\bar \bZ_0^{\infty}}-\hat\psi\brac{\lambda^\prime}\\
    &\le \phi\brac{\lambda}+\hat\psi\brac{\lambda}+\xi_{\hat\psi}+\phi\brac{\lambda^\prime}+\frakD\brac{\bar \bZ_0^{\lambda^\prime},\bar \bZ_0^{\infty}}-\hat\psi\brac{\lambda^\prime}.
\end{align*}
Taking maximum over $\lambda^\prime\in[0,\lambda)\bigcap \Lambda$ leads to 
\begin{align}\label{eq:phi-hat-range}
    0\le \hat\phi\brac{\lambda}\le 2\phi\brac{\lambda}+\hat \psi\brac{\lambda}+2\xi_{\hat\psi},\quad \forall\lambda\in\Lambda.
\end{align}
For any $\lambda\in[0,\hat\lambda)\bigcap \Lambda$,  we have
\begin{align*}
    \frakD\brac{\hat \bZ_0^{\hat \lambda},\bar \bZ_0^\infty}&\le \frakD\brac{\hat \bZ_0^{\hat \lambda},\hat \bZ_0^\lambda}+\frakD\brac{\hat \bZ_0^{\lambda},\bar \bZ_0^\infty}\\
    &\le \sqbrac{\hat\phi\brac{\hat \lambda}+\hat \psi\brac{\lambda}}+\sqbrac{\phi\brac{\lambda}+\hat \psi\brac{\lambda}+\xi_{\hat\psi}}&& (\text{definition of }\hat \phi) \\
    &\le \sqbrac{\hat \phi\brac{\hat\lambda}+\hat\psi\brac{\hat\lambda}}+ \sqbrac{\phi\brac{\lambda}+2\hat \psi\brac{\lambda}+\xi_{\hat\psi}}\\
    &\le \sqbrac{\hat \phi\brac{\lambda}+\hat\psi\brac{\lambda}}+ \sqbrac{\phi\brac{\lambda}+2\hat \psi\brac{\lambda}+\xi_{\hat\psi}}&&(\text{definition of }\hat \lambda)\\
    &\le 3\phi\brac{\lambda}+4\hat\psi\brac{\lambda}+3\xi_{\hat\psi}.&&(\text{\eqref{eq:phi-hat-range}})
\end{align*}
For any $\lambda\in[\hat\lambda,\infty)\bigcap \Lambda$,  we have
\begin{align*}
    \frakD\brac{\hat \bZ_0^{\hat \lambda},\bar \bZ_0^\infty}&\le\frakD\brac{\hat \bZ_0^{\hat \lambda},\bar \bZ_0^{\hat\lambda}}+\frakD\brac{\bar \bZ_0^{\hat \lambda},\bar \bZ_0^\infty}\\
     &\le \phi\brac{\hat\lambda}+\hat \psi\brac{\hat\lambda}+\xi_{\hat\psi} &&(\text{definition of }\hat\phi \text{~and~}\xi_{\hat\psi})\\
    &\le \phi\brac{\lambda}+\sqbrac{\hat\phi\brac{\hat \lambda}+\hat \psi\brac{\hat\lambda}}+\xi_{\hat\psi}\\
&\le \phi\brac{\lambda}+\sqbrac{\hat\phi\brac{ \lambda}+\hat \psi\brac{\lambda}}+\xi_{\hat\psi} &&(\text{definition of }\hat \lambda)\\
    &\le 3\phi\brac{\lambda}+2\hat\psi\brac{\lambda}+3\xi_{\hat\psi}.&&(\text{\eqref{eq:phi-hat-range}})
\end{align*}

\subsection{Proof of Lemma \ref{lem:bootstrap-quantile}}\label{pf-lem:bootstrap-quantile}
For each $i\in[n]$ we have
\begin{align*}
    \PP\brac{\bar Z^{\lambda}_{0,i}\ne \bar Z^\infty_{0,i}}&\le \PP\brac{\bar Z^{\lambda}_{0,i}\ne \wt Z^\lambda_{0,i}}+\PP\brac{\wt Z^{\lambda}_{0,i}\ne \wt Z^\infty_{0,i}}+\PP\brac{\wt Z^{\infty}_{0,i}\ne \bar Z^\infty_{0,i}}\\
    &\le \PP\brac{\wt Z^{\lambda}_{0,i}\ne \wt Z^\infty_{0,i}}+\gamma\sqbrac{ \PP\brac{\bar Z^\lambda_{0,i}\ne  Z_{0,i}^*}+ \PP\brac{\bar Z^\infty_{0,i}\ne  Z_{0,i}^*}} &&(\text{Assumption \ref{assump:par-est-general}})\\
    &\le \PP\brac{\wt Z^{\lambda}_{0,i}\ne \wt Z^\infty_{0,i}}+\gamma\sqbrac{\PP\brac{\bar Z^\lambda_{0,i}\ne  \bar Z_{0,i}^\infty}+ 2\PP\brac{\bar Z^\infty_{0,i}\ne  Z_{0,i}^*}}.
\end{align*}
Rearranging terms  leads to 
\begin{align}\label{eq:barleqwt-prob}
    \PP\brac{\bar Z^{\lambda}_{0,i}\ne \bar Z^\infty_{0,i}}
    \le \frac{1}{1-\gamma}\PP\brac{\wt Z^{\lambda}_{0,i}\ne \wt Z^\infty_{0,i}}+\frac{2\gamma}{1-\gamma}\PP\brac{\bar Z^\infty_{0,i}\ne  Z_{0,i}^*}.
\end{align}
On the other hand, we have
\begin{align}\label{eq:wtleqbar-prob}
    \PP\brac{\wt Z^{\lambda}_{0,i}\ne \wt Z^\infty_{0,i}}&\le \PP\brac{\wt Z^{\lambda}_{0,i}\ne \bar Z^\lambda_{0,i}}+\PP\brac{\bar Z^{\lambda}_{0,i}\ne \bar Z^\infty_{0,i}}+\PP\brac{\bar Z^{\infty}_{0,i}\ne \wt Z^\infty_{0,i}}\notag\\
    &\le \PP\brac{\bar Z^{\lambda}_{0,i}\ne \bar Z^\infty_{0,i}}+\gamma\sqbrac{\PP\brac{\bar Z^\lambda_{0,i}\ne  Z_{0,i}^*}+\PP\brac{\bar Z^\infty_{0,i}\ne  Z_{0,i}^*}} &&(\text{Assumption \ref{assump:par-est-general}})\notag\\
    &\le \PP\brac{\bar Z^{\lambda}_{0,i}\ne \bar Z^\infty_{0,i}}+\gamma\sqbrac{\PP\brac{\bar Z^\lambda_{0,i}\ne  \bar Z_{0,i}^\infty}+ 2\PP\brac{\bar Z^\infty_{0,i}\ne  Z_{0,i}^*}}\notag\\
    &= \brac{1+\gamma}\PP\brac{\bar Z^{\lambda}_{0,i}\ne \bar Z^\infty_{0,i}}+2\gamma\PP\brac{\bar Z^\infty_{0,i}\ne  Z_{0,i}^*}.
\end{align}
We need the following data-driven bound for two sample mean of Bernoulli r.v's, whose proof is given in Section \ref{sec-lem:data-driven-bound}.
\begin{lemma}\label{lem:data-driven-bound}
    Let $\ebrac{X_{0,i}}_{i=1}^n$ be i.i.d. \textsf{Ber}$\brac{p}$, $\ebrac{X_{1,i}}_{i=1}^n$ be i.i.d. \textsf{Ber}$\brac{q}$, and $\ebrac{X_{0,i}}_{i=1}^n$ are independent of $\ebrac{X_{1,i}}_{i=1}^n$. Suppose that $p\le \vartheta_1q+\vartheta_2$ for some $\vartheta_1,\vartheta_2> 0$, then for any $\vartheta\in(0,1],\delta\in(0,1)$ we have
\begin{align*}
    &\PP\brac{\frac{1}{n}\sum_{i=1}^nX_{0,i}\le \brac{1+\vartheta}\vartheta_1\brac{\frac{1}{n}\sum_{i=1}^nX_{1,i}}+\frac{7\brac{1+2\vartheta_1}\log\brac{2/\delta}}{2\vartheta n}+2\vartheta_2}\ge 1-\delta.
\end{align*}
\end{lemma}
Invoking Lemma \ref{lem:data-driven-bound} for any $\vartheta,\zeta\in(0,1)$, we obtain that with probability at least $1-\zeta$,
    \begin{align}\label{eq:barleqwt}
        &\frakD\brac{\bar \bZ^{\lambda}_{0}, \bar \bZ^\infty_{0}}
    \le \frac{1+\vartheta}{1-\gamma}\frakD\brac{\wt \bZ^{\lambda}_{0}, \wt \bZ^\infty_{0}}+\brac{\frac{3-\gamma}{1-\gamma}}\frac{7\log\brac{2/\zeta}}{2\vartheta n}+\frac{4\gamma}{1-\gamma}\EE\frakD\brac{\bar \bZ^\infty_{0},  \bZ_{0}^*},
\end{align}
and with probability at least $1-\zeta$,
\begin{align}\label{eq:wtleqbar}
    &   \frakD\brac{\wt \bZ^{\lambda}_{0}, \wt \bZ^\infty_{0}}\le \brac{1+\vartheta}\brac{1+\gamma}\frakD\brac{\bar \bZ^{\lambda}_{0}, \bar \bZ^\infty_{0}}+\brac{3+2\gamma}\frac{7\log\brac{2/\zeta}}{2\vartheta n}+4\gamma\EE\frakD\brac{\bar \bZ^\infty_{0},  \bZ_{0}^*}.
    \end{align}

Let $\bar Q_{\zeta}^{\lambda}:=\brac{1-\zeta}$ {population quantile} of $\frakD\brac{\bar \bZ^{\lambda,1}_{0}, \bar \bZ^{\infty,1}_{0}}$ conditional on $(\hat\mu,\hat\sigma)$.

Define the  quantile function as $Q_\zeta(F):=\inf\ebrac{z\in\RR:F(z)\ge \zeta}$ for any $\zeta\in(0,1)$ and any c.d.f. $F$ on $\RR$.  The following lemma is needed to control the deviation of $\bar Q_\zeta^\lambda$ to $\hat Q_\zeta^\lambda$. 
\begin{lemma}\label{lem:quantile-rel}
    Suppose $X, X_1,X_2,\cdots,X_N\overset{i.i.d.}{\sim} F$ where $F\brac{\cdot}$ is the c.d.f., and denote the empirical c.d.f. of $\ebrac{X_i}_{i=1}^N$ as $\hat F_N\brac{\cdot}$. For any $\zeta,\eta\in(0,1)$ we have with probability at least $1-\eta$,
    \begin{align*}
        Q_{1-\zeta}\brac{F}\le Q_{1-{\zeta}/{2}}(\hat F_N)\le Q_{1-{\zeta}/{3}}\brac{F},
    \end{align*} 
    provided that $N\gtrsim \zeta^{-2}\log\brac{1/\eta}$.
\end{lemma}

By  Lemma \ref{lem:quantile-rel} and a union bound, we can conclude that with probability at least $1-\zeta$, 
\begin{align}\label{eq:quantile-rel}
    \bar Q_{\zeta}^\lambda\le \hat Q_{\zeta/2}^\lambda\le \bar Q_{\zeta/3}^\lambda,\quad \forall \lambda\in\Lambda
\end{align}
provided that $B\ge C\zeta^{-2}\log\brac{\ab{\Lambda}/\zeta}$ for some large constant $C>0$. Hence we 

By definition of $\bar Q^\lambda_{\zeta}$ we have 
\begin{align}\label{eq:quantile-ineq}
    \PP\brac{\frakD\brac{\wt \bZ^{\lambda}_{0}, \wt \bZ^\infty_{0}}< \bar Q_\zeta^{\lambda}}\le 1-\zeta,\quad \PP\brac{\frakD\brac{\wt \bZ^{\lambda}_{0}, \wt \bZ^\infty_{0}}\le \bar Q_\zeta^{\lambda}}\ge 1-\zeta.
\end{align}
Combined with \eqref{eq:barleqwt} and \eqref{eq:quantile-rel}, we obtain that with probability at least $1-3\zeta$,
\begin{align*}
    \frakD\brac{\bar \bZ^{\lambda}_{0}, \bar \bZ^\infty_{0}}
    \le \frac{1+\vartheta}{1-\gamma}\hat Q_{\zeta/2}^{\lambda}+\brac{\frac{3-\gamma}{1-\gamma}}\frac{7\log\brac{4/\zeta}}{2\vartheta n}+\frac{4\gamma}{1-\gamma}\EE\frakD\brac{\bar \bZ^\infty_{0},  \bZ_{0}^*}.
\end{align*}
By  choosing $\vartheta=0.005$ and noting $\gamma<0.04$, we thus obtain that  with probability at least $1-\zeta$,
\begin{align}\label{eq:zeta-bound}
    \frakD\brac{\bar \bZ^{\lambda}_{0}, \bar \bZ^\infty_{0}}
    \le \hat\psi\brac{\lambda}+2200\sqbrac{\frac{\log\brac{1/\zeta}}{n}+\gamma\EE\frakD\brac{\bar \bZ^\infty_{0},  \bZ_{0}^*}}.
\end{align}
By definition of  $\xi_{\hat\psi}$ and a union bound, we conclude the proof of the first claim.

It suffices to prove the upper bound for $\hat\psi\brac{\lambda}$. 
By \eqref{eq:wtleqbar} with $\vartheta=1/2$, we obtain that with probability at least $1-\zeta/6$,
\begin{align*}
    \frakD\brac{\wt \bZ^{\lambda}_{0}, \wt \bZ^\infty_{0}}\le 3\frakD\brac{\bar \bZ^{\lambda}_{0}, \bar \bZ^\infty_{0}}+\frac{35\log\brac{12/\zeta}}{n}+4\gamma\EE\frakD\brac{\bar \bZ^\infty_{0},\bZ_{0}^*}.
\end{align*}
By Lemma \ref{lem:bernoulli-bernstein} with $\vartheta=1$, we obtain that with probability at least $1-\zeta/6$,
\begin{align*}
    \frakD\brac{\bar \bZ^{\lambda}_{0}, \bar \bZ^\infty_{0}}\le 2\EE\frakD\brac{\bar \bZ^{\lambda}_{0}, \bar \bZ^\infty_{0}}+\frac{7\log\brac{6/\zeta}}{6n}
\end{align*}
By a union bound, we conclude that with probability at least $1-\zeta/3$,
\begin{align*}
    \frakD\brac{\wt \bZ^{\lambda}_{0}, \wt \bZ^\infty_{0}}\le 6\EE\frakD\brac{\bar \bZ^{\lambda}_{0}, \bar \bZ^\infty_{0}}+\frac{40\log\brac{12/\zeta}}{n}+4\gamma\EE\frakD\brac{\bar \bZ^\infty_{0},\bZ_{0}^*}.
\end{align*}
Together with \eqref{eq:quantile-rel} and  \eqref{eq:quantile-ineq}, we can conclude that with probability at least $1-10\zeta/3$,
\begin{align}\label{eq:psi-bound}
    \hat\psi\brac{\lambda}=1.01\hat Q_{\zeta/2}^{\lambda}\le 1.01\bar Q_{\zeta/3}^{\lambda}\le 45\sqbrac{\EE\frakD\brac{\bar \bZ^{\lambda}_{0}, \bar \bZ^\infty_{0}}+\frac{\log\brac{1/\zeta}}{n}+\gamma\EE\frakD\brac{\bar \bZ^\infty_{0},\bZ_{0}^*}}.
\end{align}
The proof is then completed by adjusting the constant in front of  $\zeta$ and taking a union bound over \eqref{eq:zeta-bound} and \eqref{eq:psi-bound}.

\subsection{Proof of Lemma \ref{lem:bernoulli-bernstein}}\label{sec-lem:bernoulli-bernstein}
By Bernstein's inequality, see, e.g., Lemma 3.1 in \cite{han2024model}, we have with probability at least $1-\delta$ that 
\begin{align*}
    \frac{1}{n}\sum_{i=1}^nX_{0,i}-p\le\sqrt{\frac{2p\log\brac{1/\delta}}{n}}+\frac{2\log\brac{1/\delta}}{3n}=\inf_{\vartheta>0}\ebrac{\vartheta p+\frac{\log\brac{1/\delta}}{2\vartheta n}}+\frac{2\log\brac{1/\delta}}{3n}.
\end{align*}
This implies for any $\vartheta\in(0,1]$, we have with probability at least $1-\delta$ that 
\begin{align*}
    \frac{1}{n}\sum_{i=1}^nX_{0,i}\le \brac{1+\vartheta}p+\brac{\frac{1}{2\vartheta}+\frac{2}{3}}\frac{\log\brac{1/\delta}}{n}\le \brac{1+\vartheta}p+\frac{7\log\brac{1/\delta}}{6\vartheta n}.
\end{align*}
The lower bound can be obtained similarly and hence omitted.

\subsection{Proof of Lemma \ref{lem:lower-bound-mis-error-lam-adap}}\label{sec-lem:lower-bound-mis-error-lam-adap}
Denote $K^+=\frac{\lambda \sigma^2}{2\mu }\brac{1+\delta_n^\prime}$ and $K^-=\frac{\lambda \sigma^2}{2\mu }\brac{1-\delta_n^\prime}$.  By symmetry, it suffices for us to consider the case when $(Z^*_{0,i},Z^*_{1,i})=(1,1)$. First, note that the transfer clustering procedure would yield $\hat Z^{\lambda}_{0,i}=-1$ if 
\begin{align}\label{eq:-1+1-hat-gmm}
     X_{0,i}< X_{1,i}, \quad X_{0,i}<-\frac{\lambda\hat \sigma^2}{2\hat \mu },\quad  X_{1,i}>\frac{\lambda\hat \sigma^2}{2\hat \mu}
\end{align}
or
\begin{align}\label{eq:-1-1-hat-gmm}
        X_{0,i}+X_{1,i}<0,\quad X_{0,i}<\frac{\lambda\hat \sigma^2}{2\hat \mu },\quad X_{1,i}<\frac{\lambda\hat \sigma^2}{2\hat \mu}.
\end{align}
We have
\begin{align*}
    &\PP\brac{\ebrac{\text{Event }\eqref{eq:-1+1-hat-gmm}\text{~occurs}}\bigcap\calE_{\sf par}\bigg| (Z^*_{0,i},Z^*_{1,i})=(1,1)}\\
    &= \PP_{1,1}\brac{X_{0,i}<-\frac{\lambda\hat\sigma^2}{2\hat\mu },X_{1,i}>\frac{\lambda\hat\sigma^2}{2\hat\mu },\calE_{\sf par}}\\
    &\ge  \PP_{1,1}\brac{X_{0,i}<-K^+}\PP_{1,1}\brac{X_{1,i}>K^+}\\
    &=\Phi\brac{\frac{-\mu-K^+}{\sigma}}\brac{1-\Phi\brac{\frac{-\mu+K^+}{\sigma} }}.
\end{align*}
It remains to lower bound the following term
\begin{align*}
    &\PP\brac{\ebrac{\text{Event }\eqref{eq:-1-1-hat-gmm}\text{~occurs}}\bigcap\calE_{\sf par}\bigg| (Z^*_{0,i},Z^*_{1,i})=(1,1)}\\
    &\ge \PP_{1,1}\brac{X_{0,i}+X_{1,i}<0,  X_{0,i}<\frac{\lambda\sigma^2}{2\mu }\brac{1-\delta_n^\prime},X_{1,i}<\frac{\lambda\sigma^2}{2\mu }\brac{1-\delta_n^\prime}}\\
    &=1-\PP_{1,1}\brac{\ebrac{X_{0,i}+X_{1,i}\ge 0}\bigcup \ebrac{X_{0,i}\ge K^-}\bigcup\ebrac{X_{1,i}\ge K^-}}\\
    &=1-\PP_{1,1}\brac{X_{0,i}+X_{1,i}\ge 0}-\PP_{1,1}\brac{X_{0,i}\ge K^-}-\PP_{1,1}\brac{X_{1,i}\ge K^-}\\
    &+\PP_{1,1}\brac{\ebrac{X_{0,i}+X_{1,i}\ge 0}\bigcap \ebrac{X_{0,i}\ge K^-}}+\PP_{1,1}\brac{\ebrac{X_{0,i}+X_{1,i}\ge 0}\bigcap \ebrac{X_{1,i}\ge K^-}}\\
    &= \Phi\brac{-\frac{\sqrt{2}\mu}{\sigma}}-\PP_{1,1}\brac{\ebrac{X_{0,i}+X_{1,i}<0}\bigcap \ebrac{X_{0,i}\ge K^-}}\\
    &-\PP_{1,1}\brac{\ebrac{X_{0,i}+X_{1,i}
    < 0}\bigcap \ebrac{X_{1,i}\ge K^-}}.
\end{align*}
Notice that 
\begin{align*}
    \PP&\brac{\ebrac{X_{0,i}+X_{1,i}<0}\bigcap \ebrac{X_{0,i}\ge K^-}}\\
    &=\frac{1}{\sqrt{2\pi}\sigma}\int_{K^-}^{\infty}\PP\brac{X_{0,i}+X_{1,i}<0\mid X_{0,i}=x}\exp\brac{-\frac{\brac{x-\mu}^2}{2\sigma^2}}dx\\
    &= \frac{1}{\sqrt{2\pi}\sigma}\int_{K^-}^{\infty}\Phi\brac{\frac{-x-\mu}{\sigma}}\exp\brac{-\frac{\brac{x-\mu}^2}{2\sigma^2}}dx\\
    &\overset{\text{Lemma~}\ref{lem:approx-cdf-normal} }{\le}  \frac{1}{\sqrt{2\pi}\sigma}\int_{K^-}^{\infty}\frac{\sqrt{2/\pi }}{\dfrac{x+\mu}{\sigma} +\sqrt{\dfrac{\brac{x+\mu}^2}{\sigma^2} +\dfrac{8}{\pi } }}\exp\brac{-\frac{\brac{x+\mu}^2}{2\sigma^2}}\exp\brac{-\frac{\brac{x-\mu}^2}{2\sigma^2}}dx\\
    &= \frac{\sqrt{2/\pi }\exp\brac{-\mu^2/\sigma^2}}{\sqrt{2}\brac{\brac{K^-+\mu}/\sigma +\sqrt{\brac{K^-+\mu}^2/\sigma^2 +8/\pi }}}\int_{\sqrt{2}K^-/\sigma}^{\infty}\frac{1}{\sqrt{2\pi }}\exp\brac{-\frac{u^2}{2}}du\\
    &= \frac{1}{\sqrt{2}}  \frac{\sqrt{2/\pi }\exp\brac{-\mu^2/\sigma^2}}{\sqrt{2}\mu/\sigma +\sqrt{2\mu^2/\sigma^2 +4 }}\cdot \frac{\sqrt{2}\mu/\sigma +\sqrt{2\mu^2/\sigma^2 +4 }}{\brac{K^-+\mu}/\sigma +\sqrt{\brac{K^-+\mu}^2/\sigma^2 +8/\pi }}\Phi\brac{-\frac{\sqrt{2}K^-}{\sigma}}\\
    &\overset{\text{Lemma~}\ref{lem:approx-cdf-normal} }{\le}  
    \Phi\brac{-\frac{\sqrt{2}\mu}{\sigma}}\Phi\brac{-\frac{\sqrt{2}K^-}{\sigma}}
\end{align*}
Another trivial bound can be obtained by 
\begin{align*}
    &\PP\brac{\ebrac{\text{Event }\eqref{eq:-1-1-hat-gmm}\text{~occurs}}\bigg| (Z^*_{0,i},Z^*_{1,i})=(1,1)}\\
    &\ge \PP\brac{X_{0,i}+X_{1,i}<0,  X_{0,i}<\frac{\lambda\sigma^2}{2\mu }\brac{1-\delta_n^\prime},X_{1,i}<\frac{\lambda\sigma^2}{2\mu }\brac{1-\delta_n^\prime}}\\
    &\ge \PP\brac{X_{0,i}+X_{1,i}<0,  X_{0,i}<0,X_{1,i}<0}=\sqbrac{\Phi\brac{-\frac{\mu}{\sigma}}}^2
\end{align*}
Since $\ebrac{\text{Event }\eqref{eq:-1+1-hat-gmm}\text{~occurs}}\bigcap\ebrac{\text{Event }\eqref{eq:-1-1-hat-gmm}\text{~occurs}}=\emptyset$, we can conclude that
\begin{align*}
    &\PP\brac{\hat Z^{\lambda}_{0,i}=-1\bigg| (Z^*_{0,i},Z^*_{1,i})=(1,1)}\\
    &\ge \Phi\brac{\frac{-\mu-K^+}{\sigma}}\Phi\brac{\frac{\mu-K^+}{\sigma} }+\Phi\brac{-\frac{\sqrt{2}\mu}{\sigma}}\\
    &+\Phi\brac{-\frac{\sqrt{2}\mu}{\sigma}}\brac{1-2\Phi\brac{-\frac{\sqrt{2}K^-}{\sigma}}}\vee \sqbrac{\Phi\brac{-\frac{\mu}{\sigma}}}^2.
\end{align*}
By symmetry, a lower bound for $\PP\brac{\hat Z^{\lambda}_{0,i}=1\bigg| (Z^*_{0,i},Z^*_{1,i})=(-1,-1)}$ can be derived similarly and hence omitted.

\subsection{Proof of Lemma \ref{lem:checkbar-prob}}\label{sec-lem:checkbar-prob-proof}
Fix $i\in[n]$  and we can write
\begin{align}\label{eq:checkbar-formula}
    \begin{cases}
        X_{0,i}=\mu Z^*_{0,i}+\sigma V_{0,i}\\
        X_{1,i}=\mu Z^*_{1,i}+\sigma V_{1,i}\\
        \bar X_{0,i}=\mu Z^*_{0,i}+\sigma V_{1,i}
    \end{cases},
\end{align}
where $V_{0,i}\overset{d}{=}V_{1,i}\sim N(0,1)$ and $V_{0,i}$ is independent of $V_{1,i}$. 
\paragraph*{Bound for $\PP\brac{\bar Z^\infty_{0,i}\ne \bar Z^\lambda_{0,i}}$}
By definition and \eqref{eq:checkbar-formula}, we have
\begin{align*}
\bar Z^\infty_{0,i}&=Z^*_{0,i}\cdot \text{sign}\brac{1+\frac{\sigma}{\sqrt{2}\mu}\frac{\brac{V_{0,i}+V_{1,i}}Z^*_{0,i}}{\sqrt{2}}},\\
\brac{\bar Z^\lambda_{0,i},\bar Z^\lambda_{1,i}}&\in \argmin_{u,v\in\{\pm1\}}\ebrac{-\brac{\frac{\mu}{\sigma}}^2Z^*_{0,i}\brac{u+v}-\frac{\mu}{\sigma}\brac{uV_{0,i}+vV_{1,i}}+\lambda\II\brac{u\ne v}}.
\end{align*}
Hence we have $\bar Z^\infty_{0,i}=1$ if
\begin{align*}
    Z^*_{0,i}=1,\quad V_{0,i}+V_{1,i}>-\frac{2\mu}{\sigma},
\end{align*}
or
\begin{align*}
    Z^*_{0,i}=-1, \quad V_{0,i}+V_{1,i}>\frac{2\mu}{\sigma}.
\end{align*}
On the other hand, we  have $\bar Z^\lambda_{0,i}=-1$ if 
\begin{align*}
    V_{0,i}<  V_{1,i},\quad V_{0,i}< -\frac{\mu}{\sigma}Z^*_{0,i}- \frac{\lambda \sigma}{2\mu}, \quad V_{1,i}> -\frac{\mu}{\sigma}Z^*_{0,i}+\frac{\lambda \sigma}{2\mu},
\end{align*}
or 
\begin{align*}
    V_{0,i}+V_{1,i}<  -\frac{2\mu}{\sigma}Z^*_{0,i},\quad V_{0,i}<  -\frac{\mu}{\sigma}Z^*_{0,i}+\frac{\lambda \sigma}{2\mu},\quad V_{1,i}<-\frac{\mu}{\sigma}Z^*_{0,i} +\frac{\lambda \sigma}{2\mu}.
\end{align*}
Hence we have $\brac{ \bar Z^{\infty}_{0,i},\bar Z^{\lambda}_{0,i}}=\brac{1,-1}$ only if 
\begin{align*}
    Z^*_{0,i}=1,\quad V_{0,i}+V_{1,i}>-\frac{2\mu}{\sigma},\quad V_{0,i}< -\frac{\mu}{\sigma}- \frac{\lambda \sigma}{2\mu}, \quad V_{1,i}> -\frac{\mu}{\sigma}+\frac{\lambda \sigma}{2\mu},
\end{align*}
or
\begin{align*}
    Z^*_{0,i}=-1,\quad V_{0,i}+V_{1,i}>\frac{2\mu}{\sigma},\quad V_{0,i}< \frac{\mu}{\sigma}- \frac{\lambda \sigma}{2\mu}, \quad V_{1,i}> \frac{\mu}{\sigma}+\frac{\lambda \sigma}{2\mu}.
\end{align*}
We can conclude that 
\begin{align*}
    \PP\brac{\bar Z^\infty_{0,i}=1, \bar Z^\lambda_{0,i}=-1}&\le 2 \Phi\brac{-\frac{\mu}{\sigma}- \frac{\lambda \sigma}{2\mu}}\Phi\brac{\frac{\mu}{\sigma}-\frac{\lambda \sigma}{2\mu}}.
\end{align*}
By symmetry the bound for $\PP\brac{\bar Z^\infty_{0,i}=-1, \bar Z^\lambda_{0,i}=1}$ is  the same. Hence, we have obtained the first claim in Lemma \ref{lem:checkbar-prob}.

\paragraph*{Bound for $\PP\brac{\check  Z_{0,i}^\lambda\ne \bar Z_{0,i}^\lambda}$}
By definition and \eqref{eq:checkbar-formula}, we  have 
\begin{align*}
    \brac{\check  Z^\lambda_{0,i},\check Z^\lambda_{1,i}}&\in \argmin_{u,v\in\{\pm1\}}\left \{-\brac{\frac{\mu}{\sigma}}^2\brac{Z^*_{0,i}u +Z^*_{1,i}v}-\frac{\mu}{\sigma}\brac{uV_{0,i}+vV_{1,i}}+\lambda\II\brac{u\ne v}\right \},\\
    \brac{\bar Z^\lambda_{0,i},\bar Z^\lambda_{1,i}}&\in \argmin_{u,v\in\{\pm1\}}\ebrac{-\brac{\frac{\mu}{\sigma}}^2Z^*_{0,i}\brac{u+v}-\frac{\mu}{\sigma}\brac{uV_{0,i}+vV_{1,i}}+\lambda\II\brac{u\ne v}}.
\end{align*}
Hence we have $\check  Z^\lambda_{0,i}=1$ if 
\begin{align*}
     V_{0,i}> V_{1,i}+\frac{\mu}{\sigma}\brac{Z^*_{1,i}-Z^*_{0,i}}, \quad V_{0,i}>-\frac{\mu}{\sigma} Z^*_{0,i}+\frac{\lambda \sigma}{2 \mu },\quad  V_{1,i}<-\frac{\mu}{\sigma} Z^*_{1,i}-\frac{\lambda \sigma}{2 \mu},
\end{align*}
or
\begin{align*}
         V_{0,i}+V_{1,i}>-\frac{\mu}{\sigma}\brac{Z^*_{0,i}+Z^*_{1,i}},\quad V_{0,i}>-\frac{\mu}{\sigma} Z^*_{0,i}-\frac{\lambda \sigma}{2 \mu },\quad V_{1,i}>-\frac{\mu}{\sigma} Z^*_{1,i}-\frac{\lambda \sigma}{2 \mu}.
\end{align*}
On the other hand, we  have $\bar Z^\lambda_{0,i}=-1$ if 
\begin{align*}
    V_{0,i}<  V_{1,i},\quad V_{0,i}< -\frac{\mu}{\sigma}Z^*_{0,i}- \frac{\lambda \sigma}{2\mu}, \quad V_{1,i}> -\frac{\mu}{\sigma}Z^*_{0,i}+\frac{\lambda \sigma}{2\mu},
\end{align*}
or 
\begin{align*}
    V_{0,i}+V_{1,i}<  -\frac{2\mu}{\sigma}Z^*_{0,i},\quad V_{0,i}<  -\frac{\mu}{\sigma}Z^*_{0,i}+\frac{\lambda \sigma}{2\mu},\quad V_{1,i}<-\frac{\mu}{\sigma}Z^*_{0,i} +\frac{\lambda \sigma}{2\mu}.
\end{align*}
By definition, we have $\brac{\check Z^{\lambda}_{0,i},\bar Z^{\lambda}_{0,i}}=\brac{1,-1}$ only if $Z^*_{0,i}\ne Z^*_{1,i}$.  When $\brac{Z^*_{0,i},Z^*_{1,i}}=(1,-1)$, then (i) the first condition of $\check  Z^\lambda_{0,i}=1$ conflicts with the  conditions of $\bar Z^\lambda_{0,i}=-1$ due to the condition on $V_{0,i}$, (ii) the second condition  of  $\check  Z^\lambda_{0,i}=1$ conflicts with the first condition  of $\bar Z^\lambda_{0,i}=-1$ due to the condition on $V_{0,i}$ and conflicts with the second condition  of $\bar Z^\lambda_{0,i}=-1$ due to the condition on $V_{0,i}+V_{1,i}$.
Hence we have $\brac{\check Z^{\lambda}_{0,i},\bar Z^{\lambda}_{0,i}}=\brac{1,-1}$ only if $\brac{Z^*_{0,i},Z^*_{1,i}}=(-1,1)$ and 
\begin{align*}
    0<V_{0,i}+V_{1,i}<\frac{2\mu}{\sigma},\quad   \frac{\mu}{\sigma}-\frac{\lambda \sigma}{2\mu}<V_{0,i}<  \frac{\mu}{\sigma}+\frac{\lambda \sigma}{2\mu},\quad -\frac{\mu}{\sigma}-\frac{\lambda \sigma}{2\mu}<V_{1,i}<\frac{\mu}{\sigma}+\frac{\lambda \sigma}{2\mu},
\end{align*}

Hence we arrive at  
\begin{align*}
    &\PP\brac{\check Z^{\lambda}_{0,i}=1,\bar Z^{\lambda}_{0,i}=-1}\\
    &\le \frac{\varepsilon}{2}\cdot \PP\brac{\check Z^{\lambda}_{0,i}=1,\bar Z^{\lambda}_{0,i}=-1\mid \brac{Z^*_{0,i},Z^*_{1,i}}=(-1,1)}
    \\
    &\le \frac{\varepsilon}{2}\cdot \PP\brac{\frac{\mu}{\sigma}-\frac{\lambda \sigma}{2\mu}<V_{0,i}<  \frac{\mu}{\sigma}+\frac{\lambda \sigma}{2\mu},\quad -\frac{\mu}{\sigma}-\frac{\lambda \sigma}{2\mu}<V_{1,i}<-\frac{\mu}{\sigma}+\frac{\lambda \sigma}{2\mu}}\\
    &\le  \frac{\varepsilon}{2}\Phi\brac{-\frac{\mu}{\sigma}+\frac{\lambda \sigma}{2\mu}}\sqbrac{1-\Phi\brac{-\frac{\mu}{\sigma}-\frac{\lambda \sigma}{2\mu}}}.
\end{align*}
By symmetry, we have the same bound for $\PP\brac{\check Z^{\lambda}_{0,i}=-1,\bar Z^{\lambda}_{0,i}=1}$ and the proof for the second claim of Lemma \ref{lem:checkbar-prob} is completed.

\subsection{Proof of Lemma \ref{lem:checkhat-prob}}\label{sec-lem:checkhat-prob-proof}
For $\forall i\in[n]$, by definition we  have 
\begin{align*}
    \brac{\check  Z^\lambda_{0,i},\check Z^\lambda_{1,i}}&\in \argmin_{u,v\in\{\pm1\}}\left \{-\mu \brac{u X_{0,i}+vX_{1,i}}+\lambda\sigma^2\II\brac{u\ne v}\right \}.
\end{align*}
Hence we have $\check  Z^\lambda_{0,i}=1$ if 
\begin{align*}
     X_{0,i}> X_{1,i}, \quad X_{0,i}>\frac{\lambda \sigma^2}{2 \mu },\quad  X_{1,i}<-\frac{\lambda \sigma^2}{2 \mu},
\end{align*}
or
\begin{align*}
        X_{0,i}+X_{1,i}>0,\quad X_{0,i}>-\frac{\lambda \sigma^2}{2 \mu },\quad X_{1,i}>-\frac{\lambda \sigma^2}{2 \mu}.
\end{align*}
On the other hand, we  have $\hat Z^\lambda_{0,i}=-1$ if 
\begin{align*}
     X_{0,i}< X_{1,i}, \quad X_{0,i}<-\frac{\lambda\hat \sigma^2}{2\hat \mu },\quad  X_{1,i}>\frac{\lambda\hat \sigma^2}{2\hat \mu},
\end{align*}
or
\begin{align*}
        X_{0,i}+X_{1,i}<0,\quad X_{0,i}<\frac{\lambda\hat \sigma^2}{2\hat \mu },\quad X_{1,i}<\frac{\lambda\hat \sigma^2}{2\hat \mu}.
\end{align*}
Hence under Assumption \ref{assump:par-est} we have $\brac{\check Z^{\lambda}_{0,i},\hat Z^{\lambda}_{0,i}}=\brac{1,-1}$ only if
\begin{align*}
     X_{0,i}+X_{1,i}<0,\quad \frac{\lambda \sigma^2}{2 \mu }<X_{0,i}<\frac{\lambda\sigma^2}{2\mu }\brac{1+\delta_n},\quad X_{1,i}<-\frac{\lambda \sigma^2}{2 \mu},
\end{align*}
or 
\begin{align*}
    X_{0,i}+X_{1,i}>0,\quad -\frac{\lambda \sigma^2}{2 \mu }<X_{0,i}<-\frac{\lambda\sigma^2}{2\mu }\brac{1-\delta_n},\quad X_{1,i}>\frac{\lambda \sigma^2}{2 \mu}\brac{1-\delta_n}.
\end{align*}
This leads to 
\begin{align}\label{eq:checkhat+1-1-decomp}
    &\II\brac{\check Z^{\lambda}_{0,i}=1,\hat Z^{\lambda}_{0,i}=-1, \calE_{\sf par}}\notag\\
    &\le \II\brac{X_{0,i}+X_{1,i}<0,\quad \frac{\lambda \sigma^2}{2 \mu }<X_{0,i}<\frac{\lambda\sigma^2}{2\mu }\brac{1+\delta_n},\quad X_{1,i}<-\frac{\lambda \sigma^2}{2 \mu}}\notag\\
    &+\II\brac{X_{0,i}+X_{1,i}>0,\quad -\frac{\lambda \sigma^2}{2 \mu }<X_{0,i}<-\frac{\lambda\sigma^2}{2\mu }\brac{1-\delta_n},\quad X_{1,i}>\frac{\lambda \sigma^2}{2 \mu}\brac{1-\delta_n}}.
\end{align}
We first consider the case when $(Z_{0,i}^*,Z_{1,i}^*)=\brac{1,1}$. By Lemma \ref{lem:perturb-cdf-normal} we have the following bound for the first term in \eqref{eq:checkhat+1-1-decomp}:
\begin{align*}
    &\PP\brac{X_{0,i}+X_{1,i}<0,\quad \frac{\lambda \sigma^2}{2 \mu }<X_{0,i}<\frac{\lambda\sigma^2}{2\mu }\brac{1+\delta_n},\quad X_{1,i}<-\frac{\lambda \sigma^2}{2 \mu}\mid (Z_{0,i}^*,Z_{1,i}^*)=\brac{1,1}}\\
    &\lesssim \delta_n\brac{ \frac{\lambda}{\mu/\sigma}}\exp\brac{-\frac{\mu^2}{2\sigma^2}\brac{1-\frac{\lambda}{2\mu^2/\sigma^2}}^2}\Phi\brac{-\frac{\mu}{\sigma}-\frac{\lambda}{2\mu/\sigma}}.
\end{align*}
For the second term in \eqref{eq:checkhat+1-1-decomp}, by Lemma \ref{lem:perturb-cdf-normal} we have
\begin{align*}
    &\PP\brac{X_{0,i}+X_{1,i}>0,-\frac{\lambda \sigma^2}{2 \mu }<X_{0,i}<-\frac{\lambda\sigma^2}{2\mu }\brac{1-\delta_n},X_{1,i}>\frac{\lambda \sigma^2}{2 \mu}\brac{1-\delta_n}\mid (Z_{0,i}^*,Z_{1,i}^*)=\brac{1,1}}\\
    &\lesssim \delta^2_n\brac{ \frac{\lambda}{\mu/\sigma}}^2\exp\brac{-\frac{\mu^2}{2\sigma^2}\brac{1+\frac{\lambda\sigma^2}{2\mu^2}}^2\brac{1-2\delta_n}}\exp\brac{-\frac{\mu^2}{2\sigma^2}\brac{1-\frac{\lambda \sigma^2}{2\mu^2}}^2\brac{1-2\delta_n}}.
\end{align*}
Using Lemma \ref{lem:approx-cdf-normal} and  $\delta_n=o\brac{\brac{\mu/\sigma}^{-2}}$, we arrive at 
\begin{align*}
    \PP\brac{\check Z^{\lambda}_{0,i}=1,\hat Z^{\lambda}_{0,i}=-1\mid (Z_{0,i}^*,Z_{1,i}^*)=\brac{1,1}}\lesssim \frac{\delta_n\lambda\brac{1+\delta_n\lambda}}{\mu/\sigma}\Phi\brac{-\frac{\mu}{\sigma}-\frac{\lambda}{2\mu/\sigma}}.
\end{align*}
When $(Z_{0,i}^*,Z_{1,i}^*)=\brac{1,-1}$,  for the first term in \eqref{eq:checkhat+1-1-decomp} by Lemma \ref{lem:perturb-cdf-normal} we have
\begin{align*}
    &\PP\brac{X_{0,i}+X_{1,i}<0,\quad \frac{\lambda \sigma^2}{2 \mu }<X_{0,i}<\frac{\lambda\sigma^2}{2\mu }\brac{1+\delta_n},\quad X_{1,i}<-\frac{\lambda \sigma^2}{2 \mu}\mid (Z_{0,i}^*,Z_{1,i}^*)=\brac{1,-1}}\\
    &\lesssim \delta_n\brac{ \frac{\lambda}{\mu/\sigma}}\exp\brac{-\frac{\mu^2}{2\sigma^2}\brac{1-\frac{\lambda}{2\mu^2/\sigma^2}}^2}\Phi\brac{-\frac{\mu}{\sigma}+\frac{\lambda}{2\mu/\sigma}}.
\end{align*}
For the second term in \eqref{eq:checkhat+1-1-decomp} by Lemma \ref{lem:perturb-cdf-normal} we have
\begin{align*}
    &\PP\brac{X_{0,i}+X_{1,i}>0,-\frac{\lambda \sigma^2}{2 \mu }<X_{0,i}<-\frac{\lambda\sigma^2}{2\mu }\brac{1-\delta_n},X_{1,i}>\frac{\lambda \sigma^2}{2 \mu}\brac{1-\delta_n}\mid (Z_{0,i}^*,Z_{1,i}^*)=\brac{1,-1}}\\
    &\lesssim \delta^2_n\brac{ \frac{\lambda}{\mu/\sigma}}^2\exp\brac{-\frac{\mu^2}{2\sigma^2}\brac{1+\frac{\lambda \sigma^2}{2\mu^2}}^2\brac{1-2\delta_n}}\exp\brac{-\frac{\mu^2}{2\sigma^2}\brac{1+\frac{\lambda \sigma^2}{2\mu^2}}^2\brac{1-2\delta_n}}.
\end{align*}
Using Lemma \ref{lem:approx-cdf-normal} and  $\delta_n=o\brac{\brac{\mu/\sigma}^{-2}}$, we arrive at 
\begin{align*}
    \PP\brac{\check Z^{\lambda}_{0,i}=1,\hat Z^{\lambda}_{0,i}=-1\mid (Z_{0,i}^*,Z_{1,i}^*)=\brac{1,-1}, \calE_{\sf par}}\lesssim \delta_n\brac{ \frac{\lambda}{\mu/\sigma}}\Phi\brac{-\frac{\mu}{\sigma}+\frac{\lambda}{2\mu/\sigma}}.
\end{align*}
The case for $(Z_{0,i}^*,Z_{1,i}^*)=\brac{-1,1}$ and $(Z_{0,i}^*,Z_{1,i}^*)=\brac{-1,-1}$ are almost the same and hence omitted. We finally reach that
\begin{align*}
        \PP\brac{\check Z^{\lambda}_{0,i}=1,\hat Z^{\lambda}_{0,i}=-1, \calE_{\sf par}}\lesssim\frac{\delta_n\lambda\brac{1+\delta_n\lambda}}{\mu/\sigma}\sqbrac{\Phi\brac{-\frac{\mu}{\sigma}-\frac{\lambda}{2\mu/\sigma}}+\varepsilon\cdot \Phi\brac{-\frac{\mu}{\sigma}+\frac{\lambda}{2\mu/\sigma}}}
\end{align*}
Similarly, we can also the same bound for $\PP\brac{\check Z^{\lambda}_{0,i}=1,\hat Z^{\lambda}_{0,i}=-1}$ and conclude that 
\begin{align*}
    \PP\brac{\hat Z^{\lambda}_{0,i}\ne \check Z^{\lambda}_{0,i}}\le \frac{C\delta_n\lambda\brac{1+\delta_n\lambda}}{\mu/\sigma}\sqbrac{\Phi\brac{-\frac{\mu}{\sigma}-\frac{\lambda}{2\mu/\sigma}}+\varepsilon\cdot \Phi\brac{-\frac{\mu}{\sigma}+\frac{\lambda}{2\mu/\sigma}}}+n^{-C_{\sf par}}.
\end{align*}

\subsection{Proof of Lemma \ref{lem:assump-1-gmm}}\label{sec-slem:assump-1-gmm}
Notice that for $i\in[n]$ we can write
\begin{align*}
    \begin{cases}
        X_{0,i}=\mu Z^*_{0,i}+\sigma V_{0,i}\\
        \bar X_{0,i}=\mu Z^*_{0,i}+\sigma V_{1,i}
    \end{cases}\quad 
    \begin{cases}
        \wt X_{0,i}=\hat \mu Z^*_{0,i}+\hat \sigma V_{0,i}\\
        \wt X_{1,i}=\hat \mu Z^*_{0,i}+\hat \sigma V_{1,i}
    \end{cases}.
\end{align*}
where $V_{0,i}\overset{d}{=}V_{1,i}\sim N(0,1)$ and $V_{0,i}$ is independent of $V_{1,i}$.

We first consider the case when $\lambda=\infty$. For $\forall i\in[n]$, notice that  
\begin{align*}
\bar Z^\infty_{0,i}&=\argmin_{u\in\{\pm1\}}\left \{-\frac{\mu}{\sigma^2} u\brac{ X_{0,i}+\bar X_{0,i}}\right \}=\argmin_{u\in\{\pm1\}}\left \{-2\brac{\frac{\mu}{\sigma}}^2 uZ^*_{0,i}-\frac{\mu}{\sigma}u\brac{V_{0,i}+V_{1,i}}\right \}\\
&=\text{sign}\brac{\frac{2\mu}{\sigma}Z^*_{0,i}+V_{0,i}+V_{1,i}}=Z^*_{0,i}\cdot \text{sign}\brac{1+\frac{\sigma}{\sqrt{2}\mu}\frac{\brac{V_{0,i}+V_{1,i}}Z^*_{0,i}}{\sqrt{2}}}.
\end{align*}
Similarly, we have
\begin{align*}
    \wt Z^\infty_{0,i}&=Z^*_{0,i}\cdot \text{sign}\brac{1+\frac{\hat\sigma}{\sqrt{2}\hat \mu}\frac{\brac{V_{0,i}+V_{1,i}}Z^*_{0,i}}{\sqrt{2}}}.
\end{align*}
Denote $\eta:=\frac{\brac{V_{0,i}+V_{1,i}}Z^*_{0,i}}{\sqrt{2}}\sim N(0,1)$, then we have $\bar Z^\infty_{0,i}\ne \wt Z^\infty_{0,i}$ if the following event occurs:
\begin{align*}
    \ebrac{\eta>-\sqrt{2}\frac{\mu}{\sigma},\quad   \eta<-\sqrt{2}\frac{\hat \mu}{\hat \sigma}}\bigcup\ebrac{\eta<-\sqrt{2}\frac{\mu}{\sigma},\quad     \eta>-\sqrt{2}\frac{\hat \mu}{\hat \sigma}}.
\end{align*}
Hence we can arrive at
\begin{align}\label{eq:barwt+1-1-decomp-inf}
    \II\brac{\bar Z^{\infty}_{0,i}\ne \wt Z^{\infty}_{0,i}}&\le \II\brac{\eta>-\sqrt{2}\frac{\mu}{\sigma},\quad     \eta<-\sqrt{2}\frac{\hat \mu}{\hat \sigma}}+\II\brac{\eta<-\sqrt{2}\frac{\mu}{\sigma},\quad     \eta>-\sqrt{2}\frac{\hat \mu}{\hat \sigma}}.
\end{align}
It easy to obtain that (by modifying Lemma \ref{lem:lower-bound-mis-error-lam-adap} and utilizing Lemma \ref{lem:approx-cdf-normal})
\begin{align*}
    \PP\brac{\bar Z^\infty_{0,i}\ne  Z_{0,i}^*}\ge \Phi\brac{-\frac{\sqrt{2}\mu}{\sigma}}\gtrsim \frac{1}{\mu/\sigma}\exp\brac{-\frac{\mu^2}{\sigma^2}}.
\end{align*}
On the other hand, Lemma \ref{lem:perturb-cdf-normal}, \eqref{eq:barwt+1-1-decomp-inf} and Assumption \ref{assump:par-est} imply that 
\begin{align*}
    &\PP\brac{\bar Z^{\infty}_{0,i}\ne \wt Z^{\infty}_{0,i}}\\&\le \PP\brac{-\sqrt{2}\frac{\mu}{\sigma}<\eta<-\sqrt{2}\frac{\mu}{ \sigma}\brac{1-\delta_n}}+\PP\brac{-\sqrt{2}\frac{\mu}{\sigma}\brac{1+\delta_n}<\eta<-\sqrt{2}\frac{\mu}{ \sigma}}+n^{-C_{\sf par}}\\
    &\le \delta_n\frac{2\sqrt{2}\mu}{\sigma}\exp\brac{-\frac{\mu^2}{\sigma^2}\brac{1-2\delta_n}}+n^{-C_{\sf par}}
\end{align*}
Hence we can conclude that $    \PP\brac{\bar Z^{\infty}_{0,i}\ne \wt Z^{\infty}_{0,i}}=o\brac{\PP\brac{\bar Z^\infty_{0,i}\ne  Z_{0,i}^*}}$ provided that $\delta_n=o\brac{\brac{\mu/\sigma}^{-2}}$.
We then consider $\forall\lambda\in\Lambda$. For $\forall i\in[n]$, we  have 
\begin{align*}
    \brac{\bar Z^\lambda_{0,i},\bar Z^\lambda_{1,i}}&=\argmin_{u,v\in\{\pm1\}}\left \{-\frac{\mu}{\sigma^2} \brac{u X_{0,i}+v\bar X_{0,i}}+\lambda\II\brac{u\ne v}\right \}\\
    &=\argmin_{u,v\in\{\pm1\}}\ebrac{-\brac{\frac{\mu}{\sigma}}^2Z^*_{0,i}\brac{u+v}-\frac{\mu}{\sigma}\brac{uV_{0,i}+vV_{1,i}}+\lambda\II\brac{u\ne v}}.
\end{align*}
Hence $\bar Z^{\lambda}_{0,i}=1$ if the following event occurs:
\begin{align*}
    V_{1,i}>  -\frac{\lambda \sigma}{2\mu}-\frac{\mu}{\sigma}Z^*_{0,i},\quad V_{0,i}>  -\frac{\lambda \sigma}{2\mu}-\frac{\mu}{\sigma}Z^*_{0,i},\quad V_{0,i}+V_{1,i}>  -\frac{2\mu}{\sigma}Z^*_{0,i},
\end{align*}
or 
\begin{align*}
    V_{1,i}<   -\frac{\lambda \sigma}{2\mu}-\frac{\mu}{\sigma}Z^*_{0,i},\quad V_{0,i}>  \frac{\lambda \sigma}{2\mu}-\frac{\mu}{\sigma}Z^*_{0,i},\quad V_{0,i}>  V_{1,i}.
\end{align*}
On the other hand, we have $\wt Z^{\lambda}_{0,i}=-1$ if 
\begin{align*}
    V_{1,i}<   \frac{\lambda \hat \sigma}{2\hat \mu}-\frac{\hat \mu}{\hat \sigma}Z^*_{0,i},\quad V_{0,i}<   \frac{\lambda \hat \sigma}{2\hat \mu}-\frac{\hat \mu}{\hat \sigma}Z^*_{0,i},\quad V_{0,i}+V_{1,i}<   -\frac{2\hat \mu}{\hat \sigma}Z^*_{0,i},
\end{align*}
or 
\begin{align*}
    V_{1,i}>    \frac{\lambda \hat \sigma}{2\hat \mu}-\frac{\hat \mu}{\hat \sigma}Z^*_{0,i},\quad V_{0,i}<  -\frac{\lambda \hat \sigma}{2\hat \mu}-\frac{\hat \mu}{\hat \sigma}Z^*_{0,i},\quad V_{0,i}<   V_{1,i}.
\end{align*}
Under \eqref{eq:assump-event}, we have $\brac{\bar Z^{\lambda}_{0,i},\wt Z^{\lambda}_{0,i}}=\brac{1,-1}$ if
\begin{align*}
    &-\frac{\lambda \sigma}{2\mu}-\frac{\mu}{\sigma}Z^*_{0,i}<V_{1,i}<   \frac{\lambda \hat \sigma}{2\hat \mu}-\frac{\hat \mu}{\hat \sigma}Z^*_{0,i},\quad -\frac{\lambda \sigma}{2\mu}-\frac{\mu}{\sigma}Z^*_{0,i}<V_{0,i}<   \frac{\lambda \hat \sigma}{2\hat \mu}-\frac{\hat \mu}{\hat \sigma}Z^*_{0,i},\\
    &-\frac{2\mu}{\sigma}Z^*_{0,i}<V_{0,i}+V_{1,i}<   -\frac{2\hat \mu}{\hat \sigma}Z^*_{0,i},
\end{align*}
or
\begin{align*}
    &V_{1,i}>\max\ebrac{-\frac{\lambda \sigma}{2\mu}-\frac{\mu}{\sigma}Z^*_{0,i},\frac{\lambda \hat \sigma}{2\hat \mu}-\frac{\hat \mu}{\hat \sigma}Z^*_{0,i}},\quad -\frac{\lambda \sigma}{2\mu}-\frac{\mu}{\sigma}Z^*_{0,i}<V_{0,i}< -\frac{\lambda \hat \sigma}{2\hat \mu}-\frac{\hat \mu}{\hat \sigma}Z^*_{0,i},\\
    & V_{0,i}+V_{1,i}>-\frac{2\mu}{\sigma}Z^*_{0,i},
\end{align*}
or
\begin{align*}
    &V_{1,i}< \min\ebrac{-\frac{\lambda \sigma}{2\mu}-\frac{\mu}{\sigma}Z^*_{0,i},\frac{\lambda \hat \sigma}{2\hat \mu}-\frac{\hat \mu}{\hat \sigma}Z^*_{0,i}},\quad \frac{\lambda \sigma}{2\mu}-\frac{\mu}{\sigma}Z^*_{0,i}<V_{0,i}<   \frac{\lambda \hat \sigma}{2\hat \mu}-\frac{\hat \mu}{\hat \sigma}Z^*_{0,i},\\
    &V_{0,i}+V_{1,i}<-\frac{2\hat \mu}{\hat \sigma}Z^*_{0,i}.
\end{align*}
This leads to 
\begin{align}\label{eq:barwt+1-1-decomp}
    &\II\brac{\bar Z^{\lambda}_{0,i}=1,\wt Z^{\lambda}_{0,i}=-1,\calE_{\sf par}}\notag\\
    &\le \II\brac{-\frac{2\mu}{\sigma}Z^*_{0,i}<V_{0,i}+V_{1,i}<   -\frac{2\hat \mu}{\hat \sigma}Z^*_{0,i},~\calE_{\sf par}}\notag\\
    &+\II\brac{V_{1,i}>\frac{\lambda \hat \sigma}{2\hat\mu}-\frac{\hat\mu}{\hat\sigma}Z^*_{0,i}, \quad -\frac{\lambda \sigma}{2\mu}-\frac{\mu}{\sigma}Z^*_{0,i}<V_{0,i}< -\frac{\lambda \hat \sigma}{2\hat \mu}-\frac{\hat \mu}{\hat \sigma}Z^*_{0,i},~\calE_{\sf par}}\notag\\
    &+\II\brac{V_{1,i}<-\frac{\lambda \sigma}{2\mu}-\frac{\mu}{\sigma}Z^*_{0,i} ,\quad \frac{\lambda \sigma}{2\mu}-\frac{\mu}{\sigma}Z^*_{0,i}<V_{0,i}<   \frac{\lambda \hat \sigma}{2\hat \mu}-\frac{\hat \mu}{\hat \sigma}Z^*_{0,i},~\calE_{\sf par}}.
\end{align}
By Lemma \ref{lem:lower-bound-mis-error-lam-adap}, we have
\begin{align}\label{eq:barstar-lb}
    \Prob\brac{\bar Z^\lambda_{0,i}\ne Z^*_{0,i}}&\ge \sqbrac{\Phi\brac{-\frac{\mu}{\sigma}}}^2+\Phi\brac{-\frac{\mu}{\sigma}-\frac{\lambda}{2\mu/\sigma}}\Phi\brac{\frac{\mu}{\sigma}-\frac{\lambda}{2\mu/\sigma}}\notag\\
    &\gtrsim \frac{1}{\mu^2/\sigma^2}\exp\brac{-\frac{\mu^2}{\sigma^2}}+\Phi\brac{-\frac{\mu}{\sigma}-\frac{\lambda}{2\mu/\sigma}}\Phi\brac{\frac{\mu}{\sigma}-\frac{\lambda}{2\mu/\sigma}}.
\end{align}
It suffices to verify that the expectation of first three terms in \eqref{eq:barwt+1-1-decomp} is negligible compared to \eqref{eq:barstar-lb}. For the first term in \eqref{eq:barwt+1-1-decomp}, by Lemma \ref{lem:perturb-cdf-normal} we have
\begin{align*}
    \PP\brac{-\frac{2\mu}{\sigma}Z^*_{0,i}<V_{0,i}+V_{1,i}<   -\frac{2\hat \mu}{\hat \sigma}Z^*_{0,i},~\calE_{\sf par}}\lesssim \delta_n\brac{ \frac{\sqrt{2}\mu}{\sigma}}\exp\brac{-\frac{\mu^2}{\sigma^2}\brac{1-2\delta_n}}+n^{-C_{\sf par}},
\end{align*}
which is of negligible given $\delta_n=o\brac{\brac{\mu/\sigma}^{-3}}$.
We first consider the case when $Z_{0,i}^*=1$. For the second term in \eqref{eq:barwt+1-1-decomp}, we have
\begin{align}\label{eq:barwt+1-1-decomp-term-II}
    &\PP\brac{V_{1,i}>\frac{\lambda \hat \sigma}{2\hat\mu}-\frac{\hat\mu}{\hat\sigma}Z^*_{0,i},\quad -\frac{\lambda \sigma}{2\mu}-\frac{\mu}{\sigma}Z^*_{0,i}<V_{0,i}<   -\frac{\lambda \hat \sigma}{2\hat \mu}-\frac{\hat \mu}{\hat \sigma}Z^*_{0,i},~\calE_{\sf par}}\notag\\
    &\lesssim \delta_n\brac{ \frac{\lambda  \sigma}{2 \mu}+\frac{\mu}{\sigma}}\exp\brac{-\frac{1}{2}\brac{\frac{\mu}{\sigma}+\frac{\lambda \sigma}{2\mu}}^2}\Phi\brac{\frac{\mu}{\sigma}-\frac{\lambda}{2\mu/\sigma}+\delta_n\brac{ \frac{\lambda  \sigma}{2 \mu}+\frac{\mu}{\sigma}}}+n^{-C_{\sf par}}.
\end{align}
\raggedright
When $0<\lambda<2\brac{\frac{\mu}{\sigma}}^2\frac{1+\delta_n}{1-\delta_n}$, we have $\Phi\brac{\frac{\mu}{\sigma}-\frac{\lambda}{2\mu/\sigma}}\ge \Phi\brac{-\frac{2\delta_n}{1-\delta_n}\frac{\mu}{\sigma}}\gtrsim 1$ and $\Phi\brac{\frac{\mu}{\sigma}-\frac{\lambda}{2\mu/\sigma}+\delta_n\brac{ \frac{\lambda  \sigma}{2 \mu}+\frac{\mu}{\sigma}}}\le 1$. When $\lambda>2\brac{\frac{\mu}{\sigma}}^2\frac{1+\delta_n}{1-\delta_n}$,  by Lemma \ref{lem:approx-cdf-normal} and \ref{lem:perturb-cdf-normal} we have 
\begin{align*}
    \frac{\Phi\brac{\frac{\mu}{\sigma}-\frac{\lambda}{2\mu/\sigma}+\delta_n\brac{ \frac{\lambda  \sigma}{2 \mu}+\frac{\mu}{\sigma}}}}{\Phi\brac{\frac{\mu}{\sigma}-\frac{\lambda}{2\mu/\sigma}}}\lesssim \delta_n\brac{\frac{\lambda}{2\mu/\sigma}}^2.
\end{align*}
Combining two cases, we  conclude that  \eqref{eq:barwt+1-1-decomp-term-II} negligible compared to \eqref{eq:barstar-lb} given $r<2$, $\delta_n=o\brac{\brac{\frac{\mu/\sigma}{\log n}}^{2}}$ and $\lambda\lesssim \log n$. For the third term in \eqref{eq:barwt+1-1-decomp}, we have
\begin{align*}
    &\PP\brac{V_{1,i}<-\frac{\lambda \sigma}{2\mu}-\frac{\mu}{\sigma}Z^*_{0,i} ,\quad \frac{\lambda \sigma}{2\mu}-\frac{\mu}{\sigma}Z^*_{0,i}<V_{0,i}<   \frac{\lambda \hat \sigma}{2\hat \mu}-\frac{\hat \mu}{\hat \sigma}Z^*_{0,i},~\calE_{\sf par}}\\
    &\le \delta_n\brac{ \frac{\lambda  \sigma}{2 \mu}+\frac{\mu}{\sigma}}\exp\brac{-\frac{1}{2}\brac{\frac{\mu}{\sigma}-\frac{\lambda \sigma}{2\mu}}^2}\Phi\brac{-\frac{\mu}{\sigma}-\frac{\lambda}{2\mu/\sigma}}+n^{-C_{\sf par}}.
\end{align*}
which is of negligible given $\delta_n=o\brac{\brac{\mu/\sigma}^{-2}}$ by Lemma \ref{lem:approx-cdf-normal}.
We then consider $Z_{0,i}^*=-1$. For the second term in \eqref{eq:barwt+1-1-decomp}, we have
\begin{align*}
    &\PP\brac{V_{1,i}>\frac{\lambda \hat \sigma}{2\hat\mu}-\frac{\hat\mu}{\hat\sigma}Z^*_{0,i},\quad -\frac{\lambda \sigma}{2\mu}-\frac{\mu}{\sigma}Z^*_{0,i}<V_{0,i}<   -\frac{\lambda \hat \sigma}{2\hat \mu}-\frac{\hat \mu}{\hat \sigma}Z^*_{0,i},~\calE_{\sf par}}\\
    &\lesssim \delta_n\brac{ \frac{\lambda  \sigma}{2 \mu}+\frac{\mu}{\sigma}}\exp\brac{-\frac{1}{2}\brac{\frac{\mu}{\sigma}-\frac{\lambda \sigma}{2\mu}}^2}\Phi\brac{-\frac{\mu}{\sigma}-\frac{\lambda}{2\mu/\sigma}+\delta_n\brac{ \frac{\lambda  \sigma}{2 \mu}+\frac{\mu}{\sigma}}}+n^{-C_{\sf par}},
\end{align*}
which is of negligible compared to \eqref{eq:barstar-lb} given $r<2$, $\delta_n=o\brac{\brac{\frac{\mu/\sigma}{\log n}}^{2}}$ and $\lambda\lesssim \log n$ by a similar argument to \eqref{eq:barwt+1-1-decomp-term-II}. For the last term in \eqref{eq:barwt+1-1-decomp}, we have
\begin{align*}
    &\PP\brac{V_{1,i}<-\frac{\lambda \sigma}{2\mu}-\frac{\mu}{\sigma}Z^*_{0,i} ,\quad \frac{\lambda \sigma}{2\mu}-\frac{\mu}{\sigma}Z^*_{0,i}<V_{0,i}<   \frac{\lambda \hat \sigma}{2\hat \mu}-\frac{\hat \mu}{\hat \sigma}Z^*_{0,i},~\calE_{\sf par}}\\
    &\lesssim \delta_n\brac{ \frac{\lambda  \sigma}{2 \mu}+\frac{\mu}{\sigma}}\exp\brac{-\frac{1}{2}\brac{\frac{\mu}{\sigma}
    +\frac{\lambda \sigma}{2\mu}}^2}\Phi\brac{\frac{\mu}{\sigma}-\frac{\lambda}{2\mu/\sigma}}+n^{-C_{\sf par}}.
\end{align*}
which is of negligible compared to \eqref{eq:barstar-lb} provided that $r<2$,   $\delta_n=o\brac{\brac{\frac{\mu/\sigma}{\log n}}^{2}}$ and $\lambda\lesssim \log n$.  Collecting the cases for $Z_{0,i}^*=1$ and $Z_{0,i}^*=-1$, we arrive at 
\begin{align*}
        \PP\brac{\bar Z^{\lambda}_{0,i}=1,\wt Z^{\lambda}_{0,i}=-1}=o\brac{\Prob\brac{\bar Z^\lambda_{0,i}\ne Z^*_{0,i}}}.
\end{align*}
The case $\PP\brac{\bar Z^{\lambda}_{0,i}=-1,\wt Z^{\lambda}_{0,i}=1}$ is almost the same (due to symmetry) and hence omitted, we thereby complete the proof by noticing
\begin{align*}
    \PP\brac{\wt Z^{\lambda}_{0,i}\ne \bar Z^{\lambda}_{0,i}}=\PP\brac{\bar Z^{\lambda}_{0,i}=1,\wt Z^{\lambda}_{0,i}=-1}+\PP\brac{\bar Z^{\lambda}_{0,i}=-1,\wt Z^{\lambda}_{0,i}=1}.
\end{align*}

\subsection{Proof of Lemma \ref{lem:par-est}}\label{pf:par-est} 
Without loss of generality, we assume $\wt\mu>0$, i.e., $\hat\bmu^\top\bmu\ge 0$. By definition, we have
\begin{align*}
    \frac{\ab{\op{\hat\bmu}/{\hat\sigma}-\mu /{\sigma}}}{\mu/\sigma}&=\ab{\frac{\op{\hat\bmu}^2}{\hat\bmu^\top \bmu}\cdot\frac{\sigma}{\hat \sigma}-1}\le \ab{\frac{\op{\hat\bmu}^2}{\hat\bmu^\top \bmu}-1}+\ab{\frac{\sigma}{\hat \sigma}-1}\le \delta_n,
\end{align*}
where  $\ab{{ \sigma}/{\hat\sigma}-1}\le \delta_n/2$ by \Cref{assump:mul-par-est}, and 
\begin{align*}
    \ab{\frac{\op{\hat\bmu}^2}{{\hat\bmu^\top \bmu}}-1}=\ab{\frac{\op{\hat\bmu}^2-{\hat\bmu^\top \bmu}}{\hat\bmu^\top \bmu}}\le \frac{\op{\hat\bmu}\op{\hat\bmu-\bmu}}{\op{\bmu}^2-\op{\hat\bmu-\bmu}\op{\bmu}}\le \frac{\brac{1+\delta_n/4}\delta_n/4\op{\bmu}^2}{\brac{1-\delta_n/4}\op{\bmu}^2}\le \frac{\delta_n}{2}.
\end{align*}
Similarly, we could obtain the bound for the other term.

\subsection{Proof of Lemma \ref{lem:data-driven-bound}}\label{sec-lem:data-driven-bound}
Let $\upsilon:=\vartheta/\brac{\vartheta+2}$. By Lemma \ref{lem:bernoulli-bernstein}, with probability at least $1-\delta$ we have
\begin{align*}
    &\frac{1}{n}\sum_{i=1}^nX_{0,i}\le \brac{1+\upsilon}p+\frac{7\log\brac{2/\delta}}{6\upsilon n}\le \brac{1+\upsilon}\vartheta_1q+2\vartheta_2+\frac{7\log\brac{2/\delta}}{6\upsilon n},\\
    &\frac{1}{n}\sum_{i=1}^nX_{1,i}\ge  \brac{1-\upsilon}q-\frac{7\log\brac{2/\delta}}{6\upsilon n}.
\end{align*}
Hence with probability at least $1-\delta$ we have
\begin{align*}
    \frac{1}{n}\sum_{i=1}^nX_{0,i}&\le\frac{1+\upsilon}{1-\upsilon}\vartheta_1\cdot \frac{1}{n}\sum_{i=1}^nX_{1,i}+\brac{1+\frac{1+\upsilon}{1-\upsilon}\vartheta_1}\frac{7\log\brac{2/\delta}}{6\upsilon n}+2\vartheta_2\\
    &=\brac{1+\vartheta}\vartheta_1\cdot \frac{1}{n}\sum_{i=1}^nX_{1,i}+\brac{1+\frac{2}{\vartheta}}\brac{1+\brac{1+\vartheta}\vartheta_1}\frac{7\log\brac{2/\delta}}{6n}+2\vartheta_2 \\
    &\le \frac{\brac{1+\vartheta}\vartheta_1}{n}\sum_{i=1}^nX_{1,i}+\frac{7\log\brac{2/\delta}}{2\vartheta n}\brac{1+2\vartheta_1}+2\vartheta_2,
\end{align*}
where we've used $\vartheta=\frac{2\upsilon}{1-\upsilon}$ in the second equality and $\vartheta<1$ in the last inequality.

\subsection{Proof of Lemma \ref{lem:quantile-rel}} We first show  the upper bound for $Q_{1-{\zeta}/{2}}(\hat F_N)$. By definition, we have
\begin{align*}
    \PP\brac{X\le Q_{1-\zeta/3}\brac{F}}\ge 1-\zeta/3.
\end{align*}
By Hoeffding's inequality, we obtain that 
with probability at least $1-\eta/2$,
\begin{align*}
    \frac{1}{N}\sum_{i=1}^N\II\brac{X_i\le Q_{1-\zeta}\brac{F}}\ge 1-\zeta/3-C\sqrt{\frac{\log \brac{1/\eta}}{N}}\ge 1-\zeta/2,
\end{align*}
provided that $N\gtrsim \zeta^{-2}\log\brac{1/\eta}$. On that event, by definition of $Q_{1-\zeta/2}(\hat F_N)$, we have $Q_{1-\zeta/2}(\hat F_N)\le Q_{1-\zeta/3}\brac{F}$.

It remains to show the lower bound. 
By definition of $Q_{1-\zeta}( F)$, we have that 
\begin{align*}
    \PP\brac{X_i< Q_{1-\zeta}\brac{F}}\le  1-\zeta.
\end{align*}
Hoeffding's inequality tells us that  with probability at least $1-\eta/2$,
\begin{align*}
    \frac{1}{N}\sum_{i=1}^N\II\brac{X_i< Q_{1-\zeta}\brac{F}}\le \PP\brac{X< Q_{1-\zeta}\brac{F}}+C\sqrt{\frac{\log \brac{1/\eta}}{N}}\le  1-2\zeta/3<1-\zeta/2
\end{align*}
provided that $N\gtrsim \zeta^{-2}\log\brac{1/\eta}$. Hence on that event, 
\begin{itemize}
    \item Less than $1-\zeta/2$ fraction of $X_i$'s belongs to $(-\infty,Q_{1-\zeta}( F))$. 
    \item At least $1-\zeta/2$ fraction of $X_i$'s belongs to $(-\infty,Q_{1-\zeta/2}(\hat F_N)]$.  
    \item For any $\epsilon>0$, less than $1-\zeta/2$ fraction of $X_i$'s belongs to $(-\infty,Q_{1-\zeta/2}(\hat F_N)-\epsilon]$
\end{itemize}
We can conclude that  $Q_{1-\zeta}( F)\le Q_{1-\zeta/2}(\hat F_N)$.

\subsection{Proof of Lemma \ref{lem:checkbar-prob-lcm}}\label{pf-lem:checkbar-prob-lcm}
We fix $i\in[n]$ throughout the proof. Recall that $p_+=a$ and $p_+=b$, $a>b$. Let 
\begin{align*}
    L_{\sf LCM}(u):=-\log \brac{\frac{p_u}{1-p_u}} \sum_{j=1}^dX_{0,ij}-d \log \brac{1-p_u},
\end{align*}
and $\Delta_{\sf LCM}=L_{\sf LCM}(1)-L_{\sf LCM}(-1)$. Also, for any $X\in\RR$, let 
\begin{align*}
    F_\lambda(u;X):=L_{\sf LCM}(u)+\min_{v\in\ebrac{\pm 1}}\ebrac{\frac{(X-v\mu)^2}{2\sigma^2}+\lambda\II(u\ne v)}.
\end{align*}
By definition, we have $\check Z_{0,i}^\lambda=\argmin_{u\in\ebrac{\pm 1}}F_\lambda(u;X_{1,i})$ and $\bar Z_{0,i}^\lambda=\argmin_{u\in\ebrac{\pm 1}}F_\lambda(u;\bar X_{0,i})$. A direct calculation gives the following piecewise linear forms:
\begin{align*}
    F_\lambda(1;X)=\begin{cases}
        L_{\sf LCM}(1)+\frac{(X-\mu)^2}{2\sigma^2}, &X\ge -\frac{\lambda \sigma^2}{2\mu }\\
        L_{\sf LCM}(1)+\frac{(X+\mu)^2}{2\sigma^2}+\lambda , & X< -\frac{\lambda \sigma^2}{2\mu }
    \end{cases},
\end{align*}
and
\begin{align*}
     F_\lambda(-1;X)=\begin{cases}
        L_{\sf LCM}(-1)+\frac{(X+\mu)^2}{2\sigma^2}, & X\le \frac{\lambda \sigma^2}{2\mu }\\
        L_{\sf LCM}(-1)+\frac{(X-\mu)^2}{2\sigma^2}+\lambda , & X> \frac{\lambda \sigma^2}{2\mu }
    \end{cases}.
\end{align*}
Let $\Delta_\lambda(X):=F_\lambda(1;X)-F_\lambda(-1;X)$, we thereby get
\begin{align}\label{eq:Delta_lambda-exp}
    \Delta_\lambda(X)=\begin{cases}
        \Delta_{\sf LCM}-\lambda, & X> \frac{\lambda \sigma^2}{2\mu }\\
        \Delta_{\sf LCM}-\frac{2\mu}{\sigma^2}X , & -\frac{\lambda \sigma^2}{2\mu }\le X\le  \frac{\lambda \sigma^2}{2\mu }\\
        \Delta_{\sf LCM}+\lambda, & X< -\frac{\lambda \sigma^2}{2\mu }
    \end{cases}.
\end{align}
Hence we can conclude that 
\begin{align*}
    \check Z_{0,i}^\lambda=\begin{cases}
        1, & \Delta_\lambda(X_{1,i})\le 0\\
        -1, & \Delta_\lambda(X_{1,i})> 0
    \end{cases},\qquad  \bar Z_{0,i}^\lambda=\begin{cases}
        1, & \Delta_\lambda(\bar X_{0,i})\le 0\\
        -1, & \Delta_\lambda(\bar X_{0,i})> 0
    \end{cases}.
\end{align*}
Therefore, $\ebrac{\check Z_{0,i}^\lambda\ne \bar Z_{0,i}^\lambda}$ is equivalent to $\ebrac{\Delta_\lambda(X_{1,i})\Delta_\lambda(\bar X_{0,i})\le 0}$. When $\ab{\Delta_{\sf LCM}}\ge \lambda$, $\Delta_\lambda(X)$ has a constant sign for all $X$. It suffices to consider $\ab{\Delta_{\sf LCM}}< \lambda$, where the zero of $\Delta_\lambda(X)$ occurs at $X^*:=\frac{\sigma^2}{2\mu}\Delta_{\sf LCM}$. 
Notice that $\ebrac{\Delta_\lambda(X_{1,i})\Delta_\lambda(\bar X_{0,i})\le 0}$ can  occur only when $Z_{0,i}^*\ne Z_{1,i}^*$, since otherwise $X_{1,i}=\bar X_{0,i}$. 
Notice that ss
\begin{align*}
  X_{1,i}=\mu Z_{1s,i}^*+\sigma V_{1,i},\quad \bar X_{0,i}=\mu Z_{0,i}^*+\sigma V_{1,i},
\end{align*} 
for some $V_{0,i},V_{1,i}\sim N(0,1)$. Then, $(X_{1,i}-X^*)(\bar X_{0,i}-X^*)=(\sigma V_{1,i}-X^*)^2-\mu^2$.
We thus conclude that $\ebrac{\check Z_{0,i}^\lambda\ne \bar Z_{0,i}^\lambda}$ is equivalent to 
\begin{align*}
    \ebrac{\ab{\Delta_{\sf LCM}}< \lambda}\bigcap \ebrac{Z_{0,i}^*\ne Z_{1,i}^*}\bigcap \ebrac{\frac{\sigma}{2\mu}\Delta_{\sf LCM}-\frac{\mu}{\sigma}\le V_{1,i}\le \frac{\sigma}{2\mu}\Delta_{\sf LCM}+\frac{\mu}{\sigma}}.
\end{align*}
We first consider the event $ \ebrac{\ab{\Delta_{\sf LCM}}< \lambda}$. In particular, we have 
\begin{align*}
   \ebrac{\ab{\Delta_{\sf LCM}}< \lambda}&=\ebrac{\log \frac{a(1-b)}{b(1-a)} \sum_{j=1}^dX_{0,ij}\in\sqbrac{ d\log\frac{1-b}{1-a}-\lambda, d\log\frac{1-b}{1-a}+\lambda}}.
\end{align*}
Since $\sum_{j=1}^dX_{0,ij}\in[0,d]$, it suffices to consider $\lambda\in \sqbrac{0,d\brac{\log\frac{1-b}{1-a}\wedge \log\frac{a}{b}}}$. 
We thereby have for $\delta_{\lambda}\in[0,p^*\wedge\brac{1-p^*}]$,
\begin{align*}
    \ebrac{\ab{\Delta_{\sf LCM}}< \lambda}&=\ebrac{\frac{1}{d}\sum_{j=1}^dX_{0,ij}\in\sqbrac{p^*-\delta_{\lambda}, p^*+\delta_{\lambda}}}.
\end{align*}
By Chernoff bound, we get
\begin{align*}
    \PP\brac{\frac{1}{d}\sum_{j=1}^dX_{0,ij}\in\sqbrac{p^*-\delta_{\lambda}, p^*+\delta_{\lambda}}}&\le \exp\sqbrac{-d\inf_{w\in [p^*-\delta_\lambda,p^*+\delta_\lambda]}D_{\rm KL}\brac{w\mid\mid  p_{Z_{0,i}^*}}}.
\end{align*}
Notice that we can further simplify the above bound as
\begin{align}\label{eq:barcheck-final-bound}
    &\PP\brac{\frac{1}{d}\sum_{j=1}^dX_{0,ij}\in\sqbrac{p^*-\delta_{\lambda}, p^*+\delta_{\lambda}}}\notag\\
    &\le \exp\sqbrac{-d\min\ebrac{D_{\rm KL}\brac{p^*-\delta_\lambda\mid\mid  b},D_{\rm KL}\brac{p^*+\delta_\lambda\mid\mid  a}}},
\end{align}
provided that $\delta_\lambda< \min\ebrac{a-p^*,p^*-b}$. It remains to notice that $p^*$ satisfies $ D_{\rm KL}\brac{p^*\mid\mid  a}= D_{\rm KL}\brac{p^*\mid\mid  b}$ and
\begin{align*}
   a-p^*=D_{\rm KL}\brac{a\mid\mid  b}\sqbrac{\log\frac{a(1-b)}{b(1-a)}}^{-1},\qquad p^*-b={D_{\rm KL}\brac{b\mid\mid  a}}\sqbrac{\log\frac{a(1-b)}{b(1-a)}}^{-1}.
\end{align*}
The proof is completed by combining \eqref{eq:barcheck-final-bound} and  $\PP(Z_{0,i}^*\ne Z_{1,i}^*)\le \varepsilon$.

\subsection{Proof of Lemma \ref{lem:barinflam-prob-lcm}}\label{pf-lem:barinflam-prob-lcm}
We fix $i\in[n]$ throughout the proof. We adopt the notation $L_{\sf LCM}(u)$, $\Delta_{\sf LCM}$ $F_\lambda(u;X)$ and $\Delta_\lambda(X)$ defined in the proof of \Cref{lem:checkbar-prob-lcm}. By definition,  we get
\begin{align}\label{eq:Delta_inf-exp}
    \Delta_\infty(X)=F_\infty(1;X)-F_\infty(-1;X)=\Delta_{\sf LCM}-\frac{2\mu }{\sigma^2}X.
\end{align}
 By definition, $\bar Z_{0,i}^\infty\ne \bar Z_{0,i}^\lambda$ iff $\sgn( \Delta_\infty(\bar X_{0,i}))\ne \sgn ( \Delta_\lambda(\bar X_{0,i})) $. In light of \eqref{eq:Delta_inf-exp} and \eqref{eq:Delta_lambda-exp},  this is equivalent to 
 \begin{align*}
    \ebrac{\frac{2\mu}{\sigma^2}\bar X_{0,i}<\Delta_{\sf LCM}<-\lambda,\quad\text{or}\quad \lambda<\Delta_{\sf LCM}<\frac{2\mu}{\sigma^2}\bar X_{0,i}}.
 \end{align*}
We first consider the case $Z^*_{0,i}=1$, for which we have
 \begin{align*}
   \PP_{1,\pm 1}\brac{\lambda<\Delta_{\sf LCM}<\frac{2\mu}{\sigma^2}\bar X_{0,i}}& =\frac{1}{2} \PP_{1,\pm 1}\brac{ \lambda<\Delta_{\sf LCM}<\frac{2\mu}{\sigma^2}(\mu+\sigma V_{1,i})}\\
    &\le \frac{1}{2} \PP_{1,\pm 1}\brac{\Delta_{\sf LCM}>\lambda}\Phi\brac{-\frac{\sigma}{2
    \mu}\lambda+\frac{\mu}{
    \sigma}}.
 \end{align*}
 On the other hand,
  \begin{align*}
   \PP_{1,\pm1}\brac{\frac{2\mu}{\sigma^2}\bar X_{0,i}<  \Delta_{\sf LCM}<-\lambda}& =\frac{1}{2} \PP_{1,\pm 1}\brac{ \frac{2\mu}{\sigma^2}(\mu+\sigma V_{1,i})<\Delta_{\sf LCM}<-\lambda}\\
    &\le \frac{1}{2} \PP_{1,\pm 1}\brac{\Delta_{\sf LCM}<-\lambda}\Phi\brac{-\frac{\sigma}{2
    \mu}\lambda-\frac{\mu}{
    \sigma}}.
 \end{align*}
 Following the notation in  \Cref{lem:checkbar-prob-lcm}, we get
\begin{align*}
    \PP_{1,\pm 1}\brac{\Delta_{\sf LCM}>\lambda}=\PP_{1,\pm 1}\brac{\frac{1}{d}\sum_{j=1}^dX_{0,ij}<  p^*-\delta_\lambda}\le \exp\sqbrac{-d D_{\rm KL}\brac{p^*-\delta_\lambda\mid\mid a}}.
\end{align*}
On the other hand, we get
\begin{align*}
    \PP_{1,\pm 1}\brac{\Delta_{\sf LCM}<-\lambda}=\PP_{1,\pm 1}\brac{\frac{1}{d}\sum_{j=1}^dX_{0,ij}> p^*+\delta_\lambda}\le \exp\sqbrac{-d D_{\rm KL}\brac{p^*+\delta_\lambda\mid\mid a}\II\brac{\delta_\lambda>   a-p^*}}.
\end{align*}
We then consider the case $Z^*_{0,i}=-1$, for which we similarly have
 \begin{align*}
   \PP_{-1,\pm 1}\brac{\lambda<\Delta_{\sf LCM}<\frac{2\mu}{\sigma^2}\bar X_{0,i}}
    & \le \frac{1}{2} \PP_{-1,\pm 1}\brac{\Delta_{\sf LCM}>\lambda}\Phi\brac{-\frac{\sigma}{2
    \mu}\lambda-\frac{\mu}{
    \sigma}},\\
    \PP_{-1,\pm1}\brac{\frac{2\mu}{\sigma^2}\bar X_{0,i}<  \Delta_{\sf LCM}<-\lambda}
    &\le \frac{1}{2} \PP_{-1,\pm 1}\brac{\Delta_{\sf LCM}<-\lambda}\Phi\brac{-\frac{\sigma}{2
    \mu}\lambda+\frac{\mu}{
    \sigma}}.
 \end{align*}
 Moreover, we have 
\begin{align*}
    \PP_{-1,\pm 1}\brac{\Delta_{\sf LCM}>\lambda}&\le \exp\sqbrac{-d D_{\rm KL}\brac{p^*-\delta_\lambda\mid\mid b}\II(\delta_\lambda>p^*-b)},\\
    \PP_{-1,\pm 1}\brac{\Delta_{\sf LCM}<-\lambda}&\le \exp\sqbrac{-d D_{\rm KL}\brac{p^*+\delta_\lambda\mid\mid b}}.
\end{align*}

\subsection{Proof of Lemma \ref{lem:barstar-prob-lcm}}\label{pf-lem:barstar-prob-lcm}
We fix $i\in[n]$ throughout the proof. From the proof of \Cref{lem:barinflam-prob-lcm} we get
\begin{align*}
 \bar Z_{0,i}^\infty=\begin{cases}
        1, & \Delta_\infty(\bar X_{0,i})< 0\\
        -1, & \Delta_\infty(\bar X_{0,i})> 0
    \end{cases}.
\end{align*}
Consider the case $Z_{0,i}^*=1$, we then have
\begin{align*}
    \PP_{1,\pm 1}\brac{\Delta_\infty(\bar X_{0,i})> 0}&=\PP_{1,\pm 1}\brac{\Delta_{\sf LCM}-\frac{2\mu}{\sigma^2}\bar X_{0,i}>0}\\
    &=\PP_{1,\pm 1}\brac{\log \frac{a(1-b)}{b(1-a)} \sum_{j=1}^dX_{0,ij}+\frac{2\mu \bar X_{0,i}}{\sigma^2}<    d\log\frac{1-b}{1-a}}.
\end{align*}
By Markov's inequality, we get for any $t\ge 0$,
\begin{align*}
    &\PP_{1,\pm 1}\brac{\log \frac{a(1-b)}{b(1-a)} \sum_{j=1}^dX_{0,ij}+\frac{2\mu \bar X_{0,i}}{\sigma^2}<  d\log\frac{1-b}{1-a}}
    \\
    &\le e^{t d\log\frac{1-b}{1-a}}\brac{\EE e^{-t\log\frac{a(1-b)}{b(1-a)}B}}^d\brac{\EE e^{-t\cdot 2Z\mu/\sigma^2}},
\end{align*}
where $B\sim \text{Ber}(a)$, $Z\sim N(\mu,\sigma^2)$ and $B\perp Z$. Some direct calculations lead to $\EE e^{-t\log\frac{a(1-b)}{b(1-a)}B}=1-a\sqbrac{1-\brac{\frac{b(1-a)}{a(1-b)}}^t}$ and $\EE e^{-t\cdot 2Z\mu/\sigma^2}=2(t^2-t)\mu^2/\sigma^2$. 
As a result,
\begin{align*}
   \PP\brac{\log \frac{a(1-b)}{b(1-a)} \sum_{j=1}^dX_{0,ij}+\frac{2\mu \bar X_{0,i}}{\sigma^2}\ge d\log\frac{1-b}{1-a}}\le \exp\brac{-J_{a,b,\lambda}(t)},
\end{align*}
for any $t\ge 0$, where we define
\begin{align*}
    J_{a,b,\lambda}(t):=\brac{d\log\frac{1-b}{1-a}}t-d\log\sqbrac{1+a\sqbrac{\brac{\frac{b(1-a)}{a(1-b)}}^t-1}}-\frac{2\mu^2}{\sigma^2}(t^2-t).
\end{align*}
The proof is completed by taking $t=1/2$.


\bibliographystyle{plain}
\bibliography{references}

\end{document}